\documentclass[11pt,a4paper]{article}
\pdfoutput=1
\usepackage{lmodern}
\usepackage{jheppub}
\usepackage{amsmath,amsfonts,amssymb,bbm, mathtools, mathrsfs}
\usepackage{hyperref, csquotes}
\usepackage{graphicx, color, svg,subcaption}
\usepackage{tikz,pgf,tikz-cd,float}
\usepackage{physics,tensor}
\usepackage{rotating}
\usetikzlibrary{external}

\definecolor{jred}{rgb}{0.8,0,0}
\definecolor{jgreen}{rgb}{0,0.7,0}
\definecolor{jblue}{rgb}{0,0,0.8}

\usetikzlibrary{arrows,patterns,shapes,positioning,fit}
\tikzstyle{c} =	[coordinate]
\tikzstyle{vc} = [circle, draw=black, line width=1pt, inner sep=0pt, minimum size=2mm]
\tikzstyle{v} =  [circle, draw=black, line width=.2pt, fill=black, inner sep=0pt, minimum size=1.5mm]
\tikzstyle{vb} =[circle, draw=black, line width=.1pt, fill=jred, inner sep=0pt, minimum size=1mm]
\tikzstyle{vh} =[circle, draw=black, line width=.1pt, fill=jblue, inner sep=0pt, minimum size=1.5mm]
\tikzstyle{vs} =[coordinate]%[circle, draw=black, line width=.1pt, fill=jgreen, inner sep=0pt, minimum size=1mm]
\tikzstyle{e} =	[draw=jred,line width=1pt]
\tikzstyle{eb}= [draw=jgreen,line width=1pt]
\tikzstyle{es}= [dashed, line width=1pt]
\tikzstyle{ev}= [draw=jgreen,line width=1pt]
\tikzstyle{f} = [line width=0.01pt,dotted,fill=blue, fill opacity=.1]
\tikzstyle{cs}= [ellipse, draw=black, line width=.1pt, fill=jred, inner sep=1pt, minimum size=2mm]%[draw=none,fill=jred, fill opacity=.7]

\newcommand{\twomelon}{
\begin{tikzpicture}[scale=2,x=2ex,y=2ex,baseline={([yshift=-.59ex]current bounding box.center)}] 
\scriptsize{
\node [vb, label=above:]	(b1)	at (1,1)	{};
\node [vb, label=below:]	(b2)	at (1,-1)	{};
\path
    (b1)    edge    [ev,bend left=30]   (b2)
    (b1)    edge    [ev,bend right=30]  (b2);
}
\end{tikzpicture}
}

\newcommand{\twochain}{
\begin{tikzpicture}[scale=2,x=2ex,y=2ex,baseline={([yshift=-.59ex]current bounding box.center)}] 
\scriptsize{
\node [vb, label=above:]	(b1)	at (0,0)      {};
\node [vb, label=below:]	(b2)	at (1.156,0)      {};
\node [vb, label=above:]	(b3)	at (1.734,1)	  {};
\node [vb, label=below:]	(b4)	at (1.156,2)	  {};
\node [vb, label=above:]	(b5)	at (0,2)  {};
\node [vb, label=below:]	(b6)	at (-0.578,1)  {};
\path
    (b1)    edge    [ev]    (b2)
    (b2)    edge    [ev]    (b3)
    (b3)    edge    [ev]    (b4)
    (b4)    edge    [ev]    (b5)
    (b5)    edge    [ev]    (b6)
    (b6)    edge    [dashed]     (b1);
}
\end{tikzpicture}
}

\newcommand{\threemelon}{
\begin{tikzpicture}[scale=2,x=2ex,y=2ex,baseline={([yshift=-.59ex]current bounding box.center)}] 
\scriptsize{
\node [vb, label=above:]	(b1)	at (1,1)	{};
\node [vb, label=below:]	(b2)	at (1,-1)	{};
\path
    (b1)    edge    [ev]                (b2)
    (b1)    edge    [ev,bend left=30]   (b2)
    (b1)    edge    [ev,bend right=30]  (b2);
}
\end{tikzpicture}
}

\newcommand{\twomelonchain}{
\begin{tikzpicture}[scale=2,x=2ex,y=2ex,baseline={([yshift=-.59ex]current bounding box.center)}] 
\scriptsize{
\node [vb, label=above:]	(b1)	at (0,0)      {};
\node [vb, label=below:]	(b2)	at (1.156,0)      {};
\node [vb, label=above:]	(b3)	at (1.734,1)	  {};
\node [vb, label=below:]	(b4)	at (1.156,2)	  {};
\node [vb, label=above:]	(b5)	at (0,2)  {};
\node [vb, label=below:]	(b6)	at (-0.578,1)  {};
\path
    (b1)    edge    [ev,bend left=30]    (b2)
    (b1)    edge    [ev,bend right=30]    (b2)
    (b2)    edge["$c$"']    [ev]     (b3) 
    (b3)    edge    [ev,bend right=30]    (b4)
    (b3)    edge    [ev,bend left=30]    (b4)
    (b4)    edge    [ev]    (b5)
    (b5)    edge    [ev,bend left=30]    (b6)
    (b5)    edge    [ev,bend right=30]    (b6)
    (b6)    edge    [dashed]     (b1);
}
\end{tikzpicture}
}

\newcommand{\tetrahedron}{
\begin{tikzpicture}[scale=2,x=2ex,y=2ex,baseline={([yshift=-.59ex]current bounding box.center)}] 
\scriptsize{
\node [vb, label=above:]	(b1)	at (-1,-1)	{};
\node [vb, label=below:]	(b2)	at (1,-1)	{};
\node [vb, label=below:]	(b3)	at (1,1)	{};
\node [vb, label=below:]	(b4)	at (-1,1)	{};
\path
    (b1)    edge    [ev]    (b2)
    (b2)    edge    [ev]    (b3)
    (b3)    edge    [ev]    (b4)
    (b4)    edge    [ev]    (b1)
    (b1)    edge    [ev]    (b3)
    (b4)    edge    [ev]    (b2);
}
\end{tikzpicture}
}

\newcommand{\kthreethree}{
\begin{tikzpicture}[scale=2,x=2ex,y=2ex,baseline={([yshift=-.59ex]current bounding box.center)}] 
\scriptsize{
\node [vb, label=above:]	(b1)	at (-1.5,-1)	{};
\node [vb, label=below:]	(b2)	at (0,-1)	{};
\node [vb, label=below:]	(b3)	at (1.5,-1)	{};
\node [vb, label=above:]	(b4)	at (1.5,1)	{};
\node [vb, label=below:]	(b5)	at (0,1)	{};
\node [vb, label=below:]	(b6)	at (-1.5,1)	{};
\path
    (b1)    edge    [ev]    (b4)
    (b1)    edge    [ev]    (b5)
    (b1)    edge    [ev]    (b6)
    (b2)    edge    [ev]    (b4)
    (b2)    edge    [ev]    (b5)
    (b2)    edge    [ev]    (b6)
    (b3)    edge    [ev]    (b4)
    (b3)    edge    [ev]    (b5)
    (b3)    edge    [ev]    (b6);
}
\end{tikzpicture}
}

\newcommand{\fourmelon}{
\begin{tikzpicture}[scale=2,x=2ex,y=2ex,baseline={([yshift=-.59ex]current bounding box.center)}] 
\scriptsize{
\node [vb, label=above:]	(b1)	at (1,1)	{};
\node [vb, label=below:]	(b2)	at (1,-1)	{};
\path
    (b1)    edge    [ev, bend left=40]  (b2)
    (b1)    edge    [ev, bend right=40]  (b2)
    (b1)    edge    [ev,bend left=15]   (b2)
    (b1)    edge    [ev,bend right=15]  (b2);
}
\end{tikzpicture}
}

\newcommand{\threemelonchain}{
\begin{tikzpicture}[scale=2,x=2ex,y=2ex,baseline={([yshift=-.59ex]current bounding box.center)}] 
\scriptsize{
\node [vb, label=above:]	(b1)	at (0,0)      {};
\node [vb, label=below:]	(b2)	at (1.156,0)      {};
\node [vb, label=above:]	(b3)	at (1.734,1)	  {};
\node [vb, label=below:]	(b4)	at (1.156,2)	  {};
\node [vb, label=above:]	(b5)	at (0,2)  {};
\node [vb, label=below:]	(b6)	at (-0.578,1)  {};
\path
    (b1)    edge    [ev,bend left=30]    (b2)
    (b1)    edge    [ev,bend right=30]    (b2)
    (b1)    edge    [ev]    (b2)
    (b2)    edge["$c$"']    [ev]     (b3) 
    (b3)    edge    [ev,bend right=30]    (b4)
    (b3)    edge    [ev,bend left=30]    (b4)
    (b3)    edge    [ev]    (b4)
    (b4)    edge    [ev]    (b5)
    (b5)    edge    [ev,bend left=30]    (b6)
    (b5)    edge    [ev,bend right=30]    (b6)
    (b5)    edge    [ev]    (b6)
    (b6)    edge    [dashed]     (b1);
}
\end{tikzpicture}
}

\newcommand{\necklacechain}{
\begin{tikzpicture}[scale=2,x=2ex,y=2ex,baseline={([yshift=-.59ex]current bounding box.center)}] 
\scriptsize{
\node [vb, label=above:]	(b1)	at (0,0)      {};
\node [vb, label=below:]	(b2)	at (1.156,0)      {};
\node [vb, label=above:]	(b3)	at (1.734,1)	  {};
\node [vb, label=below:]	(b4)	at (1.156,2)	  {};
\node [vb, label=above:]	(b5)	at (0,2)  {};
\node [vb, label=below:]	(b6)	at (-0.578,1)  {};
\path
    (b1)    edge    [ev,bend left=30]   (b2)
    (b1)    edge    [ev,bend right=30]  (b2)
    (b2)    edge    [ev,bend left=30]   (b3)
    (b2)    edge    [ev,bend right=30]  (b3)
    (b3)    edge    [ev,bend right=30]  (b4)
    (b3)    edge    [ev,bend left=30]   (b4)
    (b4)    edge    [ev,bend right=30]  (b5)
    (b4)    edge    [ev,bend left=30]   (b5)
    (b5)    edge    [ev,bend left=30]   (b6)
    (b5)    edge    [ev,bend right=30]  (b6)
    (b6)    edge    [dashed]     (b1);
}
\end{tikzpicture}
}

\newcommand{\cvf}{
\begin{tikzpicture}[scale=2,x=2ex,y=2ex,baseline={([yshift=-.59ex]current bounding box.center)}] 
\scriptsize{
\node [vb, label=above:]	(b1)	at (-1,1)	{};
\node [vb, label=below:]	(w1)	at (-1,-1)	{};
\node [vb, label=below:]	(b2)	at (1,-1)	{};
\node [vb, label=above:]	(w2)	at (1,1)	{};
\path
\foreach \i/\j in {1/2}{
   	(w\i) edge [ev]   (b\j)
	(b\i) edge [ev]  node [c] {} (w\j)
	}
\foreach \i in {1,2}{
	(w\i)   edge 	[ev]				(b\i)
	(w\i)	edge [ev,bend left=30]	(b\i)
	(w\i)	edge [ev,bend right=30]	(b\i)
	}
;
}
\end{tikzpicture}
}

\newcommand{\cvft}{
\begin{tikzpicture}[scale=2,x=2ex,y=2ex,baseline={([yshift=-.59ex]current bounding box.center)}] 
\node [vb]	(1)	at (-1,1)	{};
\node [vb]	(2)	at (-1,-1)	{};
\node [vb]	(3)	at (1,-1)	{};
\node [vb]	(4)	at (1,1)	{};
\path
\foreach \i/\j in {1/2,2/1,3/4,4/3}{
   (\i)	edge [ev,bend left=15]	(\j)
   (\i)	edge [ev,bend left=40]	(\j)
 };
\end{tikzpicture}
}

\newcommand{\cvfn}{
\begin{tikzpicture}[scale=2,x=2ex,y=2ex,baseline={([yshift=-.59ex]current bounding box.center)}] 
\scriptsize{
\node [vb, label=above:]	(b1)	at (-1,1)	{};
\node [vb, label=below:]	(w1)	at (-1,-1)	{};
\node [vb, label=below:]	(b2)	at (1,-1)	{};
\node [vb, label=above:]	(w2)	at (1,1)	{};
\path
\foreach \i/\j in {1/2,2/1}{
   	(w\i) edge [ev,bend left=30]  (b\j)
   	(w\i) edge [ev,bend right=30] (b\j)
	}
\foreach \i in {1,2}{
	(w\i)	edge [ev,bend left=30]  (b\i)
	(w\i)	edge [ev,bend right=30]	(b\i)
	}
 ;
}
\end{tikzpicture}
}

\newcommand{\cvfs}{
\begin{tikzpicture}[scale=2,x=2ex,y=2ex,baseline={([yshift=-.59ex]current bounding box.center)}] 
\scriptsize{
\foreach \i in {72,144,216,288,360}{
\begin{scope}[rotate=\i]
\node [vb]	(\i)	at (0,1.1056)	{};
\end{scope}
}
\path
\foreach \i in {72,144,216,288,360}{
    \foreach \j in {72,144,216,288,360}{
   	    (\i) edge [ev]  (\j)
	    }
    };
}
\end{tikzpicture}
}

\newcommand{\exoticone}{
\begin{tikzpicture}[scale=2,x=2ex,y=2ex,baseline={([yshift=-.59ex]current bounding box.center)}] 
\scriptsize{
\node [vb,label={[shift={(2.3,2.2)}]$c_2$}]	(b1)	at (-1,-1)	{};
\node [vb, label=below:]	(b2)	at (1,-1)	{};
\node [vb, label=below:]	(b3)	at (1,1)	{};
\node [vb, label=below:]	(b4)	at (-1,1)	{};
\path
    (b1)    edge    [ev,bend left=30]    (b4)
    (b1)    edge    [ev,bend right=30]    (b4)
    (b1)    edge["$c_1$"']    [ev]    (b2)
    (b1)    edge    [ev]    (b3)
    (b2)    edge    [ev]    (b4)
    (b2)    edge    [ev,bend left=30]    (b3)
    (b2)    edge    [ev,bend right=30]    (b3)
    (b3)    edge    [ev]    (b4);
}
\end{tikzpicture}
}

\newcommand{\exotictwo}{
\begin{tikzpicture}[scale=2,x=2ex,y=2ex,baseline={([yshift=-.59ex]current bounding box.center)}] 
\scriptsize{
\node [vb, label=above:]	(b1)	at (-2,-1)	{};
\node [vb, label=below:]	(b2)	at (-2,1)	{};
\node [vb, label=below:]	(b3)	at (-1,0)	{};
\node [vb, label=below:]	(b4)	at (1,0)	{};
\node [vb, label=below:]	(b5)	at (2,-1)	{};
\node [vb, label=below:]	(b6)	at (2,1)	{};
\path
    (b1)    edge    [ev,bend left=30]    (b2)
    (b1)    edge    [ev,bend right=30]    (b2)
    (b1)    edge    [ev]    (b2)
    (b2)    edge    [ev]    (b3)
    (b1)    edge["$c_1$"']    [ev]    (b3)
    (b3)    edge["$c_2$"]    [ev,bend left=30]    (b4)
    (b3)    edge["$c_3$"']    [ev,bend right=30]    (b4)
    (b4)    edge    [ev]    (b5)
    (b4)    edge    [ev]    (b6)
    (b5)    edge    [ev]    (b6)
    (b5)    edge    [ev,bend left=30]    (b6)
    (b5)    edge    [ev,bend right=30]    (b6);
}
\end{tikzpicture}
}

\definecolor{timelike}{RGB}{227, 11, 91}
\definecolor{spacelike}{RGB}{0, 128, 128}
\definecolor{lightlike}{RGB}{0, 25, 150}

\newcommand{\R}{\mathbb R}
\newcommand{\C}{\mathbb C}
\newcommand{\e}{\textrm{e}}
\newcommand{\SUT}{\mathrm{SU}(2)}

\newcommand{\SL}{\text{SL$(2,\C)$}}

\renewcommand{\TH}{\text{H}^3}

\newcommand{\dloc}{d_{\mathrm{loc}}}
\newcommand{\xiloc}{\xi_{\mathrm{loc}}}
\newcommand{\xinloc}{\xi_{\mathrm{nloc}}}

\newcommand{\vbr}{\vb*{\rho}}
\newcommand{\vbk}{\vb*{k}}

\newcommand{\vbg}{\vb*{g}}
\newcommand{\vbx}{\vb*{x}}

\newcommand{\sumint}{\;\;\mathclap{\displaystyle\int}\mathclap{\textstyle\sum}\;\;\;}

\begin{document}
\title{Scale invariance beyond criticality within the mean-field analysis of tensorial field theories}

\author[a,b]{Roukaya Dekhil,}
\emailAdd{roukaya.dekhil@unifi.it}

\author[b,c,d]{Alexander F. Jercher,}
\emailAdd{alexander.jercher@campus.lmu.de}

\author[b,e,f]{Daniele Oriti,}
\emailAdd{doriti@ucm.es}

\author[b,c,d]{Andreas G. A. Pithis}
\emailAdd{andreas.pithis@uni-jena.de}

\affiliation[a]{Universit\'a degli Studi di Firenze,\\ Piazza di San Marco, 4, 50121 Firenze FI, Italy, EU}
\affiliation[b]{Munich Center for Quantum Science and Technology (MCQST),\\ Schellingstr. 4, 80799 M\"unchen, Germany, EU}
\affiliation[c]{Arnold Sommerfeld Center for Theoretical Physics,\\ Ludwig-Maximilians-Universit\"at München \\ Theresienstrasse 37, 80333 M\"unchen, Germany, EU}
\affiliation[d]{Theoretisch-Physikalisches Institut, Friedrich-Schiller-Universit\"{a}t Jena\\ Max-Wien-Platz 1, 07743 Jena, Germany, EU}
\affiliation[e]{Departamento de F\'isica Te\'orica, Facultad de Ciencias F\'isicas,\\ Universidad Complutense de Madrid, \\ Plaza de las Ciencias 1, 28040 Madrid, Spain, EU}

\affiliation[f]{Department of Physics, Shanghai University, 99 Shangda Rd, 200444, Shanghai, P.R. China}

\date{\today}

\begin{abstract}
{
We continue the series of articles on the application of Landau-Ginzburg mean-field theory to unveil the basic phase structure of tensorial field theories which are characterized by combinatorially non-local interactions. Among others, this class covers tensor field theories (TFT) which lead to a new class of conformal field theories highly relevant for investigations on the AdS/CFT conjecture. Moreover, it also encompasses models within the tensorial group field theory (TGFT) approach to quantum gravity. Crucially, in the infrared we find that the effective mass of the modes relevant for the critical behavior vanishes not only at criticality but also throughout the entire phase of non-vanishing vacuum expectation value due to the non-locality of the interactions. As a consequence, one encounters there the emergence of scale invariance on configuration space which is potentially enhanced to conformal invariance thereon.
}
\end{abstract}

\maketitle

\section{Introduction}\label{sec:Introduction}

Tensorial field theories\footnote{We utilize the inclusive term tensorial field theory here as a general term for a larger class of theories of tensor fields and do not imply that the models we consider necessarily possess tensor-invariant interactions, often associated with the label “tensorial”. With this overarching terminology, we hope that our work is accessible across relevant research communities.} with local variables are characterized by combinatorially non-local interactions. They offer a promising framework beyond local, standard-model-type quantum and statistical field theories. They generalize Kontsevich models~\cite{kontsevich1992intersection,grosse2014self,grosse2006noncommutative,Rivasseau:2007ab} from matrix to tensor fields of rank $r>2$. For colored simplicial and tensor-invariant interactions, their Feynman diagrams are in fact bijective to $r$-dimensional discrete manifolds~\cite{Gurau:2011xp,GurauBook}. Roughly speaking, such theories fall then into two different categories for which the tensor indices either stem from Lie group data or are simply $\mathbb{Z}$-valued. 

The first case allows the investigation of dynamical random and quantum geometries as in tensorial group field theory (TGFT) where the Lie group data corresponds to holonomies discretized on the Feynman graphs which turns the tensorial fields into so-called \emph{group} fields~\cite{Freidel:2005qe,Oriti:2006se,Oriti:2011jm,Carrozza:2013oiy,Carrozza:2016vsq,Gielen:2016dss}. The local arguments of these may then be motivated by discretized scalar fields typically used as a matter reference frame~\cite{Oriti:2016qtz,Li:2017uao,Gielen:2018fqv}. The perturbative expansion of the TGFT partition function corresponds to a sum over discretizations and geometries, thus yielding a quantum geometric interpretation. Thereby, TGFT models are closely related to many other quantum gravity approaches, such as loop quantum gravity (LQG)~\cite{Ashtekar:2004eh}, spin foam models~\cite{Perez:2003vx,Perez:2012wv,Engle:2023qsu,Livine:2024hhc}, simplicial gravity~\cite{Bonzom:2009hw,Baratin:2010wi,Baratin:2011tx,Baratin:2011hp,Finocchiaro:2018hks} or dynamical triangulations~\cite{Ambjorn:2013tki,Loll:2019rdj}.

%In contrast, 
In the case of tensor field theories (TFT), the \emph{tensor} fields only possess local variables
%, typically living in a flat Euclidean $\mathbb{R}^d$ space. 
which typically live in flat Euclidean space $\R^d$. Such models are also interesting for quantum gravity research since they lead to non-trivial conformal field theories (CFT)~\cite{Rosenhaus:2018dtp,Gurau:2019qag,Benedetti:2020seh,Harribey:2022esw,Gurau:2024nzv} which can in turn be used to probe the AdS/CFT conjecture~\cite{Maldacena:1997re,Maldacena:1998im,Gubser:1998bc,Witten:1998qj}.

%Clearly, for 
Given the rich connections of tensorial field theories to quantum gravity approaches it is crucial to understand their properties under coarse-graining from microscopic to macroscopic scales via renormalization. This allows to gain access to their continuum limit and their phase diagram beyond perturbation theory.
 
To this end, complex functional renormalization group (FRG) techniques known within the context of local field theory~\cite{Delamotte:2007pf,dupuis2021nonperturbative} have been applied to matrix and tensor models as well as to tensorial field theories~\cite{Sfondrini:2010zm,Eichhorn:2013isa,Eichhorn:2014xaa,Eichhorn:2017xhy,Eichhorn:2018phj,Eichhorn:2018ylk,Eichhorn:2019hsa,Castro:2020dzt,Eichhorn:2020sla,Benedetti:2015et,BenGeloun:2015ej,BenGeloun:2016kw,Benedetti:2016db,Carrozza:2016vsq,Carrozza:2016tih,Carrozza:2017vkz,BenGeloun:2018ekd,Pithis:2020sxm,Pithis:2020kio,Baloitcha:2020lha,Lahoche:2022gkz,Geloun:2023ray}. To bypass such involved analyses, a coarse account of their phase structure can be obtained via Landau-Ginzburg mean-field theory~\cite{Kopietz:2010zz,zinn2021quantum,wilson1983renormalization,hohenberg2015introduction}. It efficiently approximates the microscopic details of models in an effective theory defined from the mesoscale to the macroscale.
%which would otherwise be captured in their full-fledged non-perturbative RG study which would also capture the impact of fluctuations up to the miscroscale. 
Its application to tensorial field theories is not immediate due to the combinatorial non-locality of the interactions and the presence of both local and non-local variables. In spite of these hindrances, the implementation of the Landau-Ginzburg method to this class of theories has substantially advanced in recent years and it has been demonstrated that it is sufficient to scrutinize their basic phase properties~\cite{Pithis:2018eaq,Marchetti:2020xvf,Marchetti:2022igl,Marchetti:2022nrf}.

In this work, we continue and further develop this line of research to better understand the basic structure of the phase diagram of such theories and map the general conditions under which critical behaviour occurs therein. 
%As is well-known from this method in the local field theory context on non-compact domain, in the infrared the phase space is separated by the critical hypersurface of vanishing effective mass into a region where it is positive and negative. 
It is well-known from local field theories that in the infrared, the phase space is separated into two regions where the effective mass is positive and negative, respectively. At the interface of these two regions is the critical hypersurface of vanishing effective mass~\cite{sachs2006elements,Kopietz:2010zz,hohenberg2015introduction,zinn2021quantum}. Our novel observation here is that for tensorial field theories the effective mass of the modes relevant for the critical behavior does not only vanish at criticality but throughout the entire phase 
%defined by the 
of non-vanishing vacuum expectation value.
%of the tensorial field. 
We show that this 
%interesting complexity
structure of the phase space is a direct consequence of the non-local 
%character of their interactions 
interactions
and carefully demonstrate how the Landau-Ginzburg method must be adapted 
%and how it works 
in this scenario then. This result holds for both the TGFT and TFT cases. Such effectively massless and free theories then possess scale invariance on configuration space which, depending on the domain of the tensorial field, is further enhanced to conformal invariance thereon. We also emphasize that the extension to the TFT class of theories is also a novel aspect of our work. In particular, we find that the aforementioned result on the effective mass does hold for any $N$ of the tensor fields with rank $r\geq 2$ and for any such interaction. This is to be contrasted with previous works which investigated the phase properties and conformal symmetry of respective models mostly at rank $r=3$ and $5$ by solving their Dyson-Schwinger equations at large $N$~\cite{Witten:2016iux,Gurau:2016lzk,Klebanov:2016xxf,Klebanov:2018nfp,Giombi:2017dtl,Carrozza:2015adg,Klebanov:2018fzb,Benedetti:2018goh,Benedetti:2018ghn,Benedetti:2019eyl,Gurau:2019qag,Benedetti:2021wzt,Harribey:2022esw,Jepsen:2023pzm,Berges:2023rqa,Gurau:2024nzv}.
We conjecture that the effective masslessness and conformal invariance of the TFT models considered by us is the mean-field level equivalent of the infrared results obtained in those works. This seems to indicate that we are able to deliver the same results as the latter approach beyond large $N$, however, with the simpler (though restricted) machinery of the mean-field approximation.

The setup of this article is as follows: In Section~\ref{sec:example}, we study in detail the explicit example of a rank $4$ tensorial group field theory with local variables taking values in $\R^{\dloc}$, non-local variables in $\mathrm{U}(1)$ and quartic melonic interactions. Along this pedagogical example, we study how the effective mass vanishes and extend the mean-field approach to this case by introducing a regularization. In the sections thereafter, the argument is generalized first to TGFTs on $\text{U}(1)^r\times\mathbb{R}^{\text{d}_{\text{loc}}}$~\cite{Marchetti:2020xvf}, $\mathbb{R}^r\times\mathbb{R}^{\text{d}_{\text{loc}}}$~\cite{Marchetti:2020xvf} and $\mathrm{SL}(2,\mathbb{C})^4\times \mathbb{R}^{\text{d}_{\text{loc}}}$~\cite{Marchetti:2022igl,Marchetti:2022nrf} the latter of which corresponds to the Barrett-Crane (BC) TGFT model for Lorentzian quantum gravity in $4d$. Then the mean-field analysis is applied to tensor fields at rank $r$ on $\mathbb{R}^{\text{d}_{\text{loc}}}$ at any $N$. We introduce relevant details of the respective theory spaces \emph{in situ}. Lastly, the most general case of tensorial field theory is considered in Section~\ref{sec:The general case}. In particular, we prove that the vanishing of the effective mass for the relevant modes is a generic feature of such theories. Subsequently, we discuss how scale invariance appears in such models on configuration space and under which conditions it is further enhanced to conformal symmetry thereon. Finally, we summarize our results in
Section~\ref{sec:conclusion}, discuss the limitations of our work and propose future investigations.

\section{A simple example: Melonic TGFT on $\vb{U(1)^4\times\mathbb{\mathbf{R}}}^{\textbf{d}_{\textbf{loc}}}$}\label{sec:example}

\subsection{Model setup}\label{sec:U1 TGFT}

%Tensorial group field theories with local and non-local variables promote the tensorial indices of TFTs to be dynamic, meaning that the field is extended to be a function of group elements belonging to a given Lie group $G$. Such models are suited to study quantum geometries since the additional Lie group data allows encoding additional geometric information dressing the graph structure dual to the tensor fields. Captured by the general action in Eq.~\eqref{eq:general action}, this involves in particular a Laplacian acting on functions of the group elements belonging to $G$. The measure $\dd{g}$ corresponds in this case to the Haar measure on the group $G$~\cite{Ruehl1970}.  

The phase structure of TGFTs for Abelian groups $G=$U$(1),\R$ with and without local arguments has been studied in~\cite{Marchetti:2020xvf} and~\cite{BenGeloun:2015ej,Benedetti:2015et,Benedetti:2016db,BenGeloun:2016kw,BenGeloun:2018ekd,Pithis:2020sxm,Pithis:2020kio,Geloun:2023ray}, using Landau-Ginzburg mean-field theory and the FRG methodology, respectively. In the following, we critically examine the Landau-Ginzburg analysis to show that the resulting effective mass actually vanishes for the modes relevant for the critical behavior.

The specific TGFT model we consider in this section is defined by a real-valued field $\Phi:\mathrm{U}(1)^4\times\R^{\dloc}\longrightarrow \R$ with non-local arguments $\vb*{\theta} = (\theta_1,\dots ,\theta_4)\in\mathrm{U}(1)^4$ and local ones, $\vbx = (x_1,\dots, x_{\dloc})\in\R^{\dloc}$.\footnote{A way to motivate the $\dloc$ local $\mathbb{R}$-valued variables from the TGFT perspective is to introduce discretized scalar fields typically employed as a matter reference frame~\cite{Li:2017uao,Oriti:2016qtz}.} The action governing the field, $S[\Phi] = K[\Phi]+V[\Phi]$, consists of a kinetic and an interaction term, where the former is given by
\begin{equation}\label{eq:U1 kinetic}
K[\Phi] = \frac{1}{2}\int\limits_{\mathrm{U}(1)^4}\dd{\vb*{\theta}}\int\limits_{\R^{\dloc}}\dd{\vbx}\Phi(\vb*{\theta},\vbx)\left[\mu-\sum_{c=1}^4\Delta_\theta^c-\Delta_x\right]\Phi(\vb*{\theta},\vbx).
\end{equation}
The integration on $\mathrm{U}(1)$ is defined as the integration on the circle $S^1$ and $\dd{\vbx}$ is the standard Lebesgue measure on $\R^{\dloc}$. The Laplace operator $\Delta_\theta^c$ acts on the variable $\theta_c\in\mathrm{U}(1)$ and $\Delta_x$ denotes the Laplace operator on $\R^{\dloc}$. The parameter $\mu\in\R$ plays the role of a mass term.

The defining property of TGFTs are the combinatorially non-local interactions in the group variables which we choose in this section to be quartic melonic, pictorially represented in Fig.~\ref{figure:melon}. Explicitly, $V[\Phi]$ is defined as
\begin{equation}\label{eq:U1 interaction}
V[\Phi]= \int\dd[8]{\theta}\int\dd{\vbx}\Phi(\theta_1,\theta_2,\theta_3,\theta_4,\vbx)\Phi(\theta_5,\theta_6,\theta_7,\theta_4,\vbx)\Phi(\theta_5,\theta_6,\theta_7,\theta_8,\vbx)\Phi(\theta_1,\theta_2,\theta_3,\theta_8,\vbx),
\end{equation}
where the $\vbx\in\R^{\dloc}$ enter as local variables with point-like interactions and $\theta_c$ enter as non-local variables contracted according to the quartic melonic combinatorics. The interaction term can be more compactly denoted by introducing a trace notation~\cite{Marchetti:2020xvf}
\begin{equation}
V[\Phi] = \lambda\int\dd{\vbx}\Tr_\gamma\left[\Phi(\vb*{\theta},\vbx)^4\right],
\end{equation}
where $\lambda$ is a coupling parameter. The $\theta$-integrations are captured by $\Tr_\gamma$, where $\gamma$ denotes the vertex graph~\cite{Oriti:2014yla}, depicted in Fig.~\ref{figure:melon}, which dictates the contraction pattern of the non-local variables. The fourth power indicates that there are four fields $\Phi$ entering the interaction. In this notation, the interactions are straightforwardly generalized by choosing different vertex graphs $\gamma$ with different number of fields $n_\gamma$ and it will therefore be heavily utilized in the sections hereafter. 

\begin{figure}[h]
    \centering
        \cvf
    \caption{Pictorial representation of the quartic melonic interaction. The green half-edges indicate pairwise convolution of the non-local variables $\vb*{\theta}$ and red vertices represent the fields $\Phi(\vb*{\theta},\vbx)$. The vertex structure of local variables $\vbx$ is suppressed here which corresponds to standard point-like interaction.}
    \label{figure:melon}
\end{figure}

\subsection{Landau-Ginzburg mean-field theory}\label{sec:U1 Landau}

The Landau-Ginzburg method was originally introduced to study phase transitions in local field-theoretic descriptions of lattice systems~\cite{Kopietz:2010zz,zinn2021quantum}. Their dynamics are captured by the effective action $S[\Phi]$ which is a functional in odd and/or even powers of the field $\Phi$ and its gradient. In general, it is very difficult to exactly compute the partition function $Z$ of such systems. This issue is bypassed by Landau-Ginzburg mean-field theory, which fundamentally assumes that the scales of the system can be separated so that one can average over its microscopic details. Consequently, one works with an effective theory valid from the mesoscopic to the macroscopic regime. Hence, the field $\Phi$ corresponds to an averaged quantity which describes general features of the system like symmetries and the dimensionality of the domain. Its dynamics are typically modelled by the classical action. Thus, averaged-over microscopic details are captured by the field and values
of couplings therein. The coarse-graining of microscopically different theories oftentimes leads to the same description on larger scales, referred to as universality.

More precisely, Landau-Ginzburg mean-field theory studies the impact of quadratic fluctuations $\delta\Phi$ over uniform background field configurations $\Phi_0$, the latter of which correspond to saddle points of the classical action. One can then solve for the correlation function of those fluctuations and retrieve from it the correlation length $\xi$ which defines the scale beyond which the fluctuations fade away exponentially. It extends from the mesoscale to the macroscale and diverges at criticality. To finally check self-consistency of this approximation, one has to verify that the strength of the fluctuations relative to the background remains small up to the scale set by the correlation length. This is known as the Levanyuk-Ginzburg criterion~\cite{levanyuk1959contribution,ginzburg1961some}. For local scalar field theories on $\mathbb{R}^d$ this allows for the extraction of the upper critical dimension $d_{\text{crit}}$ beyond which this method is valid. For $d<d_{\text{crit}}$ a non-perturbative treatment via the Wilsonian renormalization group formalism is needed instead which would also capture the impact of fluctuations up to the microscale~\cite{wilson1983renormalization,dupuis2021nonperturbative}. We also refer to Ref.~\cite{Benedetti:2014gja} for a pedagogical discussion of this topic.

As demonstrated in the series of works~\cite{Pithis:2018eaq,Marchetti:2020xvf,Marchetti:2022igl,Marchetti:2022nrf,Dekhil:2024djp}, this method can be transferred to TGFTs with the main challenge being the hybrid character of such models including local and non-local variables. Hereafter, we summarize the core aspects of this construction, exemplified with the model introduced in the previous section. This will allow us to point out the mechanism of the vanishing effective mass for the relevant modes and to advance the mean-field method to this scenario. The specific construction here will then serve as the basis to generalize the observation to tensorial field theories.

\subsection{Linearization of the model}\label{sec:U1 linearization}

To start off, one computes the classical equations of motion for the field $\Phi$,
\begin{equation}
\left[\mu-\sum_{c=1}^4\Delta_\theta^c-\Delta_x\right]\Phi(\vb*{\theta},\vbx)+\lambda \fdv{V[\Phi]}{\Phi(\vb*{\theta},\vbx)} = 0,
\end{equation}
with 
\begin{equation}
\begin{aligned}
&\fdv{V[\Phi]}{\Phi(\theta_1,\dots,\theta_4,\vbx)}\\[7pt]
=& 4\lambda\int\dd{\theta_5}\dots\dd{\theta_8}\Phi(\theta_5,\theta_6,\theta_7,\theta_4,\vbx)\Phi(\theta_5,\theta_6,\theta_7,\theta_8,\vbx)\Phi(\theta_1,\theta_2,\theta_3,\theta_8,\vbx).
\end{aligned}
\end{equation}
The equations of motion are then evaluated on constant field configurations, $\Phi_0 = \mathrm{const}$, leading to the mean-field equations
\begin{equation}
\mu\Phi_0+4\lambda\Phi_0^3 V_G^4 = 0,
\end{equation}
where $V_G = 2\pi R$ arises from empty group integrations on $\mathrm{U}(1)$ with $R$ being the radius of the circle. Notice that these factors appear precisely due to the non-localities of the interaction. For $\mu>0$, the mean-field solution is simply given by $\Phi_0 = 0$. For $\mu < 0$, the minimum of the theory is instead given by the non-vanishing value
\begin{equation}\label{eq:ex mf sol}
\Phi_0 = \pm\left(\frac{\abs{\mu}}{4\lambda}\right)^{\frac{1}{2}}V_G^{-2}.
\end{equation}

To proceed with the Landau-Ginzburg description, one allows for fluctuations around the mean-field solution,
\begin{equation}
\Phi(\vb*{\theta},\vbx) = \Phi_0+\delta\Phi(\vb*{\theta},\vbx),
\end{equation}
and linearizes the equations of motion in $\delta\Phi$. This yields
\begin{equation}
\left(\mu-\sum_{c=1}^4\Delta_\theta^c-\Delta_x\right)\delta\Phi(\vb*{\theta},\vbx)+\eval{\fdv{V[\Phi]}{\Phi(\vb*{\theta},\vbx)}}_{\Phi_0+\delta\Phi} = 0,
\end{equation}
with 
\begin{equation}
\begin{aligned}
&\eval{\fdv{V[\Phi]}{\Phi(\vb*{\theta},\vbx)}}_{\Phi_0+\delta\Phi} \\[7pt]
=& 4\lambda\Phi_0^2\int\dd{\theta_5}\dots\dd{\theta_8}\left(\delta\Phi(\theta_5,\theta_6,\theta_7,\theta_4)+\delta\Phi(\theta_5,\theta_6,\theta_7,\theta_8)+\delta\Phi(\theta_1,\theta_2,\theta_3,\theta_8)\right).
\end{aligned}
\end{equation}
Inserting the mean-field solution given in Eq.~\eqref{eq:ex mf sol}, this expression can be re-written as
\begin{equation}
\eval{\fdv{V[\Phi]}{\Phi(\vb*{\theta},\vbx)}}_{\Phi_0+\delta\Phi} = -\mu\int\dd{\tilde{\vb*{\theta}}}\chi(\vb*{\theta},\tilde{\vb*{\theta}})\delta\Phi(\tilde{\vb*{\theta}},\vbx),
\end{equation}
with $\chi(\vb*{\theta},\tilde{\vb*{\theta}})$ capturing the non-local character of the interactions which can be derived as the Hessian of the interaction term, as detailed in~\cite{Marchetti:2020xvf}. In this example, it is explicitly given by
\begin{equation}
\chi(\vb*{\theta},\tilde{\vb*{\theta}}) = V_G^{-4}+\delta(\theta_4,\tilde{\theta}_4)V_G^{-3}+\delta(\theta_1,\tilde{\theta}_1)\delta(\theta_2,\tilde{\theta}_2)\delta(\theta_3,\tilde{\theta}_3)V_G^{-1}.
\end{equation}
Then, the equations of motion at first order in $\delta\Phi$ can be given the form
\begin{equation}
\int\dd{\tilde{\vb*{\theta}}}\left[\delta(\vb*{\theta},\tilde{\vb*{\theta}})\left(-\sum_c\Delta^c_{\tilde{\theta}}-\Delta_x\right)+b(\vb*{\theta},\tilde{\vb*{\theta}})\right]\delta\Phi(\tilde{\vb*{\theta}},\vbx) = 0,
\end{equation}
where $\Delta_{\tilde{\theta}}^c$ acts on the variable $\theta_c$ and with $b(\vb*{\theta},\tilde{\vb*{\theta}})$ being defined as
\begin{equation}
b({\vb*{\theta}},\tilde{\vb*{\theta}}) = \mu\left(\delta(\vb*{\theta},\tilde{\vb*{\theta}})-\chi(\vb*{\theta},\tilde{\vb*{\theta}})\right).
\end{equation}
This determines an \textit{effective} action for the perturbations $\delta\Phi$, which is given by
\begin{equation}\label{eq:ex effective action group rep}
S_{\mathrm{eff}}[\delta\Phi] = \int\dd{\vb*{\theta}}\dd{\tilde{\vb*{\theta}}}\int\dd{\vbx}\delta\Phi(\vb*{\theta},\vbx)\left[\delta(\vb*{\theta},\tilde{\vb*{\theta}})\left(-\sum_c\Delta_{\tilde{\theta}}-\Delta_x\right)+b(\vb*{\theta},\tilde{\vb*{\theta}})\right]\delta\Phi(\tilde{\vb*{\theta}},\vbx).
\end{equation}

\subsection{Vanishing effective mass}\label{sec:U1 vanishing mass}

To compute correlations of fluctuations, it is expedient to perform a Fourier transform on the total domain $\mathrm{U}(1)^4\times\R^{\dloc}$. In the non-local variables $\vb*{\theta}$, the field is a function of periodicity $2\pi R$ allowing for a decomposition in a standard Fourier series. Similarly, the Fourier transform of the local variables is the standard Fourier transform on $\R^{\dloc}$. Overall, it is thus defined as
\begin{equation}
\delta\Phi(\vb*{\theta},\vb*{x}) = \sum_{\vb*{p}\in\mathbb{Z}^4}\int\frac{\dd{\vb*{k}}}{(2\pi)^{\dloc}}\delta\Phi_{\vb*{p}}(\vb*{k})\e^{i\vb*{p}\vb*{\theta}/R}\e^{i\vb*{k}\vb*{x}}.
\end{equation}
As a result, the effective action in Eq.~\eqref{eq:ex effective action group rep} is represented as
\begin{equation}\label{eq:ex effective action spu=in rep}
S_{\mathrm{eff}}[\delta\Phi] = (2\pi R)^4\sum_{\vb*{p}}\int\frac{\dd{\vbk}}{(2\pi)^{\dloc}}\delta\Phi_{-\vb*{p}}(-\vbk)\left(\frac{1}{R^2}\vb*{p}^2+\vbk^2+b_{\vb*{p}}\right)\delta\Phi_{\vb*{p}}(\vbk),
\end{equation}
where the spectrum of the Laplace operator on $\mathrm{U(1)}$ enters as $p_c^2/R^2$ and $b_{\vb*{p}}$ is the bi-local function $b(\vb*{\theta},\tilde{\vb*{\theta}})$ in Fourier representation. It is in this representation that the role of $b_{\vb*{p}}$ as \textit{effective} mass is most apparent, 
\begin{equation}
b_{\vb*{p}} = \mu(1-\chi_{\vb*{p}}).
\end{equation}
Notice that in quartic local field theories, the effective mass is a constant and it is simply given by $b = 2\abs{\mu}$~\cite{Kopietz:2010zz,zinn2021quantum,Benedetti:2014gja}. However, as explained in detail in~\cite{Marchetti:2020xvf}, the non-local interactions that enter the expression of the effective mass through the Hessian contribution $\chi_{\vb*{p}}$ take a specific form that explicitly depends on the representation labels $\vb*{p}$. Explicitly, for the present type of quartic interactions, we find
\begin{equation}\label{eq:ex chi}
\chi_{\vb*{p}} = \prod_{c=1}^4\delta_{p_c,0}+\prod_{d\neq 4}\delta_{p_d,0}+\delta_{p_4,0},
\end{equation}
with the Kronecker-$\delta$ on $\mathrm{U}(1)$-momenta defined as
\begin{equation}
\delta_{p,p'} = \frac{1}{(2\pi R)}\int\dd{\theta}\e^{i(p-p')\theta/R}.
\end{equation}

We observe from Eq.~\eqref{eq:ex chi} that the effective mass $b_{\vb*{p}}$ contains products of projections onto zero modes where the momenta $p$ are set to zero. To elucidate the consequences of this structure, we expand the effective action in terms of zero modes, i.e., we split the $p$-sums into a contribution where $p=0$ and the rest with $p\neq 0$~\cite{Marchetti:2020xvf}. As a result, 
\begin{equation}
\begin{aligned}
S_{\mathrm{eff}}[\delta\Phi] &= (2\pi R)^4\sum_{s=0}^4\sum_{(c_1\dots c_s)}\sum_{\vb*{p}_{4-s}}\int\frac{\dd{\vbk}}{(2\pi)^{\dloc}}\delta\Phi_{-\vb*{p}_{4-s}}(-\vbk)\\[7pt]
&\times \left(\frac{1}{R^2}\sum_{c=c_{s+1}}^{c_4}p_c^2+\vbk^2+b_{c_1\dots c_s}\right)\delta\Phi_{\vb*{p}_{4-s}}(\vbk),
\end{aligned}
\end{equation}
where $s$ is the number of zero modes, $c_1\dots c_s$ are the slots where the zero modes are injected and $\vb*{p}_{4-s} = (p_{c_{s+1}},\dots p_{c_4})$ is a short-hand notation for the remaining $4-s$ non-zero momenta. The effective mass $b_{\vb*{p}}$ evaluated on $s$ zero modes in the slots $c_1\dots c_s$ is denoted by $b_{c_1\dots c_s}$ which is constant in the remaining $4-s$ variables.

At this point, we make the main observation of this work: The effective mass $b_{c_1\dots c_s}$ vanishes for particular zero mode injections. It is furthermore only positive if $s=4$ and negative in all the remaining cases. Let us look at the three cases separately in the following.

First, if $s=4$ zero modes are considered, it follows immediately from Eq.~\eqref{eq:ex chi} that $\chi_{(0,0,0,0)} = 3$. As a result, the effective mass is given by $b_{c_1\dots c_s} = 2\abs{\mu}$ which corresponds to the result obtained in local field theory. The effective action for this configuration therefore corresponds to a stable parabola opened upwards. 

Second, for $s<4$, there exist configurations of zero modes for which $\chi_{\vb*{p}} = 0$ and thus $b_{c_1\dots c_s} = -\abs{\mu} < 0$. Investigating Eq.~\eqref{eq:ex chi} closely, one finds that these correspond to $s<3$ with $p_4\neq 0$, so for instance $(p_1,p_2,0,p_4)$ or $(0,0,p_3,p_4)$. We denote the set of these configurations for a particular number of zero modes $s$ as $\bar{\mathcal{O}}_s$. In this case, the effective action takes the form of an unstable parabola that is opened downwards. The consequences of this negative effective mass on the correlations depend on the specific properties of the non-local domain. Since we consider in this section $\mathrm{U}(1)$ which is compact, $b_{c_1\dots c_s} < 0$ does not bear physical consequences, as demonstrated above. However, in Sec.~\ref{sec:non-compact limit} we study the non-compact limit $R\rightarrow \infty$ and show that $b<0$ leads to an oscillating correlation function. We argue that such configurations should be excluded when studying the critical behavior. 

Lastly, there are configurations for $s_0\leq s<4$ for which the effective mass vanishes, i.e. $b_{c_1\dots c_s} = 0$. Here, $s_0$ is the number below which the effective mass cannot vanish which is characteristic for the type of interactions considered. For the quartic melonic interaction one has $s_0=1$. The effective mass vanishes if $\chi = 1$ which, by following Eq.~\eqref{eq:ex chi}, holds for any configuration with $s=3$ or $s<3$ and $p_4 = 0$. We denote the set of these configurations for a particular number of zero modes $s$ as $\mathcal{O}_s$. In these instances, the theory becomes \textit{effectively} massless for any finite value of $\mu<0$ in the broken phase and we therefore observe the emergence of scale- or even conformal symmetry on the \textit{residual} domain $\mathrm{U}(1)^{4-s}\times\R^{\dloc}$.  We elaborate on these symmetries in Sec.~\ref{sec:Symmetries of effectively massless theories}. This feature is distinctive for the non-local interactions of the theory and we argue that this holds for general tensorial field theories, representing the main novelty of the present article.

\subsection{Correlations and Ginzburg-\textit{Q}}\label{sec:U1 Q}

Correlations of the fluctuations are captured by the two-point function 
\begin{equation}
C(\vb*{\theta},\vbx) = \langle \delta\Phi(\vb*{0},\vb*{0})\delta\Phi(\vb*{\theta},\vbx)\rangle
\end{equation}
which is defined as the inverse kinetic kernel of the effective action above. In Fourier space, the correlator is simply given by the multiplicative inverse of the kinetic kernel of the effective action in Eq.~\eqref{eq:ex effective action spu=in rep}. This yields the real space correlator defined as an integral
\begin{equation}
C(\vb*{\theta},\vbx) = \frac{1}{(2\pi R)^4}\sum_{\vb*{p}}\int\frac{\dd{\vbk}}{(2\pi)^{\dloc}}\frac{\e^{i\vb*{p}\vb*{\theta}/R}\,\e^{i\vbk\vbx}}{\frac{1}{R^2}\vb*{p}^2+\vbk^2+b_{\vb*{p}}}.
\end{equation}

To determine the behavior of fluctuations in the local and non-local variables separately, it has been suggested in~\cite{Marchetti:2020xvf} to define local and non-local correlation functions, obtained by integrating out the complementary set of variables. For local variables, we thus define
\begin{equation}
C(\vbx) = \int\dd{\vb*{\theta}}C(\vb*{\theta},\vbx),
\end{equation}
which amounts to setting the $\mathrm{U}(1)$-momenta to zero,
\begin{equation}\label{eq:U1 local correlator}
C(\vbx) = \int\frac{\dd{\vbk}}{(2\pi)^{\dloc}}\frac{\e^{i\vbk\vbx}}{\vbk^2+b_{\vb*{0}}}.
\end{equation}
Here, $b_{\vb*{0}}$ is the effective mass evaluated on four zero modes which, according to Eq.~\eqref{eq:ex chi}, results in $b_{\vb*{0}} = 2\abs{\mu}$. Following~\cite[6.566, Formula 2.]{GradshteynBook}, this integral explicitly evaluates to
\begin{equation}
C(\vbx) = \frac{2^d\pi^{\frac{d}{2}}}{(2\pi)^d \ell^{d-2}}\left(\sqrt{b_{\vb*{0}}}\ell\right)^{\frac{d-2}{2}}K_{\frac{d-2}{2}}(\sqrt{b_{\vb*{0}}}\ell),
\end{equation}
with $d\equiv\dloc$, $\ell\equiv \abs{\vbx}$ and $K_\alpha(z)$ the modified Bessel function of the second kind~\cite{GradshteynBook}. The result corresponds precisely to the correlation function of a local field theory on $\R^{\dloc}$ which is consistent with the fact that we have integrated out all the non-local variables. From the asymptotic behavior of the correlation function at large distances $\ell\gg 1$,
\begin{equation}
C(\vbx)\underset{\ell\gg 1}{\longrightarrow} \e^{-\sqrt{b_{\vb*{0}}}\ell}
\end{equation}
the local correlation length is identified as $\xiloc^{-2} = b_{\vb*{0}}$.

To obtain a correlation function in the non-local variables, the local variables are integrated out,
\begin{equation}
C(\vb*{\theta}) = \int\dd{\vbx}C(\vb*{\theta},\vbx).
\end{equation}
Since U$(1)$ is a compact group, no long-range behavior can be determined. Thus, a correlation length (if existent), can only be given by the system size, where $\xinloc = \frac{\pi}{\sqrt{6}}R < 2\pi R$ was computed explicitly in~\cite{Marchetti:2020xvf} via the second-moment method. An expansion of $C(\vb*{\theta})$ in zero modes shows
\begin{equation}\label{eq:non-loc corr U1}
C(\vb*{\theta}) = \frac{1}{(2\pi R)^4}\left[\frac{1}{b_{\vb*{0}}}+\sum_{s=0}^{3}\sum_{(c_1\dots c_s)}\sum_{\vb*{p}_{4-s}}\frac{\prod\limits_{c=c_{s+1}}^{c_4}e^{ip_c\theta_c/R}}{\frac{1}{R^2}\sum_c p_c^2+b_{c_1\dots c_s}}\right] \underset{\mu\rightarrow 0}{\longrightarrow} \frac{1}{(2\pi R)^4 b_{\vb*{0}}},
\end{equation}
which is dominated by the first term in the limit $\mu\rightarrow 0$, even if the effective mass $b_{c_1\dots c_s}$ vanishes. Notice that this is a direct consequence of the presence of the Laplace operator, which introduces a preference for small values of $\vb*{p}^2$. 

The Landau-Ginzburg analysis is completed by validating the self-consistency of the mean-field approach which is the case if the fluctuations $\delta\Phi$ remain small relative to the mean-field solution $\Phi_0$ in the considered regime. The relative size of fluctuations is quantified by the Levanyuk-Ginzburg-$Q$, defined as
\begin{equation}\label{eq:general Q}
Q = \frac{\int_{\Omega_{\xi}}\dd{\vb*{\theta}}
\dd{\vbx}\expval{\delta\Phi(\vb*{0},\vb*{0})\delta\Phi(\vb*{\theta},\vbx)}}{\int_{\Omega_\xi}\dd{\vb*{\theta}}\dd{\vbx}\Phi_0^2}.
\end{equation}
The integration domain $\Omega_\xi\subset \mathrm{U}(1)^4\times\R^{\dloc}$ is restricted by the local and non-local correlation lengths $\xiloc$ and $\xinloc$, respectively, which have been extracted above. Finally, Landau-Ginzburg mean-field theory is said to be valid near criticality if $Q\ll 1$ in the limit $\mu\rightarrow 0$.

The numerator of $Q$ is computed by extending the local integration to the whole of $\R^{\dloc}$ justified by the exponential suppression at large local distances $\ell  \gg 1$. The $\theta$-integrations are bounded by the constant non-local correlation length $\xinloc < 2\pi R$. As a result, we obtain
\begin{equation}
\int_{[0,\xinloc]^4}\dd{\vb*{\theta}}C(\vb*{\theta}) = \sum_{s=0}^4\frac{\xinloc^s}{(2\pi R)^4}\sum_{(c_1\dots c_s)}\int\dd{\vb*{\theta}_{4-s}}\sum_{\{p_{c}\}}\frac{\prod_{c}\e^{ip_c\theta_c}}{\frac{1}{R^2}\sum_{c}p_{c}^2+b_{c_1\dots c_s}}.
\end{equation}
Independent of  $b_{c_1\dots c_s}$ being negative, zero, or positive, the dominant contribution is given by the $s=4$ case. Consequently, the numerator of $Q$ is given by
\begin{equation}
 \int_{[0,\xinloc]^4}\dd{\vb*{\theta}}C(\vb*{\theta}) = \left(\frac{\xinloc}{2\pi R}\right)^4\frac{1}{b_{\vb*{0}}}.   
\end{equation}
Together with the denominator of $Q$,
\begin{equation}
\int_{\Omega_\xi}\dd{\vb*{\theta}}\dd{\vbx}\Phi_0^2 = \frac{\abs{\mu}}{4\lambda }\left(\frac{\xinloc}{2\pi R}\right)^4\xiloc^{\dloc},
\end{equation}
we finally obtain 
\begin{equation}
Q\sim \lambda\xiloc^{4-\dloc}.
\end{equation}
In the limit $\abs{\mu}\rightarrow 0$, or equivalently $\xiloc\rightarrow \infty$, the Ginzburg-$Q$ is small if the local dimension satisfies $\dloc > 4$, thus exactly reproducing the behavior of local field theories. These results are supported by~\cite{Geloun:2023ray} where in the infrared, a Gaussian fixed point is found if the local dimension takes values above the critical dimension, irrespective of the number of the group copies. In other words, the effective dimension of the model reduces to that of the local theory.

The non-local correlation length $\xinloc$ and the radius of U$(1)$ do not enter $Q$ as they refer to the compact variables that do not affect the critical behavior of the theory. In particular, the effective mass $b_{c_1\dots c_s}$ does not appear in the final expression of $Q$ which reflects two defining properties of the model:
\begin{enumerate}
    \item The group domain is compact, therefore not allowing for long-range correlations. Thus, a phase transition, characterized by a transition between long-range and short-range correlations cannot occur for the non-local variables. This is in agreement with standard results from local field theory~\cite{strocchi2005symmetry,Benedetti:2014gja,zinn2021quantum}. It is therefore also expected that for the non-Abelian $G = \SUT$, which would be a basic ingredient to relate to loop quantum gravity~\cite{Oriti:2014yla}, the same behavior would be obtained. 
    \item The kinetic term we considered includes a Laplace operator which leads to a dominance of low-spin contributions. This can be seen explicitly in the non-local correlation where the $s=4$ zero mode term dominates. Notice that this is different from the tensor field theory case treated in Sec.~\ref{sec:tensor field theories} where no such Laplacian is present and where the vanishing effective mass therefore plays a distinguished role in charting the phase structure. 
\end{enumerate}

Exploring the physical consequences of a negative or vanishing effective mass has been obstructed by the two points listed above. In the next section, we therefore study in detail the non-compact limit of the present $\mathrm{U}(1)$-model and extend the Landau-Ginzburg method to vanishing effective masses. This is intended to serve as a guiding example for the more general cases of TGFTs on non-compact group domains and tensor field theories, considered in Secs.~\ref{sec:general TGFT on R} and~\ref{sec:tensor field theories}.

\subsection{The non-compact limit}\label{sec:non-compact limit}

In this section, we study the non-compact limit of the theory defined in the previous section. More explicitly, we study the limit of infinite radius of $\mathrm{U}(1)$, $R\rightarrow \infty$, resulting in a tensorial theory defined on $\R^4\times \R^{\dloc}$.

Clearly, the behavior of the local correlation function $C(\vb*{x})$ is unaffected by the non-compact limit, showing an asymptotic exponential suppression with a correlation length $\xiloc^{-2} = b_{\vb*{0}}$.

In contrast, the non-local correlation function given in Eq.~\eqref{eq:non-loc corr U1} explicitly depends on the variables $\vb*{\theta}$. In the non-compact limit, we perform a continuum approximation of the discrete sums over the $p_c$,
\begin{equation}
\frac{1}{(2\pi R)}\sum_p f(p/R) \longrightarrow \int\frac{\dd{\tilde{p}}}{2\pi}f(\tilde{p}),
\end{equation}
with $\tilde{p} = p/R$, where the above approximation is valid for any function $f$. In this case, the non-local correlation function in Eq.~\eqref{eq:non-loc corr U1} reads as
\begin{equation}
C(\vb*{\theta}) = \sum_{s=0}^{4}\sum_{(c_1,\dots, c_s)}C_{c_1\dots c_s}(\vb*{\theta}_{4-s}) = \sum_{s=0}^{4}\sum_{(c_1,\dots, c_s)}\frac{1}{(2\pi R)^s}\int\frac{\dd{\tilde{\vb*{p}}}_{4-s}}{(2\pi)^{4-s}}\frac{\prod_v \e^{i\tilde{p}_{c_v}\theta_{c_v}}}{\sum_v\tilde{p}_{c_v}^2+b_{c_1\dots c_s}}.
\end{equation}
Notice that the residual non-local correlation function $C_{c_1\dots c_s}$ characterizing the $4-s$ zero modes contribution is similar to the local correlation function in Eq.~\eqref{eq:U1 local correlator} with the difference being that the effective mass here is given by $b_{c_1\dots c_s}$ and not by $b_{\vb*{0}}$. As a result, the explicit form of the effective mass $b_{c_1\dots c_s}$ evaluated on $s$ zero modes is important. The values that $b_{c_1\dots c_s}$ takes are the same as those discussed in Sec.~\ref{sec:U1 vanishing mass} and are unaffected by the non-compact limit which is only performed in the $4-s$ residual variables $\vb*{p}_{4-s}$. 

At four zero modes, the non-local correlation function is a constant that scales as $b_{\vb*{0}}^{-1}$, which is consistent with the result of the previous section.

For $s<4$ zero modes and negative effective mass, that is $(c_1\dots c_s)\in\bar{\mathcal{O}}_s$, the non-local correlation function $C(\vb*{\theta})$ exhibits an oscillatory behavior with polynomial decay in the regime of large distances, $\abs{\vb*{\theta}}\gg 1$. This behavior is reminiscent of anti-ferromagnets which exhibit a vanishing vacuum expectation value and long-range correlations at all scales $\mu$. The corresponding effective action is given by a downward opened parabola which leads to an unstable Gaussian as the partition function. In conclusion, these modes do not affect the critical behavior around $\mu\rightarrow 0$, characterized by a transition between short-range and long-range correlations. 
%It has therefore been argued in~\cite{Marchetti:2020xvf} to exclude these modes from the analysis of the critical behavior. 
As we discuss below, results from an FRG analysis of the present model~\cite{Geloun:2023ray} justify this exclusion a posteriori. 

In the case of $s_0\leq s<4$ zero modes that are injected at arguments $(c_1,\dots, c_s)\in\mathcal{O}_s^\gamma$, the effective mass vanishes. Instead of computing the correlation function by direct integration, we introduce a regularization of the effective mass, given by\footnote{Notice that the introduction of a regularization for the effective mass is similar to working with a vanishing effective mass and introducing cutoffs for the $Q$-integration. Because such cutoffs carry a dimension, a similar relation in terms of the parameter $\mu$ must be given in this case. The resulting Ginzburg-$Q$ turns out to be the same.}
\begin{equation}
b = \lim\limits_{\epsilon\rightarrow 0^+}\epsilon\abs{\mu}.
\end{equation}
We regularize the effective mass with the inclusion of $\mu$ in order to have a dimensionless small regulator $\epsilon$. As a result, the non-local correlation function exhibits an asymptotic exponential behavior in the $4-s$ residual variables $\vb*{\theta}_{4-s}\in\R^{4-s}$ and we extract the correlation length $\xinloc^{-2} = \epsilon\abs{\mu}$. In the limit $\epsilon\rightarrow 0$, the non-local correlation length diverges even for finite $\mu < 0$.
%, indicating the conformal invariance of the residual theory. 
%For the following computation of $Q$, we exclude all those zero modes $(c_1,\dots,c_s)\in\bar{\mathcal{O}}_s^\gamma$ for which $b_{c_1\dots c_s}<0$.

As the remaining step of this section, we compute the Ginzburg-$Q$ as defined in Eq.~\eqref{eq:general Q} in the non-compact limit. In this case, the integration range for computing $Q$ is given by $\Omega_\xi = [-\xinloc,\xinloc]^4\times[-\xiloc,\xiloc]^{\dloc}\subset\R^{4}\times\R^{\dloc}$. For the numerator of Eq.~\eqref{eq:general Q}, we find
\begin{equation}
\int_{\Omega_\xi}\dd{\vb*{\theta}}\dd{\vbx}C(\vb*{\theta},\vbx) = \int\dd{\vb*{\theta}}C(\vb*{\theta}) = \sum_{s = s_0}^4\left(\frac{\xinloc}{2\pi R}\right)^s\sum_{(c_1,\dots, c_s)\in\mathcal{O}_s^\gamma}\frac{1}{b_{c_1\dots c_s}}.   
\end{equation}
where we took into consideration that the local correlation function exhibits an asymptotic exponential suppression. For the denominator, we insert the background mean-field solution in the large-$R$ limit, yielding
\begin{equation}
\int\dd{\vb*{\theta}}\dd{\vbx}\Phi_0^2 = \left(\frac{\xinloc}{2\pi R}\right)^4\xiloc^{\dloc}\frac{\abs{\mu}}{4\lambda}.
\end{equation}
Combing the obtained expressions for the denominator and numerator, the parameter $Q$ evaluates to
\begin{equation}
Q\sim \lambda\xiloc^{-\dloc}\sum_{s = s_0}^{4}\left(\frac{\xinloc}{2\pi R}\right)^{s-4}\sum_{(c_1,\dots,c_s)\in\mathcal{O}_s^\gamma}\frac{1}{b_{c_1\dots c_s}}.
\end{equation}
%\begin{equation}
%    Q\sim \lambda^{\frac{2}{n_\gamma-2}}\abs{\mu}^{-\frac{2}{n_\gamma-2}}\xiloc^{-\dloc}\sum_{s = s_0}^{r}\left(\frac{\xinloc}{2\pi R}\right)^{s-r}\sum_{(c_1,\dots,c_s)\in\mathcal{O}_s^\gamma}\frac{1}{b_{c_1\dots c_s}}.
%\end{equation}
%
Taking the limit of large radius $R$ before the limit $\epsilon,\mu\rightarrow 0$, the sum over the number of zero modes is dominated by the smallest summand, here being the $s_0$-term, which in particular implies that $b_{c_1\dots c_s} = \epsilon\abs{\mu}$. As elaborated in Sec.~\ref{sec:U1 vanishing mass}, for the quartic melonic interaction one has $s_0 = 1$.
%We focus on this case and investigate its physical implications.  
After absorbing the radius into the coupling $\lambda$, we define the re-scaled coupling as 
\begin{equation}
\bar{\lambda} =  (2\pi R)^{4-s_0} \lambda.
\end{equation}
%\begin{equation}
%\bar{\lambda} =  (2\pi R)^{\frac{(r-s_0)(n_\gamma-2)}{2}} \lambda,  
%\end{equation}
%
The Ginzburg-$Q$ parameter expressed in terms of $\epsilon$ and $\mu$ is finally given by
\begin{equation}
Q\sim \bar{\lambda}\epsilon^{-1+\frac{4-s_0}{2}}\abs{\mu}^{-\frac{1}{2}\left(4-\dloc-(4-s_0)\right)}.
\end{equation}
%\begin{equation}
%    Q\sim \bar{\lambda}^{\frac{2}{n_\gamma-2}}\epsilon^{-1+\frac{r-s_0}{2}}\abs{\mu}^{-\frac{1}{2}\left(2\frac{n_\gamma}{n_\gamma-2}-\dloc-(r-s_0)\right)}.
%\end{equation}
%
It is important to mention that if the following inequalities
\begin{equation}
4-\dloc-(4-s_0)\leq 0,\qquad -1+\frac{4-s_0}{2}\geq 0   
\end{equation}
%begin{equation}\label{eq:ineq Q TGFT R}
%2\frac{n_\gamma}{n_\gamma-2}-\dloc-(r-s_0)\leq 0,\qquad -1+\frac{r-s_0}{2}\geq 0,
%\end{equation}
%
are satisfied, the limits taken in $\epsilon$ and $\mu$ commute and $Q\rightarrow 0$ in the limits $\epsilon,\mu\rightarrow 0$. A critical dimension of the total domain $\dloc+(4-s_0)|_{\mathrm{crit}} = 4$ is identified. This is in agreement with the results for local quartic field theories which is expected since for a fixed number of zero modes $s_0$, the theory is effectively local on the domain $\R^{4-s_0}\times\R^{\dloc}$. Moreover, these results agree with the FRG analysis conducted in~\cite{Geloun:2023ray}, where in the infrared a Gaussian fixed point is found if the dimension of the total domain, $\dloc+(4-s_0)$, is above the critical dimension.

The computation of the Ginzburg-$Q$ in the non-compact limit concludes this section. Along the lines of a melonic TGFT on $\mathrm{U}(1)^4\times\R^{\dloc}$ we have presented an example for two novelties of the present work. First, the effective mass is non-positive for $s<4$ zero modes. Terms of negative effective mass are neglected in studying the critical behavior for the reasons given above, such that the relevant contributions to the correlation function and the Ginzburg-$Q$ are those of vanishing effective mass. Second, we have provided a regularization method to advance the Landau-Ginzburg analysis to the case of vanishing effective mass. In the next two sections, we show that the arguments developed in this section generalize to general tensorial field theories. For the sake of performing explicit computations, a set of exemplary models is studied in the upcoming sections.

\section{Towards Landau-Ginzburg theory of general tensorial field theories}\label{sec:towards general TFTs}

\subsection{TGFT on $G^r\times\R^{\dloc}$ with arbitrary single interaction}\label{sec:general TGFT}

In the following, we generalize the mean-field results of the previous section to fields of arbitrary rank $r$ and living on any Lie group $G$ and consider a single but arbitrary combinatorially non-local interaction. The field is thus defined as $\Phi:G^r\times\R^{\dloc}\longrightarrow \R$ with the non-local variables now collectively denoted as $\vbg = (g_1,\dots, g_r)\in G^r$. The kinetic term is the same as in Eq.~\eqref{eq:U1 kinetic} with the $\mathrm{U}(1)$-integrations and Laplacians replaced by the corresponding objects on $G^r$. 

To define the single but arbitrary interaction term $V[\Phi]$, we introduce a vertex graph $\gamma$ as in Ref.~\cite{Oriti:2014yla}, in which every vertex $v\in\gamma$ represents a field and every edge a non-local variable. Thus, $\gamma$ captures the contraction pattern of the non-local variables. In Fig.~\ref{figure:melon}, we encountered such a vertex graph for the quartic melonic interaction at rank $r=4$. Further examples of $r=4$ vertex graphs are shown in Fig.~\ref{figure:interactions} and more general graphs are depicted in Tab.~\ref{tab:chis}. 

\begin{figure}[h]
    \centering
        \cvft~~~~~~~~~
        \cvf~~~~~~~~~
        \cvfn~~~~~~~~~
        \cvfs
    \caption{From left to right: diagrammatic representation of vertex graphs $\gamma$ corresponding to double-trace melonic, melonic, necklace and simplicial interactions for rank-$4$ tensorial fields. The green half-edges indicate pairwise convolution of the non-local variables $\boldsymbol{g}$ and red vertices represent the fields $\Phi(\boldsymbol{g},\boldsymbol{x})$. The vertex structure of local variables $\boldsymbol{x}$ is suppressed here, which corresponds to standard point-like interaction.}
    \label{figure:interactions}
\end{figure}

Associated with this graph is a vertex set $\mathcal{V}_\gamma$ with cardinality $\abs{\mathcal{V}_\gamma} = n_\gamma$ that corresponds to the power of the fields in the interaction. We assume every $v\in\mathcal{V}_\gamma$ to have the same valency (given by the rank $r$) and we exclude loops, i.e. edges that start and end at the same vertex. Furthermore, we introduce the set $\mathcal{A}_v$ as the set of vertices $v'$ adjacent to $v$. These structures, as we will see below, are employed when computing the Hessian of the action within the Landau-Ginzburg approach. Given such a graph $\gamma$, the interaction term is then given by 
\begin{equation}
\label{eq: non-local-interaction}
V_\gamma[\Phi] = \lambda \int\dd{\vb*{x}}\Tr_\gamma\left[\Phi(\vbg,\vb*{x})^{n_\gamma}\right],
\end{equation}
where the trace denotes the integration over $\frac{r n_\gamma}{2}$ elements in $G$ according to $\gamma$. In Eq.~\eqref{eq:U1 interaction}, an explicit example for this notation has been given for the case of a quartic melonic interaction at $r=4$.

%The distinction between local and non-local variables is apparent in this form: The $n_\gamma$ fields are all evaluated at the same point $\vb*{x}\in\mathcal{M}$ and thus considered as local, while the variables $\vb*{g}\in G$ follow the non-local contraction pattern of $\gamma$. 

To appreciate the implications of Eq.~\eqref{eq: non-local-interaction}, we illustrate its quantum geometric interpretation in the context of TGFTs. Fundamental excitations of a rank-$r$ TGFT are interpreted as $(r-1)$-simplices which constitute the fundamental building blocks of discretized geometries. In this picture, interactions govern the combinatorial gluing of such elements to generate $r$-dimensional cellular complexes. Simplicial interactions, as depicted on the very right of Fig.~\ref{figure:interactions}, establish a direct connection of TGFTs to quantum gravity approaches such as LQG, spin foam models, simplicial gravity or dynamical triangulations. Tensor-invariant interactions, such as the three left diagrams in Fig.~\ref{figure:interactions}, arise from colorizing the group fields~\cite{Gurau:2009tw} and integrating out all but one color~\cite{Bonzom:2012hw}. This type of interactions is heavily studied in tensor models and tensor field theories as the generated Feynman graphs are bijective to cellular pseudo-manifolds~\cite{Gurau:2010nd,Gurau:2011xp,Bonzom:2012hw}.

Notice that $V_\gamma$ consists of a single interaction. A straightforward generalization consists of a sum of multiple interaction terms, all of which enter the action with a different coupling $\lambda_\gamma$. As it has been shown in~\cite{Marchetti:2020xvf,Marchetti:2022nrf}, the mean-field description admits a simple inclusion of multiple interactions as long as their power in fields, $n_\gamma$, is equal. We discuss such an extension briefly in Sec.~\ref{sec:multiple interactions}. In the most generic case of multiple interactions capturing different degrees $n_\gamma$, the mean-field equations are possibly complicated polynomial equations to which a solution might be difficult to find. In the remainder of this work we focus on the simpler case and give emphasis when going beyond it.

Following the steps detailed in Sec.~\ref{sec:U1 linearization}, one computes the classical equations of motions,
\begin{equation}
    \left[\mu+-\sum_{c=1}^r\Delta^c_g-\Delta_x\right]\Phi(\vbg,\vbx)+\lambda\sum_{v\in\mathcal{V}_\gamma}\Tr_{\gamma/v}\left[\Phi(\vbg,\vb*{x})^{n_\gamma - 1}\right] = 0,
\end{equation}
where the sum over vertices arises from the product rule when taking the functional derivative of $V_\gamma[\Phi]$. Here, $\Tr_{\gamma/v}$ denotes the contraction of those non-local variables which are not affected by the functional derivative of the field at vertex $v$. Evaluating the equations of motion on constant field configurations, $\Phi_0=\mathrm{const.}$, its solutions are given by $\Phi_0 = 0$ for $\mu>0$ and
\begin{equation}\label{eq:general mf sol}
\Phi_0 = \left(\frac{\abs{\mu}}{\lambda n_\gamma}\right)^{\frac{1}{n_\gamma-2}}\mathrm{V}_G^{-\frac{r}{2}},
\end{equation}
for $\mu<0$. In the example of $G=\mathrm{U}(1)$, $V_G=2\pi R$ has been given as the volume of the circle. Now, $V_G$ denotes the result of an empty $G$-integral, 
\begin{equation}
    \int_{G}\dd{g} = \mathrm{V}_G.
\end{equation}
If $G$ is non-compact, volume factors necessarily diverge. Instead of restricting to compact $G$, we regularize such divergences via a cut-off $L$ in the non-compact coordinates on $G$ and take the limit of $L\rightarrow \infty$ at the end of the computation. It is in this way that physically interesting models such as the spacelike Barrett-Crane model defined on $G = \SL$ with additional constraints, which we will discuss later in Sec.~\ref{sec:BC}, are contained within the ensuing analysis.\footnote{In~\cite{Marchetti:2022igl,Marchetti:2022nrf}, the non-compact domain $G=\SL$ is regularized to Spin$(4)$ and the non-compact limit is taken at the end of the computation, similar to the previous section for the groups $\R$ and $\mathrm{U}(1)$. This procedure turns out to be equivalent to our proposal of working with regularized volume factors from the outset.} 

Introducing fluctuations around the mean-field solution, $\Phi(\vbg,\vbx) = \Phi_0+\delta\Phi(\vbg,\vbx)$, the equations of motion are linearized in $\delta\Phi$ which yields
\begin{equation}
\int\dd{\tilde{\vbg}}\left[\delta(\vbg,\tilde{\vbg})\left(-\sum_c \Delta^c_{\tilde{g}}-\Delta_x\right)+b(\vbg,\tilde{\vbg})\right]\delta\Phi(\tilde{\vbg},\vbx) = 0,
\end{equation}
where $b(\vbg,\tilde{\vbg}) = \mu(\delta(\vbg,\tilde{\vbg})+\chi(\vbg,\tilde{\vbg}))$ with $\chi$ obtained from linearizing the interaction term in $\delta\Phi$. Notice that $\chi$ is computed in precisely the same way as shown in Sec.~\ref{sec:U1 linearization} and is given as the sum of products of $\delta$-functions on the group together with corresponding volume factors. Examples of $\chi$ in Fourier space for different vertex graphs are displayed in Tab.~\ref{tab:chis}.  

Using the Peter-Weyl decomposition for compact $G$ or the Plancherel decomposition for non-compact $G$, functions in $L^2(G)$ are decomposed into unitary irreducible representations of $G$ with representation matrices $D^{(j)}(g)$ which are often referred to as Wigner matrices~\cite{Ruehl1970}. The details of the labels $j$, in particular whether they are discrete or continuous, depend on the specific properties of the group. Taking the example of the previous section with $G=\mathrm{U}(1)$ or $\R$, the $j$ correspond either to the discrete momenta $p\in\mathbb{Z}$ on the circle or the continuous momenta $p\in\R$, respectively. In Sec.~\ref{sec:BC}, we consider $G=\SL$ where the $j$ are given by real numbers denoted as $\rho\in\R$. Together with the Fourier transform on $\R$, the fluctuation $\delta\Phi\in L^2(G^r\times\R^{\dloc})$ is expanded as
\begin{equation}
\delta\Phi(\vbg,\vbx) = \sumint\!\!{\raisebox{-2.5mm}{\scalebox{0.8}[0.8]{$\vb*{j}$}}}\;\int\frac{\dd{\vbk}}{(2\pi)^{\dloc}} \delta\Phi_{\vb*{j}}(\vbk)\prod_{c=1}^r D_G^{(j_c)}(g_c) \e^{i\vbk\vbx},
\end{equation}
where the details of the measure on the variables $\vb*{j}$ depends again on the details of $G$. In this decomposition, the effective action takes the form\footnote{Here, $-\vb*{j}$ denote the dual labels to $\vb*{j}$, which is a consequence of $\delta\Phi$ being real-valued.}
\begin{equation}
S_{\mathrm{eff}}[\delta\Phi] = \sumint\!\!{\raisebox{-2.5mm}{\scalebox{0.8}[0.8]{$\vb*{j},\vbk$}}}\; \delta\Phi_{\vb*{j}}(\vbk)\left[\sum_c\lambda_g(j_c)+\vbk^2+b_{\vb*{j}}\right]\delta\Phi_{-\vb*{j}}(-\vbk),
\end{equation}
with $\lambda_g(j_c)$ the eigenvalues of the Laplace operator $-\Delta^c_g$ acting on the Wigner matrices $D^{(j_c)}(g_c)$. The effective mass in Fourier representation is given by $b_{\vb*{j}}=\mu\left(1-\chi_{\vb*{j}}\right)$ with the function $\chi_{\vb*{j}}$ being of the general form
\begin{equation}
\chi_{\vb*{j}} = \sum_{s=0}^r\sum_{(c_1\dots c_s)}\tilde{\chi}^\gamma_{c_1\dots c_s}\prod_{c=c_1}^{c_s}\delta_{j_c,j_0},
\end{equation}
with $\tilde{\chi}$ combinatorial coefficients that depend on the graph structure encoded in $\gamma$ being are either zero or one~\cite{Marchetti:2020xvf}. The $\delta_{j_c,j_0}$ represents a Kronecker-$\delta$ on $G$, singling out the value $j_c=j_0$ for which $\lambda_g(j_0) = 0$. Following the designation of the previous section, we refer to this value as the \emph{zero mode}, obtained from the projection
\begin{equation}\label{eq:zero mode projection}
\frac{1}{\mathrm{V}_G}\int\dd{g}D_G^{(j)}(g) = \delta_{j,j_0}.
\end{equation}
For $G$ being a compact Lie group, the label $j_0$ corresponds to the trivial representation which is part of the decomposition of functions on $L^2(G)$ and is associated with the constant function. In the case of non-compact manifolds, such as $\R$ or $\SL$, the constant function is not square integrable, and therefore, the trivial representation is not part of the decomposition of the $L^2$-space into unitary irreducible representations. Thus, defining the symbol $\delta_{j,j_0}$ requires either a regularization via compactification, as proposed in~\cite{Marchetti:2020xvf,Marchetti:2022igl,Marchetti:2022nrf}, or a careful extension of the $L^2$-space to the space of hyperfunctions~\cite{Ruehl1970,hormander2015analysis}. The latter contains in particular constant field configurations which correspond to the trivial representation. For a TGFT model describing $4d$ Lorentzian quantum gravity with $G=\SL$ this is applied in Refs.~\cite{Marchetti:2020xvf,Marchetti:2022igl,Marchetti:2022nrf,Dekhil:2024djp}. 

The sum/integrals over the labels $j$ can be split into zero mode contributions
\begin{equation}
\begin{aligned}
S_{\mathrm{eff}}[\delta\Phi] &= \sum_{s=0}^r\sum_{(c_1\dots c_s)}\sumint\!\!{\raisebox{-2.5mm}{\scalebox{0.8}[0.8]{$\vb*{j}_{r-s}$}}}\;\int\frac{\dd{\vbk}}{(2\pi)^{\dloc}} \delta\Phi_{\vb*{j}_{r-s}}(\vbk)\\[7pt]
&\times\left[\sum_{c=c_{s+1}}^{c_r}\lambda_g(j_c)+\vbk^2+b_{c_1\dots c_s}\right]\delta\Phi_{-\vb*{j}_{r-s}}(-\vbk),
\end{aligned}
\end{equation}
where $b_{c_1,\dots c_s}$ is the effective mass with $s$ zero modes injected into the arguments $(c_1,\dots c_s)$. At $s=r$, the effective mass evaluates to $b_{c_1\dots c_s} = \abs{\mu}(n_\gamma-2)$ which is consistent with the results of local field theories. For $s<r$, there exist zero mode injections at $(c_1,\dots c_s)\in\bar{\mathcal{O}}_s^\gamma$ for which $b_{c_1\dots c_s} < 0$. In particular, if the function $\chi$ vanishes for such configurations, one finds $b_{c_1\dots c_s} = -\abs{\mu}$. For $s_0\leq s<r$ zero modes injected at $(c_1,\dots c_s)\in\mathcal{O}_s^\gamma$, the effective mass vanishes. Notice that for every graph $\gamma$, one can define a characteristic number $s_0$ below which $b_{c_1\dots c_s} < 0$~\cite{Marchetti:2020xvf}. For rank $4$, we find for instance $s_0 = 0$ for double-trace melonic, $s_0=1$ for melonic, $s_0=2$ for necklace and $s_0=3$ for simplicial interactions. 

The behavior of the effective mass observed in Sec.~\ref{sec:U1 vanishing mass} generalizes to an arbitrary interaction characterized by a vertex graph $\gamma$. We prove this statement in full generality in Sec.~\ref{sec:Vanishing effective mass for general non-local interactions}. Altogether, this forms the main result of the present work.

The remaining elements of the Landau-Ginzburg analysis consist of computing local and non-local correlation functions, extracting a correlation length and studying the behavior of the Ginzburg-$Q$ in the limit $\mu\rightarrow 0$. The local correlation function is equal to the one given in Eq.~\eqref{eq:U1 local correlator}, leading to a local correlation length of $\xiloc^{-2} = b_{\vb*{0}}$. Furthermore, if $G$ is compact, the non-local correlation function shows the same behavior as for the $\mathrm{U}(1)$ case in Eq.~\eqref{eq:non-loc corr U1}. In particular, it is dominated by the $s=r$ zero mode contribution in the limit $\mu\rightarrow 0$, i.e. $C(\vbg) \rightarrow b_{\vb*{0}}^{-1}$. As a result, the Ginzburg-$Q$ is given by 
\begin{equation}
Q\sim\lambda^{\frac{2}{n_\gamma-2}}\xiloc^{4-\dloc},
\end{equation}
similar to the quartic melonic $\mathrm{U}(1)$ theory. The only remnant of the fixed but arbitrary interaction considered in this section is the degree of the interaction $n_\gamma$ which determines the scaling of $Q$ in $\lambda$.

For non-compact groups, the asymptotic scaling of the non-local correlation function and thus the form of $Q$ strongly depends on the specifics of $G$ and its unitary irreducible representations $D^{(j)}(g)$. Therefore, we restrict ourselves in the following to two important examples of non-compact $G$, being first $G=\R$ and second $G=\SL$. For the second example, we assume further structure of the TGFT in the form of closure and simplicity constraints on the field domain to render it into a quantum geometric model.

\subsubsection{Non-compact \textit{G} = $\R$}\label{sec:general TGFT on R}

On $G=\R$, the results of Sec.~\ref{sec:non-compact limit} generalize to the arbitrary interactions. More precisely, the non-local correlation function is given by
\begin{equation}
C(\vbg) = \sum_{s=0}^{r}V_G^{-s}\sum_{(c_1,\dots, c_s)}\int\frac{\dd{\vb*{p}}_{4-s}}{(2\pi)^{4-s}}\frac{\prod_c \e^{ip_cg_c}}{\sum_c p_c^2+b_{c_1\dots c_s}},
\end{equation}
from which we extract a non-local correlation length which is given by $\xinloc^{-2} = \epsilon\abs{\mu}$ for $s_0\leq s<r$ zero mode injections at $(c_1,\dots c_s)\in\mathcal{O}_s^\gamma$ or $\xinloc^{-2} = b_{\vb*{0}} = \abs{\mu}(n_\gamma-2) > 0$ for $s=r$ zero modes. As discussed above, we exclude contributions with negative effective mass arising from zero mode injections at $(c_1,\dots c_s)\in\bar{\mathcal{O}}_s^\gamma$. 

Using the mean-field solutions of Eq.~\eqref{eq:general mf sol}, where $V_G = 2\pi R$ with $R\rightarrow \infty$, the Ginzburg-$Q$ evaluates to
\begin{equation}
    Q\sim \lambda^{\frac{2}{n_\gamma-2}}\abs{\mu}^{-\frac{2}{n_\gamma-2}}\xiloc^{-\dloc}\sum_{s = s_0}^{r}\left(\frac{\xinloc}{2\pi R}\right)^{s-r}\sum_{(c_1,\dots,c_s)\in\mathcal{O}_s^\gamma}\frac{1}{b_{c_1\dots c_s}},
\end{equation}
which is dominated by the $s_0$-summand if the limit $R\rightarrow \infty$ is taken first. Absorbing the divergent factor into the coupling,
\begin{equation}
 \bar{\lambda} =  (2\pi R)^{\frac{(r-s_0)(n_\gamma-2)}{2}} \lambda,   
\end{equation}
the Ginzburg-$Q$ for a TGFT on $\R^r\times\R^{\dloc}$ with a single arbitrary interaction is given by
\begin{equation}\label{eq:Q non-compact limit}
    Q\sim \bar{\lambda}^{\frac{2}{n_\gamma-2}}\epsilon^{-1+\frac{r-s_0}{2}}\abs{\mu}^{-\frac{1}{2}\left(2\frac{n_\gamma}{n_\gamma-2}-\dloc-(r-s_0)\right)}.
\end{equation}
If the following inequalities are satisfied

\begin{equation}\label{eq:ineq Q TGFT R}
2\frac{n_\gamma}{n_\gamma-2}-\dloc-(r-s_0)\leq 0,\qquad -1+\frac{r-s_0}{2}\geq 0,
\end{equation}
the limits in $\epsilon$ and $\mu$ commute and $Q\rightarrow 0$. The critical dimension, above which mean-field theory is valid, is identified as $\dloc+(r-s_0) = \frac{2n_\gamma  }{n_\gamma-2}$ which is consistent with the results found in~\cite{Geloun:2023ray} from an FRG analysis.

Notice that the second inequality, $r-s_0\geq 2$, is a condition on the vertex graph $\gamma$ which in particular excludes the simplicial case where $s_0 = r-1$. In these instances, $Q$ depends on the order in which we take the limits. More precisely, if the limit $\mu\rightarrow 0$ is performed first, $Q$ vanishes if one is above the critical dimension. If instead the limit $\epsilon\rightarrow 0$ is performed first, $Q$ diverges and the mean-field approximation breaks down. In~\cite{Marchetti:2022igl}, the authors find that for a quintic simplicial TGFT on $\R^{\mathrm{d}_{\mathrm{nloc}}}\times\R^{\dloc}$, $Q$ diverges in the limit $\mu\rightarrow 0$. This result suggests that for a vanishing effective mass, the limit $\epsilon \rightarrow 0$ should be taken first and that mean-field theory is not valid in this case. To go beyond mean-field theory, clearly a non-perturbative treatment will be required.

For finite $\mu < 0$ and the second inequality of Eq.~\eqref{eq:ineq Q TGFT R} being satisfied, $Q$ vanishes in the limit $\epsilon\rightarrow 0$ and the mean-field ansatz provides a reliable approximation. The phase structure of this section's model has been studied via FRG methods in~\cite{Geloun:2023ray}, finding a Wilson-Fisher-like fixed point below the critical value of the dimension of the total dimension. 
%(if $\zeta = 1$). 
Again, the analysis therein is non-perturbative and considers all zero mode contributions. The results found here still may be tentatively seen as indicating the existence of such a fixed point with $\mu_* < 0$.   

\paragraph{Closure constraint.} In TGFTs with quantum geometric interpretation, an additional constraint on the domain is often considered, commonly referred to as the closure constraint. Imposed via group averaging, this constraint yields a symmetry of the field under diagonal group action affecting the non-local degrees of freedom only, i.e. $\Phi(\vbg h,\vbx) = \Phi(\vbg,\vbx)$ which equivalently corresponds to the closure of the fluxes dual to the group variables ~\cite{Baratin:2010nn}. As detailed in~\cite{Marchetti:2020xvf}, the effect of such a constraint is a reduction in the rank of the group copies by one, i.e. $r\rightarrow r-1$, which can be straightforwardly implemented for the $Q$-parameter above. Notice that for the spacelike BC model studied in the next section, no such shift occurs due to extending the domain by a timelike normal vector, see also~\cite{Marchetti:2022igl}. 

\subsubsection{The spacelike Barrett-Crane model}\label{sec:BC}

The Barrett-Crane model, originally formulated in~\cite{Barrett:1999qw,Perez:2000ec,Perez:2000ep}, is defined on $r=4$ copies of the double cover of the Lorentz group $G = \SL$ and additional constraints.\footnote{In the following, we restrict to the rank-$4$ case and refer the reader to~\cite{Oriti:2003wf} for a higher-dimensional generalization.} Closure and simplicity constraints, collectively referred to as the geometricity constraints, are defining factors for the model. They guarantee that the BC model provides a tentative GFT quantization of first-order Palatini gravity in $4d$ with Lorentzian signature. To ensure a covariant and commuting imposition of the geometricity constraints, an extended formulation including timelike normal vectors has been developed in~\cite{Baratin:2011tx,Jercher:2021bie}. These advances later led to the formulation of the causally complete BC model, meaning that it includes normal vectors with all signatures, i.e. they are of spacelike, lightlike, and timelike type~\cite{Jercher:2022mky}. 

In this section, we restrict to timelike normal vectors and thus spacelike tetrahedra and extend the mean-field analysis commenced in Refs.~\cite{Marchetti:2022igl,Marchetti:2022nrf} to the case of vanishing effective mass for the relevant modes. We refer to~\cite{Dekhil:2024djp} for a mean-field treatment of the complete BC model. 

The domain of the perturbation $\delta\Phi$ is $\left(\SL^4\times\TH\right)\times\R^{\dloc}$ with $\TH\ni X$ denoting the two-sheeted hyperboloid and $\R^{\dloc}$ being the domain of the local variables. Employing unitary representations of $\SL$, the Fourier decomposition of the field $\delta\Phi$ satisfying the geometricity constraints is given by
\begin{equation}
\Phi(\vbg,X,\vbx) = \left[\prod_{c=1}^4\int_\R\dd{\rho_c}^2\sum_{j_c,m_c}D^{(\rho_c,0)}_{j_cm_c00}(g_c X)\right]\int\frac{\dd{\vbk}}{(2\pi)^{\dloc}}\e^{i\vbk\vbx}\Phi^{\vbr}_{\vb*{j}\vb*{m}}(\vbk),
\end{equation}
where the $D^{(\rho,\nu)}_{jmln}(g)$ are $\SL$-Wigner matrices in the $(\rho,\nu)\in\R\times\mathbb{Z}/2$ representation  with $\nu = 0$ due to simplicity. The magnetic indices take values $j,l\in\{\abs{\nu},\abs{\nu}+1,\dots\}$ and $m\in\{-j,\dots,j\}$ and $n\in\{-l,\dots,l\}$. For further details, we refer the reader to the appendices of Refs.~\cite{Jercher:2021bie,Jercher:2022mky,Marchetti:2022igl}.

%Closure and simplicity constraints are summarized by
%
%\begin{subequations}
%\begin{align}
%\Phi(\vbg h^{-1},h\cdot X,\vbx) &= \Phi(\vbg,X,\vbx),\qquad \forall\;h\in\SL,\label{eq:closure}\\[7pt]
%\Phi(\vbg \vb*{u},X,\vbx) &= \Phi(\vbg,X,\vbx),\qquad \forall\vb*{u}\in\SUT_X^4,\label{eq:simplicity}
%\end{align}  
%end{subequations}
%
%where in Eq.~\eqref{eq:closure}, the action of $h$ on $\vbg$ is diagonal and $h\cdot X$ is understood as the Lorentz transformation of $X\in\R^{1,3}$ by $h$. In Eq.~\eqref{eq:simplicity}, the stabilizer subgroup $\SUT_X\cong \SUT$ acts on every $\SL$-element separately.  

%The kinetic part of the action acting on such a field is given by Eq.~\eqref{eq:general action} for the respective choice of domain. Here, we set $\zeta = 1$ and comment on $\zeta < 1$ at the end. Importantly, when considering the interaction $V[\Phi]$, the normal vectors are integrated out separately and are thus independent of the non-local contraction pattern $\gamma$~\cite{Jercher:2021bie}. 

Due to the non-compactness of $\SL$, the uniform field contains infinite volume factors $\mathrm{V}_G = \lim_{L\rightarrow\infty}\e^{2L/a}$ which are understood to be regularized by a cutoff $L$ in the non-compact direction.\footnote{This scaling follows from the form of the Haar measure on $\SL$, given by $\dd{g} = \frac{1}{\pi a}\dd{u}\dd{v}\dd{\eta}\sinh^2(\frac{\eta}{a})$, with $\dd{u},\dd{v}$ normalized measures on $\SUT$ and $\eta$ parametrizing the non-compact direction.} Here, $a$ is the skirt radius of the hyperbolic part of $\SL$. The background solutions are then given by
\begin{equation}\label{eq:BC mf solution}
\Phi_0 = \left(\frac{\abs{\mu}}{\lambda n_\gamma}\right)^{\frac{1}{n_\gamma -2}}\mathrm{V}_G^{-\frac{r}{2}}\mathrm{V}_G^{-\frac{n_\gamma-1}{n_\gamma-2}},
\end{equation}
where the additional last volume factor arises from the presence of the auxiliary normal vector variables.

Repeating the same steps as in the previous sections, while carefully treating the geometricity constraints, one obtains the correlation function of fluctuations
\begin{equation}
C(\vbg,\vbx) = \left[\prod_c\int\dd{\rho_c}\rho_c^2\sum_{j_c,m_c}\right]\int\frac{\dd{\vbk}}{(2\pi)^{\dloc}}\frac{\prod_c D^{(\rho_c,0)}_{j_c m_c j_c m_c}(g_c)\e^{i\vbk\vbx}}{\frac{1}{a^2}\sum_c\left(\rho_c^2+1\right)+\vbk^2+b_{\vbr}},
\end{equation}
where we identified the eigenvalues of the Laplacian acting on $\SL$ in the $(\rho_c,0)$ representation as $\frac{1}{a^2}(\rho_c^2+1)$. The effective mass term $b_{\vbr}$ contains Kronecker-$\delta$'s on the zero modes which, for $\SL$, are given by $\rho = i$. For details on the definition of these $\delta$'s and on the inclusion of the trivial representation $\rho = i$ in the Peter-Weyl decomposition, we refer the reader to~\cite{Ruehl1970} and~\cite{Marchetti:2022igl,Dekhil:2024djp}. 

By integrating out the local variables $\vbx\in\R^{\dloc}$, we obtain the non-local correlation function, which is extended to include zero modes,
\begin{equation}
C(\vbg) = \sum_{s=0}^r \mathrm{V}_G^{-s}\sum_{(c_1,\dots, c_s)}C^s(\vbg_{r-s}).
\end{equation}
To extract a correlation length from the asymptotic behavior of $C^s$, a notion of distance in the non-compact direction on $\SL$ is required. Performing a Cartan-decomposition of $\SL$~\cite{Ruehl1970}, the non-local correlation function can be decomposed as
%
%\begin{equation}\label{eq:Cartan decomp g}
%g = u\e^{\frac{\eta}{2a}\sigma_3}v,\qquad \dd{g} = \frac{1}{\pi}\dd{u}\dd{v}\dd{\eta}\sinh^2(\eta/a),
%\end{equation}
%
%where $u,v\in\SUT$, we extract this distance to be the non-compact variable $\eta\in\R_+$, related to the rapidity of Lorentz boosts. The Cartan decomposition induces a decomposition of $\SL$-Wigner matrices
%
%\begin{equation}\label{eq:Cartan decomp D}
%D^{(\rho,\nu)}_{jmln}(g) = \sum_{q=-\mathrm{min}(j,l)}^{\mathrm{min}(j,l)}D^{j}_{mq}(u)d^{(\rho,\nu)}_{jlq}(\eta)D^{l}_{qn}(v),
%\end{equation}
%
%with $d^{(\rho,\nu)}$ being referred to as reduces Wigner matrix. Then, we split the residual correlation function $C^s$ into an $\SUT$ and a boost part,
%
\begin{equation}
C^s(\vbg_{r-s}) = \left[\prod_{c=c_{s+1}}^{c_r}\sum_{j_c m_c}D^{j_c}_{m_c m_c}(v_c^{-1}u_c)\right]C^s(\vb*{\eta}_{r-s})_{\vb*{j}_{r-s}\vb*{m}_{r-s}},
\end{equation}
where the $\eta_c\in\R_+$ are boosts parametrizing the non-compact direction of $\SL$. The remaining boost part then reads as
\begin{equation}
C^s(\vb*{\eta}_{r-s}) = \left[\prod_{c=c_{s+1}}^{c_r}\int\dd{\rho_c}\rho_c^2\right]\frac{\prod_c d^{(\rho_c,0)}_{j_c j_c m_c}(\eta_c)}{\frac{1}{a^2}\sum_c\left(\rho_c^2+1\right)+b_{c_1\dots c_s}}.
\end{equation}
Using the large $\eta$ asymptotics of the reduced $\SL$-Wigner matrix~\cite{Marchetti:2022igl},
\begin{equation}
d^{(\rho,0)}_{jjm}(\eta)\underset{\eta\gg 1}\varsigma(\rho,j,m)\e^{\frac{\eta}{a}(1+i\rho)},
\end{equation}
it has been shown in~\cite{Marchetti:2022igl} that under the reasonable assumption of the isotropic limit $\eta_c\equiv\eta\gg 1$ the residual correlation function has a limiting value
\begin{equation}
    \eval{C^s(\vb*{\eta}_{r-s})}_{\eta_c\equiv\eta}\longrightarrow \exp\left(-(r-s)\left(1+\sqrt{1+\frac{a^2b_{c_1\dots c_s}}{r-s}}\right)\frac{\eta}{a}\right).
\end{equation}
Notice, that this does not imply that the correlation length is set by the skirt radius $a$, as suggested for a related case in~\cite{Guruswamy:1994pi}. Instead, the correlation length is extracted from the re-scaled correlation function $\tilde{C}^s$, that is, the product of the correlation function with the Jacobian determinant on the two-sheeted hyperboloid. As a result, we find
\begin{equation}
\eval{\tilde{C}^s(\vb*{\eta}_{r-s})}_{\eta_c\equiv\eta}\longrightarrow\e^{-\frac{1}{2}ab_{c_1\dots c_s}\eta},
\end{equation}
from which we extract
\begin{equation}
    \xinloc = \frac{1}{ab_{c_1\dots c_s}}.
\end{equation}
The explicit form of $\xinloc$ therefore depends on whether the effective mass is negative, zero or positive. In particular, $b_{c_1\dots c_s}<0$, obtained from $(c_1,\dots, c_s)\in\bar{\mathcal{O}}_s^\gamma$, leads to an exponential divergence of the residual correlation function and thus long-range correlations for any $\mu$. As argued above, to study the critical behavior, we neglect these modes in the following. As discussed in the previous sections, we regularize a vanishing effective mass via $b_{c_1\dots c_s} = \epsilon\abs{\mu}$ with $\epsilon\rightarrow 0$.

The integration domain for the Ginzburg-$Q$ is given by $\SL_{\xinloc}^4\times [-\xiloc,\xiloc]^{\dloc}$, where $\SL_{\xinloc}$ contains a restricted range of the non-compact variable $\eta\in[0,\xinloc]$. For the numerator and denominator of $Q$, we respectively obtain
\begin{equation}
\int_{\Omega_\xi}\dd{\vbg}\dd{\vbx}C(\vbg,\vbx) = \sum_{s=s_0}^r \e^{2s\frac{\xinloc-L}{a}}\sum_{(c_1,\dots, c_s)\in\mathcal{O}^\gamma_s}\frac{1}{b_{c_1\dots c_s}},
\end{equation}
and
\begin{equation}
\int_{\Omega_\xi}\dd{\vbg}\dd{\vbx}\Phi_0^2 = \e^{-4\frac{n_\gamma-1}{n_\gamma-2}\frac{L}{a}}\e^{2r\frac{\xinloc-L}{a}}\xiloc^{\dloc}\left(\frac{\abs{\mu}}{\lambda n_\gamma}\right)^{\frac{2}{n_\gamma-2}},
\end{equation}
where we injected the background mean-field solution given in Eq.~\eqref{eq:BC mf solution}. Altogether, the $Q$-parameter scales as
\begin{equation}
Q\sim \lambda^{\frac{2}{n_\gamma - 2}}\e^{4\frac{n_\gamma-1}{n_\gamma-2}\frac{L}{a}}\abs{\mu}^{-\frac{2}{n_\gamma-2}}\xiloc^{-\dloc}\sum_{s=s_0}^r\e^{-2(r-s)\frac{\xinloc-L}{a}}\sum_{(c_1,\dots,c_s)}\frac{1}{b_{c_1\dots c_s}}.
\end{equation}
Taking the large $L$-limit first, the contribution from $s_0$ zero modes dominates including in particular the case of vanishing of the effective mass $b_{c_1\dots c_s} = \epsilon\abs{\mu}$. Furthermore, we introduce the re-scaled coupling
\begin{equation}
\bar{\lambda } = \e^{\left(2(n_\gamma-1)+(r-s_0)(n_\gamma-2)\right)\frac{L}{a}}\lambda.
\end{equation}
Then, the Ginzburg-$Q$ finally evaluates to
\begin{equation}\label{eq:BC Q}
Q\sim \bar{\lambda}^{\frac{2}{n_\gamma-2}}\e^{-\frac{2(r-s_0)}{a^2\epsilon\abs{\mu}}}\epsilon^{-1}\abs{\mu}^{-\frac{1}{2}\left(2\frac{n_\gamma}{n_\gamma-2}-\dloc\right)}.
\end{equation}

Since $r>s_0$, we observe that $Q\rightarrow 0$ irrespective of the order in which the limits $\epsilon\rightarrow 0$ and $\mu\rightarrow 0$ are taken. This is a direct consequence of the non-compact hyperbolic part of $\SL$ yielding the exponential fall-off behavior. Therefore, the mean-field prescription is valid at criticality, which reproduces the findings of~\cite{Marchetti:2022nrf,Marchetti:2022igl}. More than that, $Q\rightarrow 0$ even for finite $\mu$ so that the approximation is valid in the entire phase with $\Phi_0\neq 0$. This can be summarized by the critical dimension being infinite for field theories living on $\SL$. Moreover, as pointed out in~\cite{Marchetti:2022nrf}, the exponential suppression effect of $Q$ suggests that the whole phase diagram is already described by the two phases of the mean-field regime even beyond this approximation.

In~\cite{Marchetti:2022igl,Marchetti:2022nrf}, the limit of infinite skirt radius $a$ has been considered which turns the hyperbolic part of $\SL$ into $\R^3$. Taking this limit before evaluating the integrals entering $Q$, the exponential factors reduce to polynomial ones. Inserting the local correlation length $\xi_{\mathrm{loc}}^{-2}=\epsilon\abs{\mu}$ the results of Sec.~\ref{sec:general TGFT on R} are indeed recovered. In particular, the $Q$-parameter of this section attains the form given in Eq.~\eqref{eq:Q non-compact limit}.

These results lend further support to the existence of a meaningful continuum gravitational regime in TGFT
quantum gravity via a phase transition to a non-trivial vacuum (condensate) state which can be described by mean-field theory. This is also relevant for the spin foam approach and lattice quantum gravity via their close relations to TGFT.

\subsubsection{Extension to multiple interactions}\label{sec:multiple interactions} 

The results of the previous two sections can be extended to the case where a sum of interaction terms is considered. That is, the interaction term in Eq.~\eqref{eq: non-local-interaction} is generalized to~\cite{Marchetti:2020xvf,Marchetti:2022igl}
\begin{equation}
V[\Phi] = \sum_\gamma \lambda_\gamma\int\dd{\vbx}\Tr_\gamma\left[\Phi(\vbg,\vbx)^{n_\gamma}\right].
\end{equation} 
In the most general case, where the number of vertices $n_\gamma$ varies for different $\gamma$, the mean-field equation is a polynomial equation of a potentially large degree which therefore might not exhibit closed analytical solutions. Therefore, we restrict in the following to the simpler case of interactions of the same degree $n_\gamma\equiv n$ for all graphs $\gamma$ under consideration. The mean-field background solution and the effective mass generalize to
\begin{equation}
\Phi_0 = \left(\frac{\abs{\mu}}{n\sum_\gamma\lambda_\gamma}\right)^{\frac{1}{n-2}},\qquad b_{\vb*{j}} = \abs{\mu}\left(\sum_\gamma\tilde{\lambda}_\gamma\chi^\gamma_{\vb*{j}}-1\right),
\end{equation}
with~\cite{Marchetti:2020xvf}
\begin{equation}
\tilde{\lambda}_\gamma = \frac{\lambda_\gamma}{\sum_{\gamma'}\lambda_\gamma'}.
\end{equation}
Here, $\chi^\gamma_{\vb*{j}}$ is associated to the graph $\gamma$ and computed as above. In particular, the inequality $\chi^\gamma_{\vb*{j}}\leq 1$ holds for all $\gamma$ if $s<r$ zero modes are considered. As a consequence, if $\lambda_\gamma$ has the same sign for all $\gamma$, then also in this case the effective mass is zero or negative for $s<r$ zero modes. More precisely, one finds a vanishing effective mass for
\begin{equation}
s_0 \leq s<r,\qquad  \text{and}\qquad (c_1,\dots, c_s)\in\bigcap\limits_\gamma\mathcal{O}_s^\gamma,
\end{equation}
where $s_0 \equiv  \max_\gamma\{s_0^\gamma\} $ and $s_0^\gamma$ is defined individually for every graph $\gamma$. If there are couplings of mixed sign instead, then $\sum_\gamma\tilde{\lambda}_\gamma\chi_{\vb*{j}}^\gamma$ can in principle be larger than one also for $s<r$ zero modes, resulting in a positive effective mass.\footnote{We thank J. Th\"{u}rigen for pointing out this possibility.} The introduction of multiple interaction terms therefore introduces additional complexity to the phase diagram of the theory where the relative signs and size of the interaction couplings marks different phases. We leave a thoroughe analysis of this observation to future investigations. 

Expressions for the Ginzburg-$Q$ do not change substantially and only contain an additional sum over graphs $\gamma$, see~\cite{Marchetti:2020xvf} for further details. Taking the limit of large group volume first, the contribution of the graph with smallest $s_0^\gamma$ dominates.

\subsection{Tensor field theories}\label{sec:tensor field theories}

Tensor field theories are usual field theories in an existing ambient space but with tensorial properties. Promoted to quantum fields they provide non-trivial examples for conformal field theories and are thus relevant for quantum gravity via the AdS/CFT conjecture. In particular, the Sachdev-Ye-Kitaev (SYK)~\cite{Sachdev:1992fk,kitaev2015simple,Maldacena:2016hyu} model provides a simple nearly conformal dual to a nearly AdS$_2$ black hole~\cite{Maldacena:2016upp}. It was noticed that its large-$N$ limit is dominated by melonic diagrams and can be linked to a tensor quantum mechanical model, the Gurau-Witten model~\cite{Witten:2016iux,Gurau:2016lzk,Klebanov:2016xxf,Klebanov:2018nfp,Kim:2019upg}. This development spurred the interest to also investigate the large-$N$ limit of tensor field theories in higher dimensions. These yield a new class of so-called melonic CFTs as non-trivial infrared fixed points of the renormalization group flow~\cite{Giombi:2017dtl,Carrozza:2015adg,Klebanov:2018fzb,Benedetti:2018goh,Benedetti:2018ghn,Benedetti:2019eyl,Gurau:2019qag,Benedetti:2021wzt,Harribey:2022esw,Jepsen:2023pzm,Berges:2023rqa,Gurau:2024nzv}. While these results are very intriguing it should be emphasized that they are obtained for models for tensor fields with a set rank and at large $N$.  

The goal of this section is to go beyond this and explore the phase structure of related bosonic tensor field theories for fields with arbitrary rank $r\geq 2$ and of any $N$ in the infrared using the Landau-Ginzburg mean-field method instead.%While this critically widens the scope of analysis, we again not that this method only provides a coarse picture of the phase structure of field theories. 
\footnote{Note that while the tensor fields are local in their arguments, the combinatorial pattern of the index contractions is non-local so that their Feynman expansion yields graphs dual to cellular complexes. The index data in this section is considered static while we allowed it to propagate in the previous sections.}

The tensor field theories considered hereafter are bosonic and defined on $\mathbb{Z}_N^r\times\R^{\dloc}$ with $N$ being the size of the tensor. They can be understood as TGFTs with $G$ being a finite group that has no Lie structure. Correspondingly, there is no Laplacian on $G=\mathbb{Z}_N$, meaning that the tensor indices are non-dynamical. The action of a tensor field $\Phi_{\vb*{n}}(\vbx)$ is given by
\begin{equation}\label{eq:TFT action}
S[\Phi] = \frac{1}{2}\sum_{\vb*{n}\in\mathbb{Z}_N^r}\int\dd{\vbx}\Phi_{\vb*{n}}(\vbx)\left(\mu-\Delta_x\right)\Phi_{\vb*{n}}(\vbx)+\lambda\int\dd{\vbx}\Tr_{\gamma}\left[\Phi_{\vb*{n}}(\vbx)^{n_\gamma}\right].
\end{equation}
Here, the symbol $\Tr_\gamma$ indicates a contraction of tensor indices according the vertex graph $\gamma$, characterizing the combinatorially non-local interaction. 

The solution of the mean-field equations is given in Eq.~\eqref{eq:general mf sol} with $\mathrm{V}_G = N$.
%which in this particular case yields
%
%\begin{equation}
%\Phi_0 = \left(\frac{\abs{\mu}}{\lambda n_\gamma}\right)^{\frac{1}{n_\gamma-2}}N^{-\frac{r}{2}}.
%\end{equation}
%
%Here, the volume factors $\mathrm{V}_G$ are  simply given as the cardinality of $\mathbb{Z}_N$, so $\mathrm{V}_G = N$.
Perturbations around the mean-field, $\Phi_{\vb*{n}}(\vbx) = \Phi_0 +\delta\Phi_{\vb*{n}}(\vbx)$, are governed by the effective action
\begin{equation}
S_{\mathrm{eff}}[\delta\Phi] = \sum_{\vb*{n},\vb*{n}'\in\mathbb{Z}_N^r}\int\dd{\vbx}\delta\Phi_{\vb*{n}}(\vbx)\left(-\Delta_x\delta_{\vb*{n},\vb*{n}'}+b_{\vb*{n},\vb*{n}'}\right)\delta\Phi_{\vb*{n}'}(\vbx),
\end{equation}
with an effective mass given by
\begin{equation}
b_{\vb*{n},\vb*{n}'} = \mu\left(\delta_{\vb*{n},\vb*{n}'}-\chi_{\vb*{n},\vb*{n}'}\right).
\end{equation}
The Fourier space representation of the perturbed field is given by
\begin{equation}
\delta\Phi_{\vb*{n}}(\vbx) = \frac{1}{N^r}\sum_{\vb*{p}\in \tilde{\mathbb{Z}}_N^r}\int\frac{\dd{\vbk}}{(2\pi)^{\dloc}}\e^{i\vb*{p}\vb*{n}}\e^{i\vbk\vbx}\delta\Phi_{\vb*{p}}(\vbk),
\end{equation}
where $\tilde{\mathbb{Z}}_N^r = \{0,\frac{2\pi}{N},\dots, \frac{2\pi(N-1)}{N}\}$ can be seen as the lattice reciprocal to the real-space lattice $\mathbb{Z}_N^r$. In this space, the effective action reads
\begin{equation}
S_{\mathrm{eff}}[\delta\Phi] = \frac{1}{N^r}\sum_{\vb*{p}}\int\frac{\dd{\vbk}}{(2\pi)^{\dloc}}\delta\Phi_{\vb*{p}}(\vbk)\left(\vbk^2+b_{\vb*{p}}\right)\delta\Phi_{N-\vb*{p}}(-\vbk).
\end{equation}
%
%where $\tilde{\mathbb{Z}}_N^r = \{0,\frac{2\pi}{N},\dots, \frac{2\pi(N-1)}{N}\}$ can be seen as the lattice reciprocal to the real-space lattice $\mathbb{Z}_N^r$. 

The correlation function in momentum space is obtained by the inversion of the kinetic kernel,
%, yielding
%
%\begin{equation}
%C_{\vb*{p},\vb*{p}'}(\vbk,\vbk') = \frac{1}{\vbk^2+b_{\vb*{p}}}N^r\delta_{\vb*{p},N-\vb*{p}'}(2\pi)^{\dloc}\delta(\vbk+\vbk').
%\end{equation}
%
and as a result, the correlator in real space simply yields
\begin{equation}
C_{\vb*{n},\vb*{n}'}(\vbx,\vbx') = \frac{1}{N^r}\sum_{\vb*{p}}\int\frac{\dd{\vbk}}{(2\pi)^{\dloc}}\e^{i\vb*{p}(\vb*{n}-\vb*{n}')}\e^{ik(\vbx-\vbx')}\frac{1}{\vbk^2+b_{\vb*{p}}},
\end{equation}
which we expand in terms of zero modes as
\begin{equation}
C_{\vb*{n},\vb*{n}'}(\vbx,\vbx') = \frac{1}{N^r}\sum_{s=0}^r\sum_{(c_1\dots c_r)}N^{r-s}\prod_{c=c_{s+1}}^{c_r}\delta_{n_c,n_c'}C_{c_1\dots c_s}(\vbx,\vbx'),
\end{equation}
with residual correlation function given by
\begin{equation}\label{eq:residual correlation TFT}
C_{c_1\dots c_s}(\vbx,\vbx') = \int\frac{\dd{\vbk}}{(2\pi)^{\dloc}}\frac{\e^{i\vbk(\vbx-\vbx')}}{\vbk^2+b_{c_1\dots c_s}}.
\end{equation}
As discussed in detail in the previous sections, the effective mass $b_{c_1\dots c_s}$ determines three different regimes where it is positive, zero, or negative, respectively. In the following, we study these three different cases separately and compute explicitly the corresponding correlation function. This then lays the ground to derive the $Q$-parameter in this class of theories, where we devote special attention to the case of vanishing effective mass and its further implications. 

In the case where $s=r$, the effective mass is given by $b\equiv b_{c_1\dots c_s} = \abs{\mu}(n_\gamma -2) > 0$ yielding a correlation function
\begin{equation}\label{eq:residual correlation TFT s=r}
    C_{c_1\dots c_s}(\vbx,\vbx') = \int\frac{\dd{\vbk}}{(2\pi)^{d}}\frac{\e^{i\vbk(\vbx-\vbx')}}{\vbk^2+b_{c_1\dots c_s}} = \frac{2^d\pi^{\frac{d}{2}}}{(2\pi)^d \ell^{d-2}}\left(\sqrt{b}\ell\right)^{\frac{d-2}{2}}K_{\frac{d-2}{2}}(\sqrt{b}\ell),
\end{equation}
with  $d\equiv \dloc$ and $\ell \equiv \abs{\vbx-\vbx'}$. The result is that of a correlation function on $\R^{\dloc}$ which is consistent with the fact that on $s=r$ zero modes, the residual correlator is constant in the tensor indices such that the information of the non-localities is entirely absent.

For $s<r$ zero modes injected at $(c_1,\dots ,c_s)\in\bar{\mathcal{O}}_s^\gamma$, the residual correlation function is given as in Eq.~\eqref{eq:residual correlation TFT s=r} but with a negative mass, $b<0$. This implies that $\sqrt{b} = i\sqrt{\abs{b}}$ leading to an oscillatory behavior of $C_{c_1\dots c_s}$. As discussed previously, we exclude these sets of zero mode insertions in the following.

For $s_0\leq s <r$ zero modes injected at $(c_1,\dots, c_s)\in\mathcal{O}_s^\gamma$, the effective mass vanishes. We regularize it as $b_{c_1\dots c_s} = \epsilon\abs{\mu}$ and simply use the formula above for positive effective mass to find
\begin{equation}
    C_{c_1\dots c_s}(\vbx,\vbx') = \frac{2^d\pi^{\frac{d}{2}}}{(2\pi)^d \ell^{d-2}}\left(\sqrt{\epsilon\abs{\mu}}\ell\right)^{\frac{d-2}{2}}K_{\frac{d-2}{2}}(\sqrt{\epsilon\abs{\mu}}\ell),
\end{equation}
with the limit $\epsilon\rightarrow 0$ implicitly understood. Using the properties of the modified Bessel function $K_\alpha(z)$, one can show that if the limit is taken explicitly, the correlation function correctly satisfies the scaling behavior $C\sim \ell^{2-d}$ for $d\neq 2$. From the correlation function above, one can furthermore extract a regulated correlation length $\xi^{-2} = \epsilon\abs{\mu}$ which diverges in the limit $\epsilon\rightarrow 0$ even for finite $\mu$.
%, reflecting the conformal invariance of the system beyond criticality. 

In the next paragraph, we compute the Ginzburg-$Q$ parameter for those configurations on which the effective mass vanishes. Let us remark that the case of $s=r$ zero modes is covered by the standard literature on local field theory where $Q$ is given by
\begin{equation}
\eval{Q}_{s=r}\sim\lambda^{\frac{2}{n_\gamma-2}}\xi^{2\frac{n_\gamma}{n_\gamma-2}-\dloc}.
\end{equation}
For quartic interactions, $n_\gamma = 4$, one obtains the well-known result for the upper critical dimension $d_{\mathrm{crit}} = 4$ above which the mean-field approach is valid~\cite{Kopietz:2010zz,zinn2021quantum,Benedetti:2014gja}. 

Proceeding with a vanishing effective mass, which we regulate as above, the integration domain of the Ginzburg-$Q$ in Eq.~\eqref{eq:general Q} is given by $\Omega_\xi = [-\xi,\xi]^{\dloc}$ such that the denominator evaluates to
\begin{equation}
\sum_{\vb*{n},\vb*{n}'}\int_{\Omega_\xi}\dd{\vbx}\Phi_0^2 = \left(\frac{\abs{\mu}}{\lambda n_\gamma}\right)^{\frac{2}{n_\gamma-2}}N^r\xi^{\dloc}.
\end{equation}
In the numerator of Eq.~\eqref{eq:general Q}, we use the fact that for $\epsilon \rightarrow 0$, the integration domain extends to all of $\R^{\dloc}$. As a result, 
\begin{equation}
\sum_{\vb*{n},\vb*{n}'}\int_{\Omega_\xi}\dd{\vbx}C_{\vb*{n},\vb*{n}'}(\vbx,\vb*{0}) = N^r\sum_{s=s_0}^{r-1}\sum_{(c_1,\dots,c_s)\in\mathcal{O}_s^\gamma}\int\dd{\vbx}C_{c_1\dots c_s}(\vbx,\vb*{0})\sim\frac{N^r}{\abs{\mu}\epsilon},
\end{equation}
and $Q$  evaluates to
\begin{equation}
    Q\sim \lambda^{\frac{2}{n_\gamma-2}}\epsilon^{-1}\xi^{-\dloc}\abs{\mu}^{-\frac{n_\gamma}{n_\gamma-2}}.
\end{equation}
The behavior of $Q$ in the limits $\mu\rightarrow 0$ and $\epsilon \rightarrow 0$ depends in principle on the order in which these limits are taken. Since $\mu,\xi$, and $\epsilon$  are parameters that depend on each other, we can express $Q$ just in terms of $\mu$ and $\epsilon$ to find
\begin{equation}\label{eq:Q TFT}
Q\sim \lambda^{\frac{2}{n_\gamma-2}}\epsilon^{\frac{\dloc}{2}-1}\abs{\mu}^{-\frac{1}{2}\left(2\frac{n_\gamma}{n_\gamma-2}-\dloc\right)},
\end{equation}
Clearly if $\dloc\geq 2\frac{n_\gamma}{n_\gamma-2}$ and $n_\gamma \geq 3$ it follows that $\dloc\geq 2$. In this case, the two limits commute and we find that $Q\ll 1$ in the limit $\epsilon,\mu\rightarrow 0$. This result agrees with that of local field theories in that one finds a critical dimension of $d_{\mathrm{crit}} = 2\frac{n_\gamma}{n_\gamma -2}$. Our result is furthermore backed by the non-perturbative FRG studies in~\cite{Geloun:2023ray}, where cyclic-melonic tensor field theories in the large-$N$ limit are considered. Therein, a Gaussian fixed point in the infrared is found above the critical dimension which is consistent with our results here. We interpret this agreement as a support for the reliability of the regularization scheme for the vanishing effective mass employed here.

If $\dloc > 2$, the residual massless effective theory provides a good approximation of the full theory beyond criticality, i.e. $\lim_{\epsilon\rightarrow 0}Q = 0$ for finite $\mu < 0$. In~\cite{Geloun:2023ray}, a Wilson-Fisher type fixed point is found for $2< d< d_{\mathrm{crit}}$ in the infrared, using FRG methods and taking into account all zero mode contributions $0\leq s\leq r$. This fixed point is an interacting fixed point with negative mass $\mu_* < 0$. Clearly, the FRG results are non-perturbative and therefore provide a much more sophisticated picture of the phase structure compared to the Gaussian approximation used here. Still, the similarity with results found here tentatively suggests that the Landau-Ginzburg method indicates the existence of such a Wilson-Fisher-like fixed point. 

\paragraph{TFTs with imaginary tetrahedral coupling.} We have shown in the Sec.~\ref{sec:multiple interactions} that the mean-field analysis is straightforwardly generalized to a sum of interactions of the same degree. Furthermore, no reality assumptions were imposed on the coupling parameters $\lambda_\gamma$. Consequently, the O$(N)^3$-invariant tensor field theory model with double-trace, pillow, and imaginary tetrahedral coupling, considered in~\cite{Benedetti:2020yvb,Benedetti:2019eyl,Benedetti:2019ikb}, is captured by the analysis of this section. Indeed, the critical dimension of $d_{\mathrm{crit}} =4$ (for $n_\gamma = 2$) is in agreement with the results found therein. A notable difference is that our results, in particular the Ginzburg-$Q$, are independent of $N$ and thus hold beyond the large-$N$ limit. 

\section{The general case}\label{sec:The general case}

The structure of the theories we have considered so far in this work is that of bosonic scalar field theories of hybrid type with both local and non-local variables. Common to all these examples is the behavior of the effective mass which, depending on the combinatorics of the interaction, evaluates to non-positive values for $s<r$ zero modes. In fact, as we prove in Sec.~\ref{sec:Vanishing effective mass for general non-local interactions}, this behavior is present for any combinatorially non-local interactions. In this section, we therefore set up the most general model to which the arguments of the previous sections apply. 

\subsection{Model setup}\label{sec:General model setup}

The theory is defined by real-valued fields $\Phi:G^{\times r}\times \mathcal{M}\longrightarrow \R$ with non-local arguments, $\vbg\in G^r$, and local ones, $\vbx\in\mathcal{M}$. The field domains $G$ and $\mathcal{M}$ are respectively $d_G$- and $\dloc$-dimensional smooth manifolds, endowed with metric tensors $h_G$ and $h_{\mathcal{M}}$.\footnote{Although $G$ is assumed to be a smooth metric manifold, the results of this section apply as well to the reduced case of $G=\mathbb{Z}_N$ as we have shown in Sec.~\ref{sec:tensor field theories}.} The kinetic term of the action is given by
\begin{equation}\label{eq:general kinetic term}
K[\Phi] = \frac{1}{2}\int\limits_{G^{r}}\dd{\vb*{g}}\int\limits_{\mathcal{M}}\dd{\vb*{x}}\Phi(\vbg,\vbx)\left[\mu+\left(-\sum_{c=1}^r\Delta^c_g-\Delta_{\mathcal{M}}\right)^{\zeta}\right]\Phi(\vbg,\vbx),
\end{equation}
where the integration measures are the volume elements associated with the metrics $h_G$ and $h_{\mathcal{M}}$. Similarly, the Laplace operators $\Delta_g$ and $\Delta_{\mathcal{M}}$ are defined in terms of the respective metrics and carry a minus sign to ensure a positive spectrum. The presence of the Laplacian on $G$ indicates that the non-local variables are propagating degrees of freedom which leads to the class of TGFTs. Its omission leads to tensor field theories~\cite{Gurau:2019qag,Benedetti:2020seh,Gurau:2024nzv}. 

Note that we introduced an additional parameter $0 < \zeta\leq 1$ by means of which one can model not only standard short-range propagation $(\zeta = 1$), but also long-range propagation ($\zeta < 1$)~\cite{Fisher:1972zz,Benedetti:2020seh,Gurau:2024nzv}. It leads to modified scaling exponents, which we explicitly discuss in Sec.~\ref{sec:Symmetries of effectively massless theories}. Furthermore, the scaling of the Ginzburg-$Q$ is altered by the presence of $\zeta < 1$. As an example, the $Q$-parameter of a tensor field theory with arbitrary interactions, given in Eq.~\eqref{sec:tensor field theories} for $\zeta = 1$, changes to
\begin{equation}
Q_\zeta \sim \lambda^{\frac{2}{n_\gamma-2}}\epsilon^{\frac{\dloc}{2\zeta}-1}\abs{\mu}^{-\frac{1}{2\zeta}\left(2\zeta\frac{n_\gamma}{n_\gamma-2}-\dloc\right)}.
\end{equation}
from which one extracts a modified critical dimension of $\dloc = 2\zeta\frac{n_\gamma}{n_\gamma - 2}$.

The interactions we consider are structurally the same as in Eq.~\eqref{eq: non-local-interaction}, namely local in the variables $\vbx\in\mathcal{M}$ and non-local in the variables $\vbg\in G^r$. The combinatorial non-localities are captured by a vertex graph $\gamma$, examples of which are depicted in Tab.~\ref{tab:chis}. We have shown in Sec.~\ref{sec:multiple interactions} that the single interaction can be straightforwardly generalized to a sum of interactions of the same degree $n_\gamma\equiv n$. For simplicity of the presentation, we keep in this section a single interaction. 

Evaluating the equations of motion on constant field configurations, $\Phi_0=\mathrm{const}.$, yields the mean-field equations, the solutions of which are given in Eq.~\eqref{eq:general mf sol}. Fluctuations $\delta\Phi(\vbg,\vbx)$ around the solutions $\Phi_0$ are governed by the effective action, 
\begin{equation}
\label{eq: effective_action}
S_\mathrm{eff}[\delta\Phi] = \int\dd{\vbg}\dd{\tilde{\vbg}}\int\dd{\vbx}\delta\Phi(\vbg,\vbx)\left[\delta(\vbg,\tilde{\vbg})\left(-\sum_c\Delta^c_{\tilde{g}}-\Delta_{\mathcal{M}}\right)^{\zeta}+b(\vbg,\tilde{\vbg})\right]\delta\Phi(\tilde{\vbg},\vbx),
\end{equation}
where the Laplace operator $\Delta_{\tilde{g}}^c$ acts on the variable $\tilde{g}_c$ and with $b$ the effective mass, $b(\vbg,\tilde{\vbg}) = \mu\left(\delta(\vbg,\tilde{\vbg})-\chi(\vbg,\tilde{\vbg})\right)$. Here, $\delta(\vbg,\tilde{\vbg})$ is the $\delta$-distribution on $G^{r}$. The bi-local function $\chi$ is extracted from the Hessian of the interaction term and is computed as shown in detail in Sec.~\ref{sec:U1 linearization}. 

%In the setup of this section, we keep the assumptions on the model as general as possible. However, for all the examples provided in Sec.~\ref{sec:Application} that we are going to study explicitly, the manifold $\mathcal{M} = \R^{\dloc}$ and $G$ is assumed to be a Lie group manifold rendering these models tensorial GFTs with additional local variables. Interaction graphs $\gamma$ of a tensor-invariant type~\cite{Gurau:2011tj,Bonzom:2012hw,GurauBook} are of particular interest for modern tensor models and tensor field theories~\cite{Gurau:2019qag} while simplicial graphs directly relate to quantum gravity approaches such as LQG~\cite{Ashtekar:2004eh}, spin foams models~\cite{Perez:2003vx,Perez:2012wv,Engle:2023qsu,Livine:2024hhc}, GFTs~\cite{Freidel:2005qe,Oriti:2006se,Gurau:2011xp,Carrozza:2013oiy}, simplicial gravity path integrals~\cite{Bonzom:2009hw,Baratin:2010wi,Baratin:2011tx,Baratin:2011hp,Finocchiaro:2018hks} and dynamical triangulations~\cite{Loll:1998aj,Ambjorn:2012jv,Jordan:2013sok,Loll:2019rdj,Jercher:2022mky}. We also present more exotic examples of graphs $\gamma$ in Tab.~\ref{tab:chis} which are not part of either of these two classes, underlining the generality of the statements we are about to derive.  

Using harmonic analysis on the domains $G^r$ and $\mathcal{M}$, we \emph{formally} define a decomposition of the field $\delta\Phi$ in terms of eigenfunctions of the Laplace operators which form an orthogonal basis of $L^2\left(G^r\times\mathcal{M}\right)$. Details and explicit expressions of this decomposition depend on the specific choices of $G$ and $\mathcal{M}$, for which we presented detailed examples in Secs.~\ref{sec:example} and~
\ref{sec:towards general TFTs}. To that end, we introduce the set of eigenfunctions $D^{(j_c)}_G(g_c)$ and $D^{(\vbk)}(\vbx)$ satisfying
\begin{equation}
-\Delta_g^c D_G^{(j_c)}(g_c) = \lambda_g(j_c)D_G^{(j_c)}(g_c),\qquad -\Delta_{\mathcal{M}}D_{\mathcal{M}}^{(\vbk)}(\vbx) = \lambda_{\mathcal{M}}(\vbk)D_{\mathcal{M}}^{(\vbk)}(\vbx),  
\end{equation}
where $j_c$ and $\vbk$ are discrete or continuous, depending on whether $G$ and $\mathcal{M}$ are compact or non-compact, respectively. The $\lambda_g$ and $\lambda_{\mathcal{M}}$ are the respective eigenvalues and functions of the $j_c$ and $\vbk$. Then, the fluctuation field $\delta\Phi\in L^2\left(G^r\times\mathcal{M}\right)$ is decomposed in this basis as
\begin{equation}
\delta\Phi(\vbg,\vbx) = \sumint\!\!{\raisebox{-2.5mm}{\scalebox{0.8}[0.8]{$\vb*{j},\vbk$}}}\; \delta\Phi(\vb*{j},\vbk)\prod_{c=1}^r D_G^{(j_c)}(g_c) D_{\mathcal{M}}^{(\vbk)}(\vbx),
\end{equation}
where the details of the measures on $\vb*{j}$ and $\vbk$ depend again on the choice and properties of $G$ and $\mathcal{M}$, respectively. In this decomposition, the effective action of Eq.~\eqref{eq: effective_action} reads
\begin{equation}
S_{\mathrm{eff}}[\delta\Phi] = \sumint\!\!{\raisebox{-2.5mm}{\scalebox{0.8}[0.8]{$\vb*{j},\vbk$}}}\; \delta\Phi_{\vb*{j}}(\vbk)\left[\left(\sum_c\lambda_g(j_c)+\lambda_{\mathcal{M}}(\vbk)\right)^{\zeta}+b_{\vb*{j}}\right]\delta\Phi_{-\vb*{j}}(-\vbk),
\end{equation}
where $-\vb*{j}$ and $-\vbk$ are dual to the labels $\vb*{j}$ and $\vbk$, respectively. 

The effective mass contains projections onto zero modes $j=j_0$ via integrals of the form 
\begin{equation}
\frac{1}{V_G}\int\dd{g}D^{(j)}(g) = \delta_{j,j_0},
\end{equation}
and we refer the reader to Sec.~\ref{sec:general TGFT} for a detailed discussion of the definition of this integral for non-compact $G$. The effective action can be split into zero mode contributions
\begin{equation}
\begin{aligned}
S_{\mathrm{eff}}[\delta\Phi] &= \sum_{s=0}^r\sum_{(c_1\dots c_s)}\sumint\!\!{\raisebox{-2.5mm}{\scalebox{0.8}[0.8]{$\vb*{j}_{r-s},\vbk$}}}\;\delta\Phi_{\vb*{j}_{r-s}}(\vbk)\\[7pt]
&\times\left[\left(\sum_{c=c_{s+1}}^{c_r}\lambda_g(j_c)+\lambda_{\mathcal{M}}(\vbk)\right)^{\zeta}+b_{c_1\dots c_s}\right]\delta\Phi_{-\vb*{j}_{r-s}}(-\vbk),
\end{aligned}
\end{equation}
where $b_{c_1\dots c_s}$ is the effective mass with zero modes injected at $(c_1,\dots c_s)$. 

With the example of a melonic TGFT in Sec.~\ref{sec:example}, a non-positive effective mass has been observed for $s<r$ zero modes. To which extent the correlations of fluctuations depend on this particular behavior, hinges on the specifics of the non-local domain $G$. Therefore, a general expression of $Q$ in the limit $\mu\rightarrow 0$ cannot be given, and we refer the reader to the previous sections for explicit examples.

In the following section we prove that the effective mass is non-positive for $s<r$ zero modes for general non-local interactions which holds irrespective of the details of $G$ and $\mathcal{M}$. 

\subsection{Vanishing effective mass for general non-local interactions}\label{sec:Vanishing effective mass for general non-local interactions}

The sign of the effective mass is crucial for the behavior of fluctuations in the phase with $\Phi_0\neq 0$. In previous applications of Landau-Ginzburg theory to TGFTs~\cite{Marchetti:2020xvf,Marchetti:2022igl,Marchetti:2022nrf}, the possibility of a negative effective mass has been realized which leads to the introduction of the characteristic number of zero modes $s_0$. The new feature is that there exists a third regime in which the effective mass vanishes beyond criticality. In this section, we prove that this third regime is present generically for single non-local interactions with vertex graphs $\gamma$ that do not contain loops. Examples of such graphs and the corresponding $\chi_{\vb*{j}}$ are presented in Tab.~\ref{tab:chis}. Remarkably, this result holds in great generality irrespective of the compactness properties of the group, the inclusion of Laplace operators on $G$ and $\mathcal{M}$ and the specifics of the non-local interactions. Therefore, it bears consequences for any theory of non-local interactions as presented here and forms the basis for the symmetry analysis we conduct in Sec.~\ref{sec:Symmetries of effectively massless theories}.

To begin with, consider the functional derivative of the interaction term $V[\Phi]$ of the full theory,
\begin{equation}\label{eq:vertex derivative gen}
\fdv{V[\Phi]}{\Phi(\vbg,\vbx)} = \lambda \sum_{v\in\mathcal{V}_\gamma}\Tr_{\gamma/v}\left[\Phi(\vbg,\vbx)^{n_\gamma -1}\right].
\end{equation}
In the following, we introduce the necessary notation to study the expansion of this expression around the background mean-field solution.

The set of vertices $v'$ that are adjacent to $v$ is referred to as $\mathcal{A}_v$. Notice that $\abs{\mathcal{A}_v}\leq r$ with $\abs{\mathcal{A}_v} = r$ and $\abs{\mathcal{A}_v} = 1$ corresponding to simplicial and trace-melonic combinatorics, respectively. A pair of vertices $vv'$ can have multiple connecting links, such as in the trace-melonic case. We denote this multiplicity with $\iota_{vv'}$ which satisfies 
\begin{equation}\label{eq:iota properties}
\iota_{vv'}\leq r,\qquad\text{and}\qquad   \sum_{v'\in\mathcal{A}_v}\iota_{vv'} = r.
\end{equation}
With this notation, we label the pairs $vv'$ with $\{1,\dots,\abs{\mathcal{A}_v}\}$ and we introduce the following partitioning of the group element indices 
\begin{equation}
c_1,\dots, c_r\longrightarrow \underbrace{c^{(1)}_1,\dots, c^{(1)}_{\iota_1}},\underbrace{c^{(2)}_1,\dots,c^{(2)}_{\iota_2}},\dots,\underbrace{c^{(\abs{\mathcal{A}_v})}_1,\dots,c^{(\abs{\mathcal{A}_v})}_{\iota_{\abs{\mathcal{A}_v}}}}.
\end{equation}
The superscript in brackets denotes the pair of vertices $vv'$ while the subscript indicates which of the $\iota_{vv'}$ links the group element corresponds to. 

At this stage, the derivative of the interaction term in Eq.~\eqref{eq:vertex derivative gen} can be expanded in terms of perturbations around the mean-field solution $\Phi_0$, yielding
\begin{equation}\label{eq:vertex derivative group rep}
\begin{aligned}
&\eval{\fdv{V[\Phi]}{\Phi(\vbg,\vbx)}}_{\Phi_0+\delta\Phi}  \\[7pt]
=& -\frac{\mu}{n_\gamma}\mathrm{V}_G^{-r}\int\dd{\tilde{\vbg}}\sum_{v\in\mathcal{V}_\gamma}\left[\sum_{v'\in\mathcal{A}_v}\mathrm{V}_G^{\iota_{vv'}}\prod_{c=c^{vv'}_1}^{c^{vv'}_{\iota_{vv'}}}\delta(g_c,\tilde{g_c})+n_\gamma-\abs{\mathcal{A}_v}-1\right]\delta\Phi(\tilde{\vbg},\vbx).
\end{aligned}
\end{equation}
In this step, we inserted the background solution of Eq.~\eqref{eq:general mf sol}, leading to the presence of the parameter $\mu$ and the powers of volume factors $\mathrm{V}_G$. Notice that the summand $n_\gamma -\abs{\mathcal{A}_v} -1$ enters because the vertex $v$, connected to $\abs{\mathcal{A}_v}$ other vertices, is disconnected from $n_\gamma -\abs{\mathcal{A}_v} - 1$ vertices. In the derivative of the interaction, the perturbations associated with these vertices are integrated out fully, thus not yielding a $\delta$-function on $G$. 

As a next step, we go to representation space as described in the previous section. In Eq.~\eqref{eq:vertex derivative group rep}, this procedure yields a factor of unity for every $\delta$-function on $G$ and a projection onto the zero mode $j_0$ for every constant term with inverse volume factor $\mathrm{V}_G$, see Eq.~\eqref{eq:zero mode projection}. As a result, one obtains the function $\chi_{\vb*{j}}$,
\begin{equation}\label{eq:chi gen}
\chi_{\vb*{j}} = \frac{1}{n_\gamma}\sum_{v\in\mathcal{V}_\gamma}\left[\sum_{v'\in\mathcal{A}_v}\prod_{c=c_1^{vv'}}^{c^{vv'}_{r-\iota_{vv'}}}\delta_{j_c,j_0}+(n_\gamma-\abs{\mathcal{A}_v}-1)\prod_{c=1}^r\delta_{j_c,j_0}\right],
\end{equation}
where we have relabelled the elements $c_1^{(1)},\dots,c^{(\abs{\mathcal{A}_v})}_{\iota_{\abs{\mathcal{A}_v}}}$ accordingly. With this formula, we are now in position to study the behavior of $\chi_{\vb*{j}}$, and thus of the effective mass $b_{\vb*{j}} = \mu(1-\chi_{\vb*{j}})$, when evaluated on $s$ zero modes. 

In accordance with the end of the last section, let us first consider the case of $s = r$ zero modes. Clearly, all the Kronecker-$\delta$'s in Eq.~\eqref{eq:chi gen} evaluate to one, yielding
\begin{equation}
\eval{\chi_{\vb*{j}}}_{s=r} = \frac{1}{n_\gamma}\sum_{v\in\mathcal{V}_\gamma}\left[\sum_{v'\in\mathcal{A}_v}1 + n_\gamma -\abs{\mathcal{A}_v} - 1\right] = \frac{1}{n_\gamma}\sum_{v\in\mathcal{V}_\gamma}\left[n_\gamma - 1\right] = n_\gamma - 1.
\end{equation}
Correspondingly, the effective mass $b_{\vb*{j}}$ evaluated on $r$ zero modes is given by
\begin{equation}
\eval{b_{\vb*{j}}}_{s=r} = \mu(1-(n_\gamma - 1) )= \abs{\mu}(n_\gamma - 2)>0,
\end{equation}
which corresponds to the well-known result from the Landau-Ginzburg analysis of local field theories~\cite{Kopietz:2010zz}. 

Next, we consider $s < r$ zero modes and show that the effective mass is either negative or vanishing. Moreover, we study for which properties of the vertex graph $\gamma$ either of the two behaviors are obtained. As a first step, we notice that on $s<r$ zero modes, the product of $r$ Kronecker-$\delta$'s in Eq.~\eqref{eq:chi gen} always vanishes. Secondly, we show that the sum over $v'$ in Eq.~\eqref{eq:chi gen} evaluates to either zero or one for each $v\in\mathcal{V}_\gamma$. It follows immediately that if $r-\iota_{vv'} > s$ for all $v'\in\mathcal{A}_v$, then the product of $\delta$'s needs to vanish. Suppose instead that there exists a $v_1\in\mathcal{A}_v$, such that $r-\iota_{vv_1} \leq s$. Then, there exists a configurations with indices $c_1,\dots, c_s$ such that 
\begin{equation}
\prod_{c=c_1^{vv_1}}^{c^{vv_1}_{r-\iota_{vv_1}}}\delta_{j_c,j_0} = 1.
\end{equation}
It remains to show that this is the only summand that evaluates to one. To that end, assume that there exists a $v_2\in\mathcal{A}_v$, such that $r-\iota_{vv_2}\leq s-(r-\iota_{vv_1})$. This implies that
\begin{equation}
\iota_{vv_1}+\iota_{vv_2} \geq 2r-s > r,
\end{equation}
which is a contradiction with Eq.~\eqref{eq:iota properties}. Hence, no such $v_2$ can exist. Overall, we find
\begin{equation}
\sum_{v'\in\mathcal{A}_v}\prod_{c=c_1^{vv'}}^{c^{vv'}_{r-\iota_{vv'}}}\delta_{j_c,j_0}\in\{0,1\}\quad \forall \;v\in\mathcal{V}_\gamma.
\end{equation}
As a result, we obtain the following inequality for $\chi_{\vb*{j}}$,
\begin{equation}
\eval{\chi_{\vb*{j}}}_{s<r} = \frac{1}{n_\gamma}\sum_{v\in\mathcal{V}_\gamma}\sum_{v'\in\mathcal{A}_v}\prod_c\delta_{j_c,j_0}\leq \frac{1}{n_\gamma}\sum_{v\in\mathcal{V}_\gamma}1\leq 1,
\end{equation}
resulting in an inequality for $b_{\vb*{j}}$,
\begin{equation}
\eval{b_{\vb*{j}}}_{s<r} = \abs{\mu}(\chi_{\vb*{j}}-1) \leq 0.
\end{equation}
This is a novel result as it shows that one cannot have a positive effective mass for such non-local interactions. In particular, the necessity of a non-positive mass for $s<r$ is new compared to previous works on the Landau-Ginzburg analysis of TGFTs~\cite{Marchetti:2020xvf,Marchetti:2022igl,Marchetti:2022nrf}. We emphasize the generality of this observation which applies to any non-local interaction that is governed by a graph $\gamma$ of equal-valent vertices without loops. In principle, this behavior is not limited to a single interaction and we commented on the inclusion of multiple interactions of the same degree $n_\gamma$ in Sec.~\ref{sec:multiple interactions}. Determining the extent to which this observation also applies to a sum of interactions of different degree is obstructed by the mean-field equations and goes beyond the scope of our work.   

To obtain a strictly vanishing effective mass, the required minimal number of zero modes as well as the position of the slots at which the zero modes are inserted crucially depends on the vertex graph $\gamma$. For every $\gamma$, there is a minimal number of zero modes $s_0$, such that there exist configurations labelled by $(c_1,\dots,c_s)\in \mathcal{O}_s^\gamma$, with $s\geq s_0$, on which the effective mass vanishes. The set of configurations of $s < r$ zero modes for which the effective mass is negative is denoted by $\bar{\mathcal{O}}_s^\gamma$.\footnote{Notice that there can in principle be combinatorics encoded by some $\gamma$ that yield a negative effective mass even for $s>s_0$ zero modes.} With this distinction, we summarize the three different behaviors of the effective mass as
\begin{equation}
b_{c_1\dots c_s} 
\begin{cases}
=\abs{\mu}(n_\gamma -2) > 0,\quad &\text{for }s=r,\\[7pt]
=0,\quad &\text{for }s_0\leq s <r   \text{ and  }(c_1,\dots, c_s)\in \mathcal{O}_s^\gamma,\\[7pt]
< 0,\quad &\text{for }s < r   \text{ and }(c_1,\dots, c_s)\in\bar{\mathcal{O}}_s^\gamma.
\end{cases}
\end{equation}

\paragraph{Examples.} In Tab.~\ref{tab:chis}, an exemplary list of vertex graphs $\gamma$, together with the corresponding functions $\chi_{\vb*{j}}$ is provided. These examples include tensor-invariant interactions~\cite{Bonzom:2012hw,GurauBook,Carrozza:2013oiy}, such as multi-traces, cyclic melons, a chain of necklaces and the utility graph $K_{3,3}$ as a sextic interaction at rank $r=3$, see for instance~\cite{Bonzom:2015kzh}. Also, we included simplicial interactions in two, three and four dimensions, corresponding to the $r=2$ chain with $n_\gamma = 3$ (which directly relates to the case of matrix field theory~\cite{kontsevich1992intersection,grosse2014self}), the tetrahedron and the $4$-simplex, respectively. In all of these cases, the minimum number of zero modes, $s_0$, can be determined below which $\chi < 1$ and thus $b < 0$. 

\begin{table}[h]
    \centering
    \begin{tabular}{c|c|l}
    \hline
    \multicolumn{3}{c}{$r=2$} \\
    \hline
        multi-trace & $\bigsqcup\limits_{i=1}^{n_\gamma/2}$ {\scriptsize \twomelon} & $(n_\gamma-2)\delta_{j_1,0}\delta_{j_2,0}+1$\\
        multi-chain & {\scriptsize \twochain} & $(n_\gamma -3)\delta_{j_1,0}\delta_{j_2,0}+\delta_{j_1,0}+\delta_{j_2,0}$ \\
    \hline
    \multicolumn{3}{c}{$r=3$} \\
    \hline
        multi-trace & $\bigsqcup\limits_{i=1}^{n_\gamma/2}$ {\scriptsize \threemelon} & $(n_\gamma-2)\prod\limits_{c=1}^3\delta_{j_c,0}$+1 \\
        cyclic $2$-melon & {\scriptsize \twomelonchain} & $(n_\gamma-3)\prod\limits_{d=1}^3\delta_{j_d,0}+\delta_{j_c,0}+\prod\limits_{b\neq c}\delta_{j_b,0}$\\
         tetrahedron & {\scriptsize \tetrahedron} & $\sum\limits_{c=1}^3\prod\limits_{b\neq c}\delta_{j_b,0}$ \\
         $K_{3,3}$ & {\scriptsize \kthreethree} & $2\prod\limits_{c=1}^3\delta_{j_c,0}+\sum\limits_{c=1}^3\prod\limits_{b\neq c}\delta_{j_b,0}$ \\
    \hline
    \multicolumn{3}{c}{$r=4$} \\
    \hline
        multi-trace & $\bigsqcup\limits_{i=1}^{n_\gamma/2}$ {\scriptsize \fourmelon} & $(n_\gamma-2)\prod\limits_{c=1}^4\delta_{j_c,0}+1$ \\
        cyclic $3$-melon & {\scriptsize \threemelonchain} & $(n_\gamma-3)\prod\limits_{d=1}^4\delta_{j_d,0}+\delta_{j_c,0}+\prod\limits_{b\neq c}\delta_{j_b,0}$ \\
        necklace chain & {\scriptsize \necklacechain} & $(n_\gamma-3)\prod\limits_{c=1}^4\delta_{j_c,0}+\delta_{j_1,0}\delta_{j_2,0}+\delta_{j_3,0}\delta_{j_4,0}$ \\
        $4$-simplex & {\scriptsize \cvfs} & $\sum\limits_{c=1}^4\prod\limits_{b\neq c}\delta_{j_b,0}$ \\
    \hline
    \multicolumn{3}{c}{Exotic examples} \\
    \hline
        {} & {\scriptsize \exoticone} & $\delta_{j_c,0}\delta_{j_{c'},0}+\prod\limits_{b_1\neq c_1}\delta_{j_{b_1},0}+\prod\limits_{b_2\neq c_2}\delta_{j_{b_2},0}$ \\
            {} & {\scriptsize \exotictwo} & $\frac{2}{3}\left(3\prod\limits_{c=1}^4\delta_{j_c,0}+\delta_{j_{c_1},0}+\prod\limits_{b_1\neq c_1}\delta_{j_{b_1},0}\right)$\\
            {} & {} & $+\frac{1}{3}\left(2\prod\limits_{c=1}^4\delta_{j_c,0}+\delta_{j_{c_2},0}\delta_{j_{c_3},0}+\prod\limits_{b_2\neq c_2}\delta_{j_{b_2},0}+\prod\limits_{b_3\neq c_3}\delta_{j_{b_3},0}\right)$
    \end{tabular}
    \caption{Explicit expressions of $\chi$ for different vertex graphs $\gamma$. We set $\delta_{j,j_0}\equiv \delta_{j,0}$ for notational ease. The vertex structure of local variables $\boldsymbol{x}$, is again suppressed here.}
    \label{tab:chis}
\end{table}

\subsection{Symmetries of effectively massless theories}\label{sec:Symmetries of effectively massless theories}

The results of the previous section suggest the splitting of the effective action $S_\mathrm{eff}[\delta\Phi]$ according to the contribution obtained from the three different regimes where the effective mass is negative, vanishing, and positive, respectively. For $b_{c_1\dots c_s}>0$, the action is constant in the non-local variables and reduces to a local field theory, while for $b_{c_1\dots c_s} < 0$, the action is unstable and is not sensitive to the critical behavior. Focusing instead on the contribution of vanishing effective mass, the residual theory in representation space is governed by the following action
\begin{equation}\label{eq:residual effective action}
\begin{aligned}
&\eval{S_\mathrm{eff}[\delta\Phi]}_{b = 0} = \sum_{s=s_0}^{r-1}\sum_{(c_1,\dots, c_s)\in \mathcal{O}_s^\gamma}\int\dd{\vbg_{r-s}}\int\dd{\vbx}\delta\Phi(\vbg_{r-s},\vbx)\\[7pt]
&\times \left(-\sum_{c=c_{s+1}}^{c_r}\Delta^c_{g}-\Delta_{\mathcal{M}}\right)^{\zeta}\delta\Phi(\vbg_{r-s},\vbx),
\end{aligned}
\end{equation}
where $\delta\Phi(\vbg_{r-s},\vbx)$ is short-hand notation for the field $\delta\Phi$ being constant in the entries $(c_1,\dots, c_s)\in \mathcal{O}_s^\gamma$, and depending on the local frame coordinate $\vbx$ and the remaining $r-s$ group variables $g_{c_{s+1}}, \dots,g_{c_r}$. This residual theory is effectively massless and as such, it attains symmetries that were absent in the unperturbed action for $\Phi(\vbg,\vbx)$. Notice, that in contrast to local field theories, these symmetries are not only obtained at criticality, i.e. at $\mu = 0$, but in the entire phase with $\mu < 0$, that is where the order parameter $\Phi_0\neq 0 $. In the following, we provide a classification of the resulting symmetries.

\paragraph{Scale invariance.} Local free and massless scalar field theories exhibit a scale invariance, which similarly applies to the present model. To make this explicit, we denote the total domain as $\mathcal{D} = G^r\times\mathcal{M}$ with product metric $h_{G^r}\oplus h_{\mathcal{M}}$, and write $d_\mathcal{D} = d_G(r-s)+\dloc$, $h_{\mathcal{D}}$ and $\Delta_{\mathcal{D}}$ for the dimension, metric and the Laplace operator of $\mathcal{D}$, respectively. Then, let us consider a global re-scaling of the metric $h_{\mathcal{D}}$, 
\begin{subequations}\label{eq:metric re-scaling}
\begin{align}
h_{\mathcal{D}} &\longmapsto \e^{2\omega}h_{\mathcal{D}}, \\[7pt]
\dd{\vbg}\dd{\vbx}&\longmapsto \e^{d_{\mathcal{D}}\omega} \dd{\vbg}\dd{\vbx},\\[7pt]
   \left(-\sum_{c=s+1}^r\Delta_g^c-\Delta_{\mathcal{M}}\right)^\zeta&\longmapsto \e^{-2\zeta\omega}\left(-\sum_{c=s+1}^r\Delta_g^c-\Delta_{\mathcal{M}}\right)^\zeta,\label{eq:Laplace re-scaling}
\end{align}
\end{subequations}
and the field $\delta\Phi$,
\begin{equation}\label{eq:field re-scaling}
\delta\Phi(\vbg_{r-s},\vbx)\longmapsto \e^{-\frac{d_{\mathcal{D}}-2\zeta}{2}\omega}\delta\Phi(\vbg_{r-s},\vbx),
\end{equation}
with $\omega\in\R$ a constant. Clearly, the action in Eq.~\eqref{eq:residual effective action} is invariant under such a re-scaling which defines the scaling dimension of $\delta\Phi$ as\footnote{Parametrizing $\zeta = 1-\frac{\eta}{2}$, it is clear that the presence of $\zeta$ is similar to an anomalous dimension $\eta$~\cite{WipfRG,hohenberg2015introduction}, which modifies the scaling dimension as
\begin{equation}\label{eq:scaling dim eta}
\Delta = \frac{d_G(r-s)+\dloc - 2+\eta}{2}.
\end{equation}}
\begin{equation}\label{eq:scaling dim zeta}
\Delta = \frac{d_G(r-s)+\dloc-2\zeta}{2}.
\end{equation}

We emphasize that the scaling transformations in Eq.~\eqref{eq:metric re-scaling} need to be performed simultaneously on the entire product manifold $G^r\times\mathcal{M}$ with the induced metric $h_G\oplus h_{\mathcal{M}}$. In particular, given Laplace operators on both domains, one does not have a scale invariance on only one of the factors, i.e. on $G^r$ or on $\mathcal{M}$. Furthermore, we notice that the introduction of different $\zeta$-parameters for the respective Laplacians, i.e. $(-\Delta_G)^{\zeta_{\mathrm{nloc}}}$ and $(-\Delta_{\mathcal{M}})^{\zeta_\mathrm{loc}}$, immediately breaks scale invariance. This can be seen from Eq.~\eqref{eq:Laplace re-scaling}, where the scaling factor $\e^{2\omega}$ cannot be factorized for $\zeta_{\mathrm{nloc}}\neq\zeta_{\mathrm{loc}}$. The explicit breaking of scale symmetry would not only apply to the residual theory in the $\Phi_0\neq 0$ phase but also to the original theory in Eq.~\eqref{eq:general kinetic term} at criticality, $\mu = 0$. This would clearly inhibit the existence of a phase transition which justifies the form of the kinetic term in Eq.~\eqref{eq:general kinetic term}.

\paragraph{Conformal invariance.}  Scale invariance of a field theory is a necessary but not a sufficient condition for conformal invariance. In spacetime-based field theories, the equivalence of these two symmetries has been proven for dimension $d=2$ under certain assumptions~\cite{Polchinski:1987dy,Zamolodchikov:1986gt}. In $d > 2$, the relation between these two symmetries is subject of active research and we refer the reader to~\cite{Jackiw:2011vz,Nakayama:2013is} for reviews. 

For the residual effective action in Eq.~\eqref{eq:residual effective action}, conformal invariance can be studied by enhancing the global re-scalings of Eqs.~\eqref{eq:metric re-scaling} and~\eqref{eq:field re-scaling} with $\omega\in\R$ to local re-scalings by assuming $\omega$ to be a real-valued smooth function on the domain $G^r\times\mathcal{M}$. Such transformations are referred to as Weyl transformations. Together with passive diffeomorphism (i.e. coordinate transformation) invariance, the invariance under Weyl transformations implies conformal invariance~\cite{Nakahara:2003nw}. Importantly, the converse is not necessarily true and thus, the absence of Weyl invariance does not imply the absence of conformal invariance. We refer the reader to~\cite{Nakayama:2013is} for further details on the relation between Weyl and conformal invariance.

In the following, we set $\zeta = 1$ and comment on the inclusion of $\zeta$ in the analysis at the end of the section.  If $\omega$ is now a function on $\mathcal{D}$, then 
\begin{equation}
\Delta_{\mathcal{D}}\delta\Phi\longmapsto \e^{-2\omega}\e^{-\frac{d_\mathcal{D}-2}{2}\omega}\left[\Delta_{\mathcal{D}}\delta\Phi -\frac{d_\mathcal{D}-2}{2}\left(\Delta_{\mathcal{D}}\omega+\frac{d_{\mathcal{D}-2}}{2} h_{\mathcal{D}}(\dd{\omega},\dd{\omega})\right)\delta\Phi\right],
\end{equation}
Clearly, in the case where the field's domain $\mathcal{D}$ is curved, the action in Eq.~\eqref{eq:residual effective action} is not invariant under Weyl transformation. To ensure the presence of this symmetry, we notice that the Ricci scalar $R_{\mathcal{D}}$ on $\mathcal{D}$ transforms under Weyl transformations as
\begin{equation}
R_{\mathcal{D}}\longmapsto 2(d_{\mathcal{D}}-1)\e^{-2\omega}\left[R_{\mathcal{D}}-\Delta_{\mathcal{D}}\omega-\frac{d_{\mathcal{D}-2}}{2} h_{\mathcal{D}}(\dd\omega,\dd\omega)\right].
\end{equation}
As a result, a Weyl- and therefore conformally invariant residual effective action would be given by
\begin{equation}
\begin{aligned}
&\eval{S^{\mathrm{inv}}_{\mathrm{eff}}}_{b=0} = \sum_{s=s_0}^{r-1}\sum_{(c_1,\dots, c_s)\in\mathcal{O}_s^\gamma}\int\dd{\vbg_{r-s}}\dd{\vbx}\delta\Phi(\vbg_{r-s},\vbx)(-\Delta_{\mathcal{D}}^{\mathrm{conf}})\delta\Phi(\vbg_{r-s},\vbx),
\end{aligned}
\end{equation}
with the conformal Laplacian
\begin{equation}
-\Delta_{\mathcal{D}}^{\mathrm{conf}} = -\Delta_{\mathcal{D}} + \frac{d_{\mathcal{D}}-2}{4(d_{\mathcal{D}}-1)}R_{\mathcal{D}}.
\end{equation}

From these considerations, we can draw the following conclusions:
\begin{itemize}
    \item For $\mathcal{M} = \R^{\dloc}$ and $G$ being either $\R$ or U$(1)$, the effectively massless theory is indeed conformally invariant. 
    \item It is a result proven in~\cite{Zamolodchikov:1986gt,Polchinski:1987dy} that for local field theories on Minkowski spacetime, a two-dimensional scale-invariant theory is conformally invariant if the theory is unitary and Poincar\'{e} invariant~\cite{Nakayama:2013is}. In principle, the dimension of the total domain can be two, $d_{\mathcal{D}} = d_G(r-s)+\dloc = 2$. However, unitarity requires the definition of a time direction whose interpretation is not straightforward in the present setting. In fact, this missing element in the analysis is a common feature of non-local field theories, especially those that are prominent in background-independent quantum gravity approaches, where the notion of spacetime is a relationally-emergent quantity from pre-geometric degrees of freedom. In the current setting, these are precisely the elements $\vbg\in G$~\cite{Oriti:2018dsg}. 
    
    Moreover, this case is potentially related to the case of matrix quantum mechanics~\cite{Gross:1990pa}. We leave it to future research to explore whether ($d_{\mathcal{D}} = 2$)-dimensional scale invariant and non-local field theories are also conformally invariant.
    \item For $\mathcal{M} = \R^{\dloc}$ and $G = \SL$, the residual massless theory in Eq.~\eqref{eq:residual effective action} does not exhibit Weyl invariance since $\SL$ has curvature. Notice that $\SL$ is topologically given as $S^3\times\TH$, and thus, its Ricci scalar is a constant. Hence, Weyl invariance would be ensured by adding a constant term quadratic in $\delta\Phi$.
\end{itemize}

Examining the symmetry properties of the residual effective action for $\zeta < 1$ is obscured by the fact that in this case, the Laplacians act as fractional derivative operators. One conceivable modification of Eq.~\eqref{eq:residual effective action} that ensures conformal symmetry would be to take a power of the conformal Laplacian, i.e. to introduce $(-\Delta_{\mathcal{D}}^{\mathrm{conf}})^{\zeta}$. We leave it to future investigations, if and under which conditions such long-range theories show conformal symmetry.

To summarize, we observed the emergence of effectively massless theories in the mean-field approximation of field theories with local and non-local variables in the $\mu\leq 0$ phase. We have studied under which conditions of the vertex graph $\gamma$ the effective mass vanishes. Furthermore, we studied the resulting symmetries finding scale invariance and under certain conditions also conformal invariance. Importantly, all of these observations crucially hinged on the mean-field approximation the validity of which we have studied in several explicit scenarios throughout this work. 
%It was therefore essential to scrutinize the validity of this method. Following the standard procedure of Landau-Ginzburg theory, we computed the parameter $Q$, defined in Eq.~\eqref{eq:general Q}, and checked under which conditions $Q\ll 1$ holds true. The precise form of correlation functions on $\mathcal{D}$ and thus of the Ginzburg-$Q$ depends on the specific details of $G$ and $\mathcal{M}$. For definiteness, we considered in the subsequent section examples of theories the phase structure of which has been studied via standard methods like in the case of tensor field theories~\cite{Gurau:2019qag,Benedetti:2020seh,Harribey:2022esw} or via the FRG methodology like in the case of TGFTs on U$(1)$ and $\R$ with and without additional local variables~\cite{BenGeloun:2015ej,BenGeloun:2016kw,Pithis:2020kio,Pithis:2020sxm,Geloun:2023ray} and via the Landau-Ginzburg mean-field approach also for more involved quantum geometric models~\cite{Pithis:2018eaq,Marchetti:2020xvf,Marchetti:2022igl,Marchetti:2022nrf}.
We note that the vanishing effective mass of the relevant modes is a new observation of this work and that its impact on the mean-field analysis has been made explicit.

%Thereafter, we consider the Lorentzian Barrett-Crane TGFT model~\cite{Perez:2000ec,Barrett:1999qw,Jercher:2021bie,Jercher:2022mky} with $G=\SL$ and additional constraints and advance its mean-field analysis performed in~\cite{Marchetti:2022nrf,Marchetti:2022igl}. This model provides a TGFT quantization of first-order Palatini gravity and the application of Landau-Ginzburg theory to this model is particularly interesting since it allows us to verify phase transitions therein towards a non-perturbative vacuum state with a tentative continuum geometric interpretation.

\section{Conclusion and discussion}\label{sec:conclusion}

The main objective of this article was to further develop the application of Landau-Ginzburg mean-field theory to tensorial field theories which allows the investigation of their basic phase structure and the realization of phase transitions therein. 

To this aim, building on previous works~\cite{Pithis:2018eaq,Marchetti:2020xvf,Marchetti:2022igl,Marchetti:2022nrf}, we carried over the Landau-Ginzburg method from the context of local statistical field
theories to tensorial field theories which are marked by their combinatorial non-local interactions. In particular, we improved on these works by carefully analyzing the structure of their phase space in the infrared. We do this by reference to the effective mass which is not a constant, in contrast to usual local field theories. We explicitly show that non-local interactions generically lead to a regime of a vanishing effective mass for the modes relevant to describe the critical behavior in the infrared. We emphasize that this result holds in great generality irrespective of the specifics of the field domain and the type of non-local interactions. In effect, such theories therefore become massless and free in this regime providing the necessary condition for scale invariance. Moreover, we discussed under which conditions this symmetry is enhanced to conformal invariance on the residual configuration space. In the case of the TFT models analyzed, we conjecture that our findings, valid for any $N$, are the mean-field level pendant of 
%those results on their conformal invariance extracted via their Dyson-Schwinger equations at large $N$.
the non-perturbative results extracted from the Dyson-Schwinger equations at large $N$ in the infrared, see~\cite{Gurau:2019qag,Benedetti:2020seh} for reviews. 

In particular, we explain in detail the mechanism behind the vanishing of the effective mass. To this end, we employ a mode-by-mode expansion of the mean-field two-point correlation function. 
%This expansion allows us to evaluate it also on the zero modes which in turn determine its critical behavior.
Depending on the type of interaction, we determine the number of zero modes for which the effective mass is positive, negative, and for which it vanishes. For the latter case, we further develop the application of the Landau-Ginzburg method by introducing a regularized mass that can be sent to zero towards the end of our analysis.

With this improvement, we compute the Ginzburg $Q$-parameter which allows us to quantify the strength of linearized perturbations over the mean-field background. From this quantity, we derive the critical dimension of several specific models investigated in this article, which explicitly depends on the combinatorics of the interactions with their respective minimum number of zero modes $s_0$. Our results are perfectly consistent with those previously obtained in this series of papers. However, they go beyond them by accounting for the vanishing of the effective mass and the consequential appearance of scale or conformal invariance.

We point out that despite the necessary simplifications of the mean-field setting, our findings agree with those obtained with more complex methods, especially with those employing FRG~\cite{Pithis:2020kio,Pithis:2020sxm,Geloun:2023ray}. Indeed, this underlines the effectiveness of the Landau-Ginzburg method for understanding the basic phase properties of such tensorial field theories. In particular, it proves to be very useful to towards the clarification of the continuum limit of realistic TGFT models for Lorentzian quantum gravity. 

In the following, we close by discussing the limitations of our work and possible future research directions. Clearly, our results critically hinge on the simplifying assumptions of Landau-Ginzburg mean-field theory. In particular, the projection onto 
%uniformized
constant field configurations could be lifted in future works to study the linearized fluctuations over other types of backgrounds. This would also help to better understand the relation of our findings on the TFT models to those gained with the Dyson-Schwinger equations. Apart from the case of TGFT models on a hyperbolic domain, below the critical dimension, non-perturbative methods are certainly required to unveil a more detailed account of the phase structure of tensorial field theories, as summarized for instance in~\cite{Carrozza:2016vsq,Gurau:2019qag,Benedetti:2020seh,Harribey:2022esw}. 

The vanishing of the effective mass for the relevant modes and its regularization is relevant for the Landau-Ginzburg mean-field analysis~\cite{Dekhil:2024djp} of the complete Barrett-Crane GFT model~\cite{Jercher:2022mky} and also for the corresponding analysis of the EPRL model as well as for other TGFT models for $4d$ Lorentzian quantum gravity. Since the FRG methodology is a direct non-perturbative extension of the Landau-Ginzburg setting~\cite{Delamotte:2007pf,Kopietz:2010zz}, our results should also have a bearing on the FRG analysis of corresponding models if the Wetterich equation is expanded around a non-trivial and constant vacuum, instead of a standard analysis around the trivial vacuum as in~\cite{Pithis:2020kio,Pithis:2020sxm,Geloun:2023ray}. More precisely, the analog of the vanishing effective mass should be observed also on the right-hand side of the latter equation which suggests that the regularization of the mass term employed in this work might then also be useful there.  

To better understand the physical implication and the precise technical details under which the emergence of conformal symmetry takes place in the models explored here, it could be of interest to investigate whether the so-called Landau-Ginzburg/CFT correspondence~\cite{Vafa:1988uu,Lerche:1989uy} also applies to tensorial field theories. It was already proven that such a correspondence holds for certain locally interacting scalar field theories. It states that the infrared fixed points of a given Landau-Ginzburg model with a polynomial interaction term can be associated with a two-dimensional
rational conformal field theory with a given central charge \cite{Camacho:2019dvi}. From the perspective of the current results of this article, it will require a careful analysis of the enhancement of the scale invariance, that occurred from the vanishing of the effective mass, in order to end up with an effective CFT in the infrared. Such investigation could also help to illuminate the origin of the conformal symmetry in TFTs relevant to quantum gravity research within the context of the AdS/CFT conjecture~\cite{Gurau:2019qag,Benedetti:2020seh}.

It is furthermore important to better understand the physical interpretation of the emergent scale and conformal symmetry on the residual configuration space. A particularly interesting but still feasible ground for explorations in this direction are TGFT models with cosmological interpretation, the phenomenology of which is actively studied using standard field-theory methods in a mean-field approximation~\cite{Gielen:2013kla,Gielen:2016dss,Oriti:2016acw,Pithis:2019tvp,Oriti:2024qav}. In particular, it could be interesting to check if this symmetry relates to scale and conformal invariance on superspace and minisuperspace in classical general relativity extracted among others with the Eisenhart-Duval lift method~\cite{Duval:2024eod,Cariglia:2014dwa,Cariglia:2015bla,Cariglia:2016oft,BenAchour:2017qpb,Cariglia:2018mos,BenAchour:2019ufa,BenAchour:2020njq,Achour:2021lqq,BenAchour:2022fif,BenAchour:2023dgj}. Along with the fact that the mean-field analysis allows us to transition to the picture of an emergent spacetime (when considering such quantum gravity approaches), this inquiry would provide the basis to explore whether the additional term in the kinetic part of the action required to ensure a conformal enhancement is indeed correspondent to a non-minimal coupling to the Ricci scalar, or on a more speculative ground, generating an $f(R)$ theory \cite{De_Felice_2010, CLIFTON20121,Sotiriou_2006}. %This of course requires a careful study of our model and such premises. 

\subsection*{Acknowledgements}

The authors thank J. Ben Achour, L. Marchetti, R. Schmieden, and J. Thürigen for insightful discussions. DO acknowledges financial support from the ATRAE programme of the Spanish Government, through the grant PR28/23 ATR2023-145735. DO and AGAP acknowledge funding from the Deutsche Forschungsgemeinschaft (DFG, German Research Foundation) research grants OR432/3-1 and OR432/4-1 and the John-Templeton Foundation via research grant 6242. AFJ acknowledges support by the DFG under Grant No 406116891 within the Research Training Group RTG 2522/1 and under Grant No 422809950. RD, AFJ, and AGAP are grateful for the generous financial support by the MCQST via the seed funding Aost 862933-9 granted to AGAP and the seed funding Aost 862981-8 granted to Jibril Ben Achour by the DFG under Germany’s Excellence Strategy – EXC-2111 – 390814868. AGAP in particular acknowledges funding by the DFG under the author’s project number 527121685 as a Principal Investigator. 

\appendix

\addcontentsline{toc}{section}{References}
\bibliographystyle{JHEP}
\bibliography{references.bib}

\providecommand{\href}[2]{#2}\begingroup\raggedright\begin{thebibliography}{100}

\bibitem{kontsevich1992intersection}
M.~Kontsevich, \emph{{Intersection theory on the moduli space of curves and the matrix Airy function}}, {\emph{Communications in Mathematical Physics} {\bfseries 147} (1992) 1}.

\bibitem{grosse2014self}
H.~Grosse and R.~Wulkenhaar, \emph{{Self-dual noncommutative $\phi^4$-theory in four dimensions is a non-perturbatively solvable and non-trivial quantum field theory}}, {\emph{Communications in Mathematical Physics} {\bfseries 329} (2014) 1069}.

\bibitem{grosse2006noncommutative}
H.~Grosse and M.~Wohlgenannt, \emph{Noncommutative qft and renormalization},  in \emph{Journal of Physics: Conference Series}, vol.~53, p.~764, IOP Publishing, 2006.

\bibitem{Rivasseau:2007ab}
V.~Rivasseau, \emph{{Non-commutative Renormalization}},  \href{https://arxiv.org/abs/0705.0705}{{\ttfamily 0705.0705}}.

\bibitem{Gurau:2011xp}
R.~Gurau and J.P.~Ryan, \emph{{Colored Tensor Models - a review}}, \href{https://doi.org/10.3842/SIGMA.2012.020}{\emph{SIGMA} {\bfseries 8} (2012) 020} [\href{https://arxiv.org/abs/1109.4812}{{\ttfamily 1109.4812}}].

\bibitem{GurauBook}
R.~Gurau, \emph{{Random Tensors}}, Oxford University Press (2016).

\bibitem{Freidel:2005qe}
L.~Freidel, \emph{{Group field theory: An Overview}}, \href{https://doi.org/10.1007/s10773-005-8894-1}{\emph{Int. J. Theor. Phys.} {\bfseries 44} (2005) 1769} [\href{https://arxiv.org/abs/hep-th/0505016}{{\ttfamily hep-th/0505016}}].

\bibitem{Oriti:2006se}
D.~Oriti, \emph{{The Group field theory approach to quantum gravity}},  \href{https://arxiv.org/abs/gr-qc/0607032}{{\ttfamily gr-qc/0607032}}.

\bibitem{Oriti:2011jm}
D.~Oriti, \emph{{The microscopic dynamics of quantum space as a group field theory}},  in \emph{{Foundations of Space and Time: Reflections on Quantum Gravity}}, pp.~257--320, 10, 2011 [\href{https://arxiv.org/abs/1110.5606}{{\ttfamily 1110.5606}}].

\bibitem{Carrozza:2013oiy}
S.~Carrozza, \emph{{Tensorial methods and renormalization in Group Field Theories}}, Ph.D. thesis, Orsay, LPT, 2013.
\newblock \href{https://arxiv.org/abs/1310.3736}{{\ttfamily 1310.3736}}.
\newblock 10.1007/978-3-319-05867-2.

\bibitem{Carrozza:2016vsq}
S.~Carrozza, \emph{{Flowing in Group Field Theory Space: a Review}}, \href{https://doi.org/10.3842/SIGMA.2016.070}{\emph{SIGMA} {\bfseries 12} (2016) 070} [\href{https://arxiv.org/abs/1603.01902}{{\ttfamily 1603.01902}}].

\bibitem{Gielen:2016dss}
S.~Gielen and L.~Sindoni, \emph{{Quantum Cosmology from Group Field Theory Condensates: a Review}}, \href{https://doi.org/10.3842/SIGMA.2016.082}{\emph{SIGMA} {\bfseries 12} (2016) 082} [\href{https://arxiv.org/abs/1602.08104}{{\ttfamily 1602.08104}}].

\bibitem{Oriti:2016qtz}
D.~Oriti, L.~Sindoni and E.~Wilson-Ewing, \emph{{Emergent Friedmann dynamics with a quantum bounce from quantum gravity condensates}}, \href{https://doi.org/10.1088/0264-9381/33/22/224001}{\emph{Class. Quant. Grav.} {\bfseries 33} (2016) 224001} [\href{https://arxiv.org/abs/1602.05881}{{\ttfamily 1602.05881}}].

\bibitem{Li:2017uao}
Y.~Li, D.~Oriti and M.~Zhang, \emph{{Group field theory for quantum gravity minimally coupled to a scalar field}}, \href{https://doi.org/10.1088/1361-6382/aa85d2}{\emph{Class. Quant. Grav.} {\bfseries 34} (2017) 195001} [\href{https://arxiv.org/abs/1701.08719}{{\ttfamily 1701.08719}}].

\bibitem{Gielen:2018fqv}
S.~Gielen, \emph{{Group field theory and its cosmology in a matter reference frame}}, \href{https://doi.org/10.3390/universe4100103}{\emph{Universe} {\bfseries 4} (2018) 103} [\href{https://arxiv.org/abs/1808.10469}{{\ttfamily 1808.10469}}].

\bibitem{Ashtekar:2004eh}
A.~Ashtekar and J.~Lewandowski, \emph{{Background independent quantum gravity: A Status report}}, \href{https://doi.org/10.1088/0264-9381/21/15/R01}{\emph{Class. Quant. Grav.} {\bfseries 21} (2004) R53} [\href{https://arxiv.org/abs/gr-qc/0404018}{{\ttfamily gr-qc/0404018}}].

\bibitem{Perez:2003vx}
A.~Perez, \emph{{Spin foam models for quantum gravity}}, \href{https://doi.org/10.1088/0264-9381/20/6/202}{\emph{Class. Quant. Grav.} {\bfseries 20} (2003) R43} [\href{https://arxiv.org/abs/gr-qc/0301113}{{\ttfamily gr-qc/0301113}}].

\bibitem{Perez:2012wv}
A.~Perez, \emph{{The Spin Foam Approach to Quantum Gravity}}, \href{https://doi.org/10.12942/lrr-2013-3}{\emph{Living Rev. Rel.} {\bfseries 16} (2013) 3} [\href{https://arxiv.org/abs/1205.2019}{{\ttfamily 1205.2019}}].

\bibitem{Engle:2023qsu}
J.~Engle and S.~Speziale, \emph{{Spin Foams: Foundations}},  (2023), \href{https://doi.org/10.1007/978-981-19-3079-9_99-1}{DOI} [\href{https://arxiv.org/abs/2310.20147}{{\ttfamily 2310.20147}}].

\bibitem{Livine:2024hhc}
E.R.~Livine, \emph{{Spinfoam Models for Quantum Gravity: Overview}},  \href{https://arxiv.org/abs/2403.09364}{{\ttfamily 2403.09364}}.

\bibitem{Bonzom:2009hw}
V.~Bonzom, \emph{{Spin foam models for quantum gravity from lattice path integrals}}, \href{https://doi.org/10.1103/PhysRevD.80.064028}{\emph{Phys. Rev. D} {\bfseries 80} (2009) 064028} [\href{https://arxiv.org/abs/0905.1501}{{\ttfamily 0905.1501}}].

\bibitem{Baratin:2010wi}
A.~Baratin and D.~Oriti, \emph{{Group field theory with non-commutative metric variables}}, \href{https://doi.org/10.1103/PhysRevLett.105.221302}{\emph{Phys. Rev. Lett.} {\bfseries 105} (2010) 221302} [\href{https://arxiv.org/abs/1002.4723}{{\ttfamily 1002.4723}}].

\bibitem{Baratin:2011tx}
A.~Baratin and D.~Oriti, \emph{{Quantum simplicial geometry in the group field theory formalism: reconsidering the Barrett-Crane model}}, \href{https://doi.org/10.1088/1367-2630/13/12/125011}{\emph{New J. Phys.} {\bfseries 13} (2011) 125011} [\href{https://arxiv.org/abs/1108.1178}{{\ttfamily 1108.1178}}].

\bibitem{Baratin:2011hp}
A.~Baratin and D.~Oriti, \emph{{Group field theory and simplicial gravity path integrals: A model for Holst-Plebanski gravity}}, \href{https://doi.org/10.1103/PhysRevD.85.044003}{\emph{Phys. Rev. D} {\bfseries 85} (2012) 044003} [\href{https://arxiv.org/abs/1111.5842}{{\ttfamily 1111.5842}}].

\bibitem{Finocchiaro:2018hks}
M.~Finocchiaro and D.~Oriti, \emph{{Spin foam models and the Duflo map}}, \href{https://doi.org/10.1088/1361-6382/ab58da}{\emph{Class. Quant. Grav.} {\bfseries 37} (2020) 015010} [\href{https://arxiv.org/abs/1812.03550}{{\ttfamily 1812.03550}}].

\bibitem{Ambjorn:2013tki}
J.~Ambj\o{}rn, A.~G\"orlich, J.~Jurkiewicz and R.~Loll, \emph{{Quantum Gravity via Causal Dynamical Triangulations}},  in \emph{{Springer Handbook of Spacetime}}, A.~Ashtekar and V.~Petkov, eds., pp.~723--741 (2014), \href{https://doi.org/10.1007/978-3-642-41992-8_34}{DOI} [\href{https://arxiv.org/abs/1302.2173}{{\ttfamily 1302.2173}}].

\bibitem{Loll:2019rdj}
R.~Loll, \emph{{Quantum Gravity from Causal Dynamical Triangulations: A Review}}, \href{https://doi.org/10.1088/1361-6382/ab57c7}{\emph{Class. Quant. Grav.} {\bfseries 37} (2020) 013002} [\href{https://arxiv.org/abs/1905.08669}{{\ttfamily 1905.08669}}].

\bibitem{Rosenhaus:2018dtp}
V.~Rosenhaus, \emph{{An introduction to the SYK model}}, \href{https://doi.org/10.1088/1751-8121/ab2ce1}{\emph{J. Phys. A} {\bfseries 52} (2019) 323001} [\href{https://arxiv.org/abs/1807.03334}{{\ttfamily 1807.03334}}].

\bibitem{Gurau:2019qag}
R.G.~Gurau, \emph{{Notes on tensor models and tensor field theories}}, \href{https://doi.org/10.4171/aihpd/117}{\emph{Ann. Inst. H. Poincare D Comb. Phys. Interact.} {\bfseries 9} (2022) 159} [\href{https://arxiv.org/abs/1907.03531}{{\ttfamily 1907.03531}}].

\bibitem{Benedetti:2020seh}
D.~Benedetti, \emph{{Melonic CFTs}}, \href{https://doi.org/10.22323/1.376.0168}{\emph{PoS} {\bfseries CORFU2019} (2020) 168} [\href{https://arxiv.org/abs/2004.08616}{{\ttfamily 2004.08616}}].

\bibitem{Harribey:2022esw}
S.~Harribey, \emph{{Renormalization in tensor field theory and the melonic fixed point}}, Ph.D. thesis, Heidelberg U., 2022.
\newblock \href{https://arxiv.org/abs/2207.05520}{{\ttfamily 2207.05520}}.
\newblock 10.11588/heidok.00031883.

\bibitem{Gurau:2024nzv}
R.~Gurau and V.~Rivasseau, \emph{{Quantum Gravity and Random Tensors}},  1, 2024 [\href{https://arxiv.org/abs/2401.13510}{{\ttfamily 2401.13510}}].

\bibitem{Maldacena:1997re}
J.M.~Maldacena, \emph{{The Large N limit of superconformal field theories and supergravity}}, \href{https://doi.org/10.4310/ATMP.1998.v2.n2.a1}{\emph{Adv. Theor. Math. Phys.} {\bfseries 2} (1998) 231} [\href{https://arxiv.org/abs/hep-th/9711200}{{\ttfamily hep-th/9711200}}].

\bibitem{Maldacena:1998im}
J.M.~Maldacena, \emph{{Wilson loops in large N field theories}}, \href{https://doi.org/10.1103/PhysRevLett.80.4859}{\emph{Phys. Rev. Lett.} {\bfseries 80} (1998) 4859} [\href{https://arxiv.org/abs/hep-th/9803002}{{\ttfamily hep-th/9803002}}].

\bibitem{Gubser:1998bc}
S.S.~Gubser, I.R.~Klebanov and A.M.~Polyakov, \emph{{Gauge theory correlators from noncritical string theory}}, \href{https://doi.org/10.1016/S0370-2693(98)00377-3}{\emph{Phys. Lett. B} {\bfseries 428} (1998) 105} [\href{https://arxiv.org/abs/hep-th/9802109}{{\ttfamily hep-th/9802109}}].

\bibitem{Witten:1998qj}
E.~Witten, \emph{{Anti-de Sitter space and holography}}, \href{https://doi.org/10.4310/ATMP.1998.v2.n2.a2}{\emph{Adv. Theor. Math. Phys.} {\bfseries 2} (1998) 253} [\href{https://arxiv.org/abs/hep-th/9802150}{{\ttfamily hep-th/9802150}}].

\bibitem{Delamotte:2007pf}
B.~Delamotte, \emph{{An Introduction to the nonperturbative renormalization group}}, \href{https://doi.org/10.1007/978-3-642-27320-9_2}{\emph{Lect. Notes Phys.} {\bfseries 852} (2012) 49} [\href{https://arxiv.org/abs/cond-mat/0702365}{{\ttfamily cond-mat/0702365}}].

\bibitem{dupuis2021nonperturbative}
N.~Dupuis, L.~Canet, A.~Eichhorn, W.~Metzner, J.M.~Pawlowski, M.~Tissier et~al., \emph{The nonperturbative functional renormalization group and its applications}, {\emph{Physics Reports} {\bfseries 910} (2021) 1}.

\bibitem{Sfondrini:2010zm}
A.~Sfondrini and T.A.~Koslowski, \emph{{Functional Renormalization of Noncommutative Scalar Field Theory}}, \href{https://doi.org/10.1142/S0217751X11054048}{\emph{Int. J. Mod. Phys. A} {\bfseries 26} (2011) 4009} [\href{https://arxiv.org/abs/1006.5145}{{\ttfamily 1006.5145}}].

\bibitem{Eichhorn:2013isa}
A.~Eichhorn and T.~Koslowski, \emph{{Continuum limit in matrix models for quantum gravity from the Functional Renormalization Group}}, \href{https://doi.org/10.1103/PhysRevD.88.084016}{\emph{Phys. Rev. D} {\bfseries 88} (2013) 084016} [\href{https://arxiv.org/abs/1309.1690}{{\ttfamily 1309.1690}}].

\bibitem{Eichhorn:2014xaa}
A.~Eichhorn and T.~Koslowski, \emph{{Towards phase transitions between discrete and continuum quantum spacetime from the Renormalization Group}}, \href{https://doi.org/10.1103/PhysRevD.90.104039}{\emph{Phys. Rev. D} {\bfseries 90} (2014) 104039} [\href{https://arxiv.org/abs/1408.4127}{{\ttfamily 1408.4127}}].

\bibitem{Eichhorn:2017xhy}
A.~Eichhorn and T.~Koslowski, \emph{{Flowing to the continuum in discrete tensor models for quantum gravity}}, \href{https://doi.org/10.4171/AIHPD/52}{\emph{Ann. Inst. H. Poincare Comb. Phys. Interact.} {\bfseries 5} (2018) 173} [\href{https://arxiv.org/abs/1701.03029}{{\ttfamily 1701.03029}}].

\bibitem{Eichhorn:2018phj}
A.~Eichhorn, T.~Koslowski and A.D.~Pereira, \emph{{Status of background-independent coarse-graining in tensor models for quantum gravity}}, \href{https://doi.org/10.3390/universe5020053}{\emph{Universe} {\bfseries 5} (2019) 53} [\href{https://arxiv.org/abs/1811.12909}{{\ttfamily 1811.12909}}].

\bibitem{Eichhorn:2018ylk}
A.~Eichhorn, T.~Koslowski, J.~Lumma and A.D.~Pereira, \emph{{Towards background independent quantum gravity with tensor models}}, \href{https://doi.org/10.1088/1361-6382/ab2545}{\emph{Class. Quant. Grav.} {\bfseries 36} (2019) 155007} [\href{https://arxiv.org/abs/1811.00814}{{\ttfamily 1811.00814}}].

\bibitem{Eichhorn:2019hsa}
A.~Eichhorn, J.~Lumma, A.D.~Pereira and A.~Sikandar, \emph{{Universal critical behavior in tensor models for four-dimensional quantum gravity}}, \href{https://doi.org/10.1007/JHEP02(2020)110}{\emph{JHEP} {\bfseries 02} (2020) 110} [\href{https://arxiv.org/abs/1912.05314}{{\ttfamily 1912.05314}}].

\bibitem{Castro:2020dzt}
A.~Castro and T.~Koslowski, \emph{{Renormalization Group Approach to the Continuum Limit of Matrix Models of Quantum Gravity with Preferred Foliation}}, \href{https://doi.org/10.3389/fphy.2021.531766}{\emph{Front. in Phys.} {\bfseries 9} (2021) 114} [\href{https://arxiv.org/abs/2008.10090}{{\ttfamily 2008.10090}}].

\bibitem{Eichhorn:2020sla}
A.~Eichhorn, A.D.~Pereira and A.G.A.~Pithis, \emph{{The phase diagram of the multi-matrix model with ABAB-interaction from functional renormalization}}, \href{https://doi.org/10.1007/JHEP12(2020)131}{\emph{JHEP} {\bfseries 12} (2020) 131} [\href{https://arxiv.org/abs/2009.05111}{{\ttfamily 2009.05111}}].

\bibitem{Benedetti:2015et}
D.~Benedetti, J.~Ben~Geloun and D.~Oriti, \emph{{Functional Renormalisation Group Approach for Tensorial Group Field Theory: a Rank-3 Model}}, {\emph{JHEP} {\bfseries 03} (2015) 084} [\href{https://arxiv.org/abs/1411.3180}{{\ttfamily 1411.3180}}].

\bibitem{BenGeloun:2015ej}
J.~Ben~Geloun, R.~Martini and D.~Oriti, \emph{{Functional Renormalization Group analysis of a Tensorial Group Field Theory on ${R}^3$}}, {\emph{EPL} {\bfseries 112} (2015) 31001} [\href{https://arxiv.org/abs/1508.01855}{{\ttfamily 1508.01855}}].

\bibitem{BenGeloun:2016kw}
J.~Ben~Geloun, R.~Martini and D.~Oriti, \emph{{Functional renormalization group analysis of tensorial group field theories on $R^d$}}, {\emph{Phys. Rev. D} {\bfseries 94} (2016) 024017} [\href{https://arxiv.org/abs/1601.08211}{{\ttfamily 1601.08211}}].

\bibitem{Benedetti:2016db}
D.~Benedetti and V.~Lahoche, \emph{{Functional renormalization group approach for tensorial group field theory: a rank-6 model with closure constraint}}, {\emph{Classical And Quantum Gravity} {\bfseries 33} (2016) } [\href{https://arxiv.org/abs/1508.06384}{{\ttfamily 1508.06384}}].

\bibitem{Carrozza:2016tih}
S.~Carrozza and V.~Lahoche, \emph{{Asymptotic safety in three-dimensional SU(2) Group Field Theory: evidence in the local potential approximation}}, \href{https://doi.org/10.1088/1361-6382/aa6d90}{\emph{Class. Quant. Grav.} {\bfseries 34} (2017) 115004} [\href{https://arxiv.org/abs/1612.02452}{{\ttfamily 1612.02452}}].

\bibitem{Carrozza:2017vkz}
S.~Carrozza, V.~Lahoche and D.~Oriti, \emph{{Renormalizable Group Field Theory beyond melonic diagrams: an example in rank four}}, \href{https://doi.org/10.1103/PhysRevD.96.066007}{\emph{Phys. Rev. D} {\bfseries 96} (2017) 066007} [\href{https://arxiv.org/abs/1703.06729}{{\ttfamily 1703.06729}}].

\bibitem{BenGeloun:2018ekd}
J.~Ben~Geloun, T.A.~Koslowski, D.~Oriti and A.D.~Pereira, \emph{{Functional Renormalization Group analysis of rank 3 tensorial group field theory: The full quartic invariant truncation}}, \href{https://doi.org/10.1103/PhysRevD.97.126018}{\emph{Phys. Rev. D} {\bfseries 97} (2018) 126018} [\href{https://arxiv.org/abs/1805.01619}{{\ttfamily 1805.01619}}].

\bibitem{Pithis:2020sxm}
A.G.A.~Pithis and J.~Th\"urigen, \emph{{(No) phase transition in tensorial group field theory}}, \href{https://doi.org/10.1016/j.physletb.2021.136215}{\emph{Phys. Lett. B} {\bfseries 816} (2021) 136215} [\href{https://arxiv.org/abs/2007.08982}{{\ttfamily 2007.08982}}].

\bibitem{Pithis:2020kio}
A.G.A.~Pithis and J.~Th\"urigen, \emph{{Phase transitions in TGFT: functional renormalization group in the cyclic-melonic potential approximation and equivalence to O$(N)$ models}}, \href{https://doi.org/10.1007/JHEP12(2020)159}{\emph{JHEP} {\bfseries 12} (2020) 159} [\href{https://arxiv.org/abs/2009.13588}{{\ttfamily 2009.13588}}].

\bibitem{Baloitcha:2020lha}
E.~Baloitcha, V.~Lahoche and D.~Ousmane~Samary, \emph{{Flowing in discrete gravity models and Ward identities: a review}}, \href{https://doi.org/10.1140/epjp/s13360-021-01823-z}{\emph{Eur. Phys. J. Plus} {\bfseries 136} (2021) 982} [\href{https://arxiv.org/abs/2001.02631}{{\ttfamily 2001.02631}}].

\bibitem{Lahoche:2022gkz}
V.~Lahoche and D.O.~Samary, \emph{{Stochastic dynamics for group field theories}}, \href{https://doi.org/10.1103/PhysRevD.107.086009}{\emph{Phys. Rev. D} {\bfseries 107} (2023) 086009} [\href{https://arxiv.org/abs/2209.02321}{{\ttfamily 2209.02321}}].

\bibitem{Geloun:2023ray}
J.B.~Geloun, A.G.A.~Pithis and J.~Th\"urigen, \emph{{QFT with Tensorial and Local Degrees of Freedom: Phase Structure from Functional Renormalization}}, \href{https://doi.org/10.1063/5.0158724}{\emph{J. Math. Phys.} {\bfseries 65} (2024) 032302} [\href{https://arxiv.org/abs/2305.06136}{{\ttfamily 2305.06136}}].

\bibitem{Kopietz:2010zz}
P.~Kopietz, L.~Bartosch and F.~Sch\"utz, \emph{{Introduction to the functional renormalization group}}, vol.~798 (2010), \href{https://doi.org/10.1007/978-3-642-05094-7}{10.1007/978-3-642-05094-7}.

\bibitem{zinn2021quantum}
J.~Zinn-Justin, \emph{Quantum field theory and critical phenomena}, vol.~171, Oxford university press (2021).

\bibitem{wilson1983renormalization}
K.G.~Wilson, \emph{The renormalization group and critical phenomena}, {\emph{Reviews of Modern Physics} {\bfseries 55} (1983) 583}.

\bibitem{hohenberg2015introduction}
P.C.~Hohenberg and A.P.~Krekhov, \emph{An introduction to the ginzburg--landau theory of phase transitions and nonequilibrium patterns}, {\emph{Physics Reports} {\bfseries 572} (2015) 1}.

\bibitem{Pithis:2018eaq}
A.G.A.~Pithis and J.~Th\"urigen, \emph{{Phase transitions in group field theory: The Landau perspective}}, \href{https://doi.org/10.1103/PhysRevD.98.126006}{\emph{Phys. Rev. D} {\bfseries 98} (2018) 126006} [\href{https://arxiv.org/abs/1808.09765}{{\ttfamily 1808.09765}}].

\bibitem{Marchetti:2020xvf}
L.~Marchetti, D.~Oriti, A.G.A.~Pithis and J.~Th\"urigen, \emph{{Phase transitions in tensorial group field theories: Landau-Ginzburg analysis of models with both local and non-local degrees of freedom}}, \href{https://doi.org/10.1007/JHEP12(2021)201}{\emph{JHEP} {\bfseries 21} (2020) 201} [\href{https://arxiv.org/abs/2110.15336}{{\ttfamily 2110.15336}}].

\bibitem{Marchetti:2022igl}
L.~Marchetti, D.~Oriti, A.G.A.~Pithis and J.~Th\"urigen, \emph{{Phase transitions in TGFT: a Landau-Ginzburg analysis of Lorentzian quantum geometric models}}, \href{https://doi.org/10.1007/JHEP02(2023)074}{\emph{JHEP} {\bfseries 02} (2023) 074} [\href{https://arxiv.org/abs/2209.04297}{{\ttfamily 2209.04297}}].

\bibitem{Marchetti:2022nrf}
L.~Marchetti, D.~Oriti, A.G.A.~Pithis and J.~Th\"urigen, \emph{{Mean-Field Phase Transitions in Tensorial Group Field Theory Quantum Gravity}}, \href{https://doi.org/10.1103/PhysRevLett.130.141501}{\emph{Phys. Rev. Lett.} {\bfseries 130} (2023) 141501} [\href{https://arxiv.org/abs/2211.12768}{{\ttfamily 2211.12768}}].

\bibitem{sachs2006elements}
I.~Sachs, S.~Sen and J.~Sexton, \emph{Elements of statistical mechanics: with an introduction to quantum field theory and numerical simulation}, Cambridge University Press (2006).

\bibitem{Witten:2016iux}
E.~Witten, \emph{{An SYK-Like Model Without Disorder}}, \href{https://doi.org/10.1088/1751-8121/ab3752}{\emph{J. Phys. A} {\bfseries 52} (2019) 474002} [\href{https://arxiv.org/abs/1610.09758}{{\ttfamily 1610.09758}}].

\bibitem{Gurau:2016lzk}
R.~Gurau, \emph{{The complete $1/N$ expansion of a SYK\textendash{}like tensor model}}, \href{https://doi.org/10.1016/j.nuclphysb.2017.01.015}{\emph{Nucl. Phys. B} {\bfseries 916} (2017) 386} [\href{https://arxiv.org/abs/1611.04032}{{\ttfamily 1611.04032}}].

\bibitem{Klebanov:2016xxf}
I.R.~Klebanov and G.~Tarnopolsky, \emph{{Uncolored random tensors, melon diagrams, and the Sachdev-Ye-Kitaev models}}, \href{https://doi.org/10.1103/PhysRevD.95.046004}{\emph{Phys. Rev. D} {\bfseries 95} (2017) 046004} [\href{https://arxiv.org/abs/1611.08915}{{\ttfamily 1611.08915}}].

\bibitem{Klebanov:2018nfp}
I.R.~Klebanov, A.~Milekhin, F.~Popov and G.~Tarnopolsky, \emph{{Spectra of eigenstates in fermionic tensor quantum mechanics}}, \href{https://doi.org/10.1103/PhysRevD.97.106023}{\emph{Phys. Rev. D} {\bfseries 97} (2018) 106023} [\href{https://arxiv.org/abs/1802.10263}{{\ttfamily 1802.10263}}].

\bibitem{Giombi:2017dtl}
S.~Giombi, I.R.~Klebanov and G.~Tarnopolsky, \emph{{Bosonic tensor models at large $N$ and small $\epsilon$}}, \href{https://doi.org/10.1103/PhysRevD.96.106014}{\emph{Phys. Rev. D} {\bfseries 96} (2017) 106014} [\href{https://arxiv.org/abs/1707.03866}{{\ttfamily 1707.03866}}].

\bibitem{Carrozza:2015adg}
S.~Carrozza and A.~Tanasa, \emph{{$O(N)$ Random Tensor Models}}, \href{https://doi.org/10.1007/s11005-016-0879-x}{\emph{Lett. Math. Phys.} {\bfseries 106} (2016) 1531} [\href{https://arxiv.org/abs/1512.06718}{{\ttfamily 1512.06718}}].

\bibitem{Klebanov:2018fzb}
I.R.~Klebanov, F.~Popov and G.~Tarnopolsky, \emph{{TASI Lectures on Large $N$ Tensor Models}}, \href{https://doi.org/10.22323/1.305.0004}{\emph{PoS} {\bfseries TASI2017} (2018) 004} [\href{https://arxiv.org/abs/1808.09434}{{\ttfamily 1808.09434}}].

\bibitem{Benedetti:2018goh}
D.~Benedetti and R.~Gurau, \emph{{2PI effective action for the SYK model and tensor field theories}}, \href{https://doi.org/10.1007/JHEP05(2018)156}{\emph{JHEP} {\bfseries 05} (2018) 156} [\href{https://arxiv.org/abs/1802.05500}{{\ttfamily 1802.05500}}].

\bibitem{Benedetti:2018ghn}
D.~Benedetti and N.~Delporte, \emph{{Phase diagram and fixed points of tensorial Gross-Neveu models in three dimensions}}, \href{https://doi.org/10.1007/JHEP01(2019)218}{\emph{JHEP} {\bfseries 01} (2019) 218} [\href{https://arxiv.org/abs/1810.04583}{{\ttfamily 1810.04583}}].

\bibitem{Benedetti:2019eyl}
D.~Benedetti, R.~Gurau and S.~Harribey, \emph{{Line of fixed points in a bosonic tensor model}}, \href{https://doi.org/10.1007/JHEP06(2019)053}{\emph{JHEP} {\bfseries 06} (2019) 053} [\href{https://arxiv.org/abs/1903.03578}{{\ttfamily 1903.03578}}].

\bibitem{Benedetti:2021wzt}
D.~Benedetti, R.~Gurau, S.~Harribey and D.~Lettera, \emph{{The F-theorem in the melonic limit}}, \href{https://doi.org/10.1007/JHEP02(2022)147}{\emph{JHEP} {\bfseries 02} (2022) 147} [\href{https://arxiv.org/abs/2111.11792}{{\ttfamily 2111.11792}}].

\bibitem{Jepsen:2023pzm}
C.~Jepsen and Y.~Oz, \emph{{RG flows and fixed points of O(N)$^{r}$ models}}, \href{https://doi.org/10.1007/JHEP02(2024)035}{\emph{JHEP} {\bfseries 02} (2024) 035} [\href{https://arxiv.org/abs/2311.09039}{{\ttfamily 2311.09039}}].

\bibitem{Berges:2023rqa}
J.~Berges, R.~Gurau and T.~Preis, \emph{{Asymptotic freedom in a strongly interacting scalar quantum field theory in four Euclidean dimensions}}, \href{https://doi.org/10.1103/PhysRevD.108.016019}{\emph{Phys. Rev. D} {\bfseries 108} (2023) 016019} [\href{https://arxiv.org/abs/2301.09514}{{\ttfamily 2301.09514}}].

\bibitem{Oriti:2014yla}
D.~Oriti, J.P.~Ryan and J.~Th\"urigen, \emph{{Group field theories for all loop quantum gravity}}, \href{https://doi.org/10.1088/1367-2630/17/2/023042}{\emph{New J. Phys.} {\bfseries 17} (2015) 023042} [\href{https://arxiv.org/abs/1409.3150}{{\ttfamily 1409.3150}}].

\bibitem{levanyuk1959contribution}
A.~Levanyuk, \emph{Contribution to the theory of light scattering near the second-order phase-transition points}, {\emph{Sov. Phys. JETP} {\bfseries 9} (1959) 571}.

\bibitem{ginzburg1961some}
V.~Ginzburg, \emph{Some remarks on phase transitions of the second kind and the microscopic theory of ferroelectric materials}, {\emph{Soviet Phys. Solid State} {\bfseries 2} (1961) 1824}.

\bibitem{Benedetti:2014gja}
D.~Benedetti, \emph{{Critical behavior in spherical and hyperbolic spaces}}, \href{https://doi.org/10.1088/1742-5468/2015/01/P01002}{\emph{J. Stat. Mech.} {\bfseries 1501} (2015) P01002} [\href{https://arxiv.org/abs/1403.6712}{{\ttfamily 1403.6712}}].

\bibitem{Dekhil:2024djp}
R.~Dekhil, A.F.~Jercher and A.G.A.~Pithis, \emph{{Phase transitions in TGFT: Landau-Ginzburg analysis of the causally complete Lorentzian Barrett-Crane model}},  \href{https://arxiv.org/abs/2407.02325}{{\ttfamily 2407.02325}}.

\bibitem{GradshteynBook}
I.S.~Gradshteyn and R.~M., \emph{Table of Integrals, Series, and Products}, Academic Press (1943).

\bibitem{strocchi2005symmetry}
F.~Strocchi, \emph{Symmetry breaking}, vol.~643, Springer (2005).

\bibitem{Gurau:2009tw}
R.~Gurau, \emph{{Colored Group Field Theory}}, \href{https://doi.org/10.1007/s00220-011-1226-9}{\emph{Commun. Math. Phys.} {\bfseries 304} (2011) 69} [\href{https://arxiv.org/abs/0907.2582}{{\ttfamily 0907.2582}}].

\bibitem{Bonzom:2012hw}
V.~Bonzom, R.~Gurau and V.~Rivasseau, \emph{{Random tensor models in the large N limit: Uncoloring the colored tensor models}}, \href{https://doi.org/10.1103/PhysRevD.85.084037}{\emph{Phys. Rev. D} {\bfseries 85} (2012) 084037} [\href{https://arxiv.org/abs/1202.3637}{{\ttfamily 1202.3637}}].

\bibitem{Gurau:2010nd}
R.~Gurau, \emph{{Lost in Translation: Topological Singularities in Group Field Theory}}, \href{https://doi.org/10.1088/0264-9381/27/23/235023}{\emph{Class. Quant. Grav.} {\bfseries 27} (2010) 235023} [\href{https://arxiv.org/abs/1006.0714}{{\ttfamily 1006.0714}}].

\bibitem{Ruehl1970}
W.~Ruehl, \emph{Lorentz group and harmonic analysis}, W A Benjamin, Inc, United States (1970).

\bibitem{hormander2015analysis}
L.~H{\"o}rmander, \emph{The analysis of linear partial differential operators I: Distribution theory and Fourier analysis}, Springer (2015).

\bibitem{Baratin:2010nn}
A.~Baratin, B.~Dittrich, D.~Oriti and J.~Tambornino, \emph{{Non-commutative flux representation for loop quantum gravity}}, \href{https://doi.org/10.1088/0264-9381/28/17/175011}{\emph{Class. Quant. Grav.} {\bfseries 28} (2011) 175011} [\href{https://arxiv.org/abs/1004.3450}{{\ttfamily 1004.3450}}].

\bibitem{Barrett:1999qw}
J.W.~Barrett and L.~Crane, \emph{{A Lorentzian signature model for quantum general relativity}}, \href{https://doi.org/10.1088/0264-9381/17/16/302}{\emph{Class. Quant. Grav.} {\bfseries 17} (2000) 3101} [\href{https://arxiv.org/abs/gr-qc/9904025}{{\ttfamily gr-qc/9904025}}].

\bibitem{Perez:2000ec}
A.~Perez and C.~Rovelli, \emph{{Spin foam model for Lorentzian general relativity}}, \href{https://doi.org/10.1103/PhysRevD.63.041501}{\emph{Phys. Rev. D} {\bfseries 63} (2001) 041501} [\href{https://arxiv.org/abs/gr-qc/0009021}{{\ttfamily gr-qc/0009021}}].

\bibitem{Perez:2000ep}
A.~Perez and C.~Rovelli, \emph{{3+1 spinfoam model of quantum gravity with space - like and time - like components}}, \href{https://doi.org/10.1103/PhysRevD.64.064002}{\emph{Phys. Rev. D} {\bfseries 64} (2001) 064002} [\href{https://arxiv.org/abs/gr-qc/0011037}{{\ttfamily gr-qc/0011037}}].

\bibitem{Oriti:2003wf}
D.~Oriti, \emph{{Spin foam models of quantum space-time}},  other thesis, 11, 2003, [\href{https://arxiv.org/abs/gr-qc/0311066}{{\ttfamily gr-qc/0311066}}].

\bibitem{Jercher:2021bie}
A.F.~Jercher, D.~Oriti and A.G.A.~Pithis, \emph{{Emergent cosmology from quantum gravity in the Lorentzian Barrett-Crane tensorial group field theory model}}, \href{https://doi.org/10.1088/1475-7516/2022/01/050}{\emph{JCAP} {\bfseries 01} (2022) 050} [\href{https://arxiv.org/abs/2112.00091}{{\ttfamily 2112.00091}}].

\bibitem{Jercher:2022mky}
A.F.~Jercher, D.~Oriti and A.G.A.~Pithis, \emph{{Complete Barrett-Crane model and its causal structure}}, \href{https://doi.org/10.1103/PhysRevD.106.066019}{\emph{Phys. Rev. D} {\bfseries 106} (2022) 066019} [\href{https://arxiv.org/abs/2206.15442}{{\ttfamily 2206.15442}}].

\bibitem{Guruswamy:1994pi}
S.~Guruswamy and P.~Vitale, \emph{{Correlation functions of a conformal field theory in three-dimensions}}, \href{https://doi.org/10.1142/S0217732396001089}{\emph{Mod. Phys. Lett. A} {\bfseries 11} (1996) 1047} [\href{https://arxiv.org/abs/hep-th/9411146}{{\ttfamily hep-th/9411146}}].

\bibitem{Sachdev:1992fk}
S.~Sachdev and J.~Ye, \emph{{Gapless spin fluid ground state in a random, quantum Heisenberg magnet}}, \href{https://doi.org/10.1103/PhysRevLett.70.3339}{\emph{Phys. Rev. Lett.} {\bfseries 70} (1993) 3339} [\href{https://arxiv.org/abs/cond-mat/9212030}{{\ttfamily cond-mat/9212030}}].

\bibitem{kitaev2015simple}
A.~Kitaev, \emph{A simple model of quantum holography (part 1 \& 2)}, {\emph{Entanglement in strongly-correlated quantum matter} (2015) 38}.

\bibitem{Maldacena:2016hyu}
J.~Maldacena and D.~Stanford, \emph{{Remarks on the Sachdev-Ye-Kitaev model}}, \href{https://doi.org/10.1103/PhysRevD.94.106002}{\emph{Phys. Rev. D} {\bfseries 94} (2016) 106002} [\href{https://arxiv.org/abs/1604.07818}{{\ttfamily 1604.07818}}].

\bibitem{Maldacena:2016upp}
J.~Maldacena, D.~Stanford and Z.~Yang, \emph{{Conformal symmetry and its breaking in two dimensional Nearly Anti-de-Sitter space}}, \href{https://doi.org/10.1093/ptep/ptw124}{\emph{PTEP} {\bfseries 2016} (2016) 12C104} [\href{https://arxiv.org/abs/1606.01857}{{\ttfamily 1606.01857}}].

\bibitem{Kim:2019upg}
J.~Kim, I.R.~Klebanov, G.~Tarnopolsky and W.~Zhao, \emph{{Symmetry Breaking in Coupled SYK or Tensor Models}}, \href{https://doi.org/10.1103/PhysRevX.9.021043}{\emph{Phys. Rev. X} {\bfseries 9} (2019) 021043} [\href{https://arxiv.org/abs/1902.02287}{{\ttfamily 1902.02287}}].

\bibitem{Benedetti:2020yvb}
D.~Benedetti, R.~Gurau and K.~Suzuki, \emph{{Conformal symmetry and composite operators in the $O(N)^{3}$ tensor field theory}}, \href{https://doi.org/10.1007/JHEP06(2020)113}{\emph{JHEP} {\bfseries 06} (2020) 113} [\href{https://arxiv.org/abs/2002.07652}{{\ttfamily 2002.07652}}].

\bibitem{Benedetti:2019ikb}
D.~Benedetti, R.~Gurau, S.~Harribey and K.~Suzuki, \emph{{Hints of unitarity at large $N$ in the $O(N)^3$ tensor field theory}}, \href{https://doi.org/10.1007/JHEP02(2020)072}{\emph{JHEP} {\bfseries 02} (2020) 072} [\href{https://arxiv.org/abs/1909.07767}{{\ttfamily 1909.07767}}].

\bibitem{Fisher:1972zz}
M.E.~Fisher, S.-k.~Ma and B.G.~Nickel, \emph{{Critical Exponents for Long-Range Interactions}}, \href{https://doi.org/10.1103/PhysRevLett.29.917}{\emph{Phys. Rev. Lett.} {\bfseries 29} (1972) 917}.

\bibitem{Bonzom:2015kzh}
V.~Bonzom, L.~Lionni and V.~Rivasseau, \emph{{Colored triangulations of arbitrary dimensions are stuffed Walsh maps}},  \href{https://arxiv.org/abs/1508.03805}{{\ttfamily 1508.03805}}.

\bibitem{WipfRG}
A.~Wipf, \emph{Statistical Approach to Quantum Field Theory}, Springer, Switzerland (2021), \href{https://doi.org/https://doi.org/10.1007/978-3-030-83263-6}{https://doi.org/10.1007/978-3-030-83263-6}.

\bibitem{Polchinski:1987dy}
J.~Polchinski, \emph{{Scale and Conformal Invariance in Quantum Field Theory}}, \href{https://doi.org/10.1016/0550-3213(88)90179-4}{\emph{Nucl. Phys. B} {\bfseries 303} (1988) 226}.

\bibitem{Zamolodchikov:1986gt}
A.B.~Zamolodchikov, \emph{{Irreversibility of the Flux of the Renormalization Group in a 2D Field Theory}}, {\emph{JETP Lett.} {\bfseries 43} (1986) 730}.

\bibitem{Jackiw:2011vz}
R.~Jackiw and S.Y.~Pi, \emph{{Tutorial on Scale and Conformal Symmetries in Diverse Dimensions}}, \href{https://doi.org/10.1088/1751-8113/44/22/223001}{\emph{J. Phys. A} {\bfseries 44} (2011) 223001} [\href{https://arxiv.org/abs/1101.4886}{{\ttfamily 1101.4886}}].

\bibitem{Nakayama:2013is}
Y.~Nakayama, \emph{{Scale invariance vs conformal invariance}}, \href{https://doi.org/10.1016/j.physrep.2014.12.003}{\emph{Phys. Rept.} {\bfseries 569} (2015) 1} [\href{https://arxiv.org/abs/1302.0884}{{\ttfamily 1302.0884}}].

\bibitem{Nakahara:2003nw}
M.~Nakahara, \emph{{Geometry, topology and physics}} (2003).

\bibitem{Oriti:2018dsg}
D.~Oriti, \emph{{Levels of spacetime emergence in quantum gravity}},  \href{https://arxiv.org/abs/1807.04875}{{\ttfamily 1807.04875}}.

\bibitem{Gross:1990pa}
D.J.~Gross, \emph{{The C = 1 matrix models}}, .

\bibitem{Vafa:1988uu}
C.~Vafa and N.P.~Warner, \emph{{Catastrophes and the Classification of Conformal Theories}}, \href{https://doi.org/10.1016/0370-2693(89)90473-5}{\emph{Phys. Lett. B} {\bfseries 218} (1989) 51}.

\bibitem{Lerche:1989uy}
W.~Lerche, C.~Vafa and N.P.~Warner, \emph{{Chiral Rings in N=2 Superconformal Theories}}, \href{https://doi.org/10.1016/0550-3213(89)90474-4}{\emph{Nucl. Phys. B} {\bfseries 324} (1989) 427}.

\bibitem{Camacho:2019dvi}
A.R.~Camacho, \emph{{On the Landau-Ginzburg/conformal field theory correspondence}},  \href{https://arxiv.org/abs/1901.05365}{{\ttfamily 1901.05365}}.

\bibitem{Gielen:2013kla}
S.~Gielen, D.~Oriti and L.~Sindoni, \emph{{Cosmology from Group Field Theory Formalism for Quantum Gravity}}, \href{https://doi.org/10.1103/PhysRevLett.111.031301}{\emph{Phys. Rev. Lett.} {\bfseries 111} (2013) 031301} [\href{https://arxiv.org/abs/1303.3576}{{\ttfamily 1303.3576}}].

\bibitem{Oriti:2016acw}
D.~Oriti, \emph{{The universe as a quantum gravity condensate}}, \href{https://doi.org/10.1016/j.crhy.2017.02.003}{\emph{Comptes Rendus Physique} {\bfseries 18} (2017) 235} [\href{https://arxiv.org/abs/1612.09521}{{\ttfamily 1612.09521}}].

\bibitem{Pithis:2019tvp}
A.G.A.~Pithis and M.~Sakellariadou, \emph{{Group field theory condensate cosmology: An appetizer}}, \href{https://doi.org/10.3390/universe5060147}{\emph{Universe} {\bfseries 5} (2019) 147} [\href{https://arxiv.org/abs/1904.00598}{{\ttfamily 1904.00598}}].

\bibitem{Oriti:2024qav}
D.~Oriti, \emph{{Hydrodynamics on (mini)superspace, or a non-linear extension of quantum cosmology}},  3, 2024 [\href{https://arxiv.org/abs/2403.10741}{{\ttfamily 2403.10741}}].

\bibitem{Duval:2024eod}
C.~Duval, M.~Henkel, P.~Horvathy, S.~Rouhani and P.~Zhang, \emph{{Schr\"odinger symmetry: a historical review}},  \href{https://arxiv.org/abs/2403.20316}{{\ttfamily 2403.20316}}.

\bibitem{Cariglia:2014dwa}
M.~Cariglia, G.W.~Gibbons, J.W.~van Holten, P.A.~Horvathy and P.M.~Zhang, \emph{{Conformal Killing Tensors and covariant Hamiltonian Dynamics}}, \href{https://doi.org/10.1063/1.4902933}{\emph{J. Math. Phys.} {\bfseries 55} (2014) 122702} [\href{https://arxiv.org/abs/1404.3422}{{\ttfamily 1404.3422}}].

\bibitem{Cariglia:2015bla}
M.~Cariglia and F.K.~Alves, \emph{{The Eisenhart lift: a didactical introduction of modern geometrical concepts from Hamiltonian dynamics}}, \href{https://doi.org/10.1088/0143-0807/36/2/025018}{\emph{Eur. J. Phys.} {\bfseries 36} (2015) 025018} [\href{https://arxiv.org/abs/1503.07802}{{\ttfamily 1503.07802}}].

\bibitem{Cariglia:2016oft}
M.~Cariglia, C.~Duval, G.W.~Gibbons and P.A.~Horvathy, \emph{{Eisenhart lifts and symmetries of time-dependent systems}}, \href{https://doi.org/10.1016/j.aop.2016.07.033}{\emph{Annals Phys.} {\bfseries 373} (2016) 631} [\href{https://arxiv.org/abs/1605.01932}{{\ttfamily 1605.01932}}].

\bibitem{BenAchour:2017qpb}
J.~Ben~Achour and E.R.~Livine, \emph{{Thiemann complexifier in classical and quantum FLRW cosmology}}, \href{https://doi.org/10.1103/PhysRevD.96.066025}{\emph{Phys. Rev. D} {\bfseries 96} (2017) 066025} [\href{https://arxiv.org/abs/1705.03772}{{\ttfamily 1705.03772}}].

\bibitem{Cariglia:2018mos}
M.~Cariglia, A.~Galajinsky, G.W.~Gibbons and P.A.~Horvathy, \emph{{Cosmological aspects of the Eisenhart\textendash{}Duval lift}}, \href{https://doi.org/10.1140/epjc/s10052-018-5789-x}{\emph{Eur. Phys. J. C} {\bfseries 78} (2018) 314} [\href{https://arxiv.org/abs/1802.03370}{{\ttfamily 1802.03370}}].

\bibitem{BenAchour:2019ufa}
J.~Ben~Achour and E.R.~Livine, \emph{{Cosmology as a CFT$_1$}}, \href{https://doi.org/10.1007/JHEP12(2019)031}{\emph{JHEP} {\bfseries 12} (2019) 031} [\href{https://arxiv.org/abs/1909.13390}{{\ttfamily 1909.13390}}].

\bibitem{BenAchour:2020njq}
J.~Ben~Achour and E.R.~Livine, \emph{{Conformal structure of FLRW cosmology: spinorial representation and the $ \mathfrak{so} $ (2, 3) algebra of observables}}, \href{https://doi.org/10.1007/JHEP03(2020)067}{\emph{JHEP} {\bfseries 03} (2020) 067} [\href{https://arxiv.org/abs/2001.11807}{{\ttfamily 2001.11807}}].

\bibitem{Achour:2021lqq}
J.B.~Achour, \emph{{Proper time reparametrization in cosmology: M\"obius symmetry and Kodama charges}}, \href{https://doi.org/10.1088/1475-7516/2021/12/005}{\emph{JCAP} {\bfseries 12} (2021) 005} [\href{https://arxiv.org/abs/2103.10700}{{\ttfamily 2103.10700}}].

\bibitem{BenAchour:2022fif}
J.~Ben~Achour, E.R.~Livine, D.~Oriti and G.~Piani, \emph{{Schr\"odinger Symmetry in Gravitational Mini-Superspaces}}, \href{https://doi.org/10.3390/universe9120503}{\emph{Universe} {\bfseries 9} (2023) 503} [\href{https://arxiv.org/abs/2207.07312}{{\ttfamily 2207.07312}}].

\bibitem{BenAchour:2023dgj}
J.~Ben~Achour, E.R.~Livine and D.~Oriti, \emph{{Schr\"odinger symmetry of Schwarzschild-(A)dS black hole mechanics}}, \href{https://doi.org/10.1103/PhysRevD.108.104028}{\emph{Phys. Rev. D} {\bfseries 108} (2023) 104028} [\href{https://arxiv.org/abs/2302.07644}{{\ttfamily 2302.07644}}].

\bibitem{De_Felice_2010}
A.~De~Felice and S.~Tsujikawa, \emph{f(r) theories}, \href{https://doi.org/10.12942/lrr-2010-3}{\emph{Living Reviews in Relativity} {\bfseries 13} (2010) }.

\bibitem{CLIFTON20121}
T.~Clifton, P.G.~Ferreira, A.~Padilla and C.~Skordis, \emph{Modified gravity and cosmology}, \href{https://doi.org/https://doi.org/10.1016/j.physrep.2012.01.001}{\emph{Physics Reports} {\bfseries 513} (2012) 1}.

\bibitem{Sotiriou_2006}
T.P.~Sotiriou, \emph{f ( r ) gravity and scalar–tensor theory}, \href{https://doi.org/10.1088/0264-9381/23/17/003}{\emph{Classical and Quantum Gravity} {\bfseries 23} (2006) 5117–5128}.

\end{thebibliography}\endgroup

\end{document}